\documentclass[prd,aps,eqsecnum]{revtex4}
\usepackage{epsfig}
\usepackage{amsfonts}
\usepackage{rotating} 
\usepackage{sparticles}
\usepackage{dcolumn}
\usepackage{bm}
\usepackage{hyperref}
\hypersetup{
    bookmarks=true,         
    unicode=false,          
    pdftoolbar=true,        
    pdfmenubar=true,        
    pdffitwindow=false,     
    pdfstartview={FitH},    
    pdftitle={My title},    
    pdfauthor={Author},     
    pdfsubject={Subject},   
    pdfcreator={Creator},   
    pdfproducer={Producer}, 
    pdfkeywords={keyword1} {key2} {key3}, 
    pdfnewwindow=true,      
    colorlinks=true,       
    linkcolor=red,          
    citecolor=cyan,        
    filecolor=magenta,      
    urlcolor=green,           
    linktocpage=true
}
\usepackage{amsmath,amssymb,amsthm,graphicx,latexsym}
\usepackage{subfigure}
\usepackage{pstricks}
\usepackage{color}

\usepackage{mathrsfs}

\newcommand{\be}{\begin{equation}}
\newcommand{\ee}{\end{equation}}
\newcommand{\bea}{\begin{eqnarray}}
\newcommand{\eea}{\end{eqnarray}}
\newcommand{\bml}{\begin{subequations}}
\newcommand{\eml}{\end{subequations}}
\newcommand{\bfig}{\begin{figure}}
\newcommand{\efig}{\end{figure}}

\newcommand{\slashed}{\hspace{-1.1ex}/}

\begin{document}

\title{Features of warped geometry in presence of Gauss-Bonnet coupling}

\author{Sayantan Choudhury$^{1}$\footnote{Electronic address: {sayanphysicsisi@gmail.com}} ${}^{}$
and Soumitra SenGupta$^{2}$
\footnote{Electronic address: {tpssg@iacs.res.in
}} ${}^{}$}
\affiliation{$^1$Physics and Applied Mathematics Unit, Indian Statistical Institute, 203 B.T. Road, Kolkata 700 108, India\\
$^2$Department of Theoretical Physics,
Indian Association for the Cultivation of Science,
2A and 2B Raja S.C. Mullick Road,
Kolkata - 700 032, India.
}

\date{\today}
\begin{abstract}
 We study the role of the Gauss-Bonnet corrections and two loop higher genus contribution to the gravity
action on the Kaluza-Klien modes and their interactions for different bulk fields which enable one to study various
phenomenological implications of string loop corrected Gauss-Bonnet modified warped geometry model in one canvas.
We have explicitly derived a phenomenological bound on the Gauss-Bonnet parameter so that the required Planck to TeV scale hierarchy can be achieved through the
warp factor in the light of recently discovered Higgs like boson at $125$ GeV. Moreover due to the presence of small perturbative Gauss-Bonnet
 as well as string loop corrections we have shown that
the warping solution can be obtained for both de-Sitter and anti-de-Sitter bulk which is quite distinct from
Randall-Sundrum scenario. Finally we have evaluated various interactions among these bulk fields and determined the coupling parameters and the Kaluza- Klien
mode masses which is crucial
to understand the phenomenology of a string two loop corrected Einstein-Gauss-Bonnet warp geometry.

.

\end{abstract}


\maketitle
\tableofcontents
\section{\bf Introduction}
\label{intro}

Warp geometry models have been extensively studied in recent years from both 
theoretical and phenomenological perspectives. The Randall Sundrum braneworld model ( RS )\cite{lisa1,lisa2},
one of the pioneering warped geometry model, was proposed to resolve the long 
standing problem in connection with the fine tuning of the mass of
Higgs ( also known as gauge hierarchy or naturalness problem ) in an otherwise successful Standard Model of elementary particles.
RS model
has been studied extensively both in the context of collider physics 
\cite{lisa3,lisa4,lisa5,lisa6,lisa7,lisa8,lisa9,lisa10,lisa11,rizo1,rizo2,rizo3,rizo4,rizo5,
rizo6,rizo7,rizo8,rizo9,rizo10,rizo11,rizo12,rizo13,rizo14,rizo15,rizo16,rizo17,rizo18,rizo19,rizo20,rizo21,rizo22,
dal1,dal2,dal3,dal4,dal5,sayan4,gold1,gold2,gold3,har,ssg4,ssg5,ssg6,ssg7,ssg8,ssg9,ssga,war} as well as
cosmological physics \cite{saki,kar,sayan1,sayan2,sayan3,safi1,safi2,safi3,roy,gary,ssg1,ssg2,ssg3,fed1,saba,gio,hana}. 
In particle phenomenology, one of the important experimental signatures of such extra dimensional models is the search of the  Kaluza-Klien ( KK ) gravitons
in pp collision leading to dilepton decays in the Large Hadron Collider (LHC) \cite{lhc}. The couplings of the zero mode as well as the higher KK modes are
determined by assuming the standard model fields to be confined on a 3-brane located at an orbifold fixed point. Such a picture is rooted in a string-inspired model
where the standard model fields being open string-excitations are localized on a 3-brane. This led to the braneworld description of extra dimensional 
models with gravity only propagating in the bulk as a closed string excitation. But apart from graviton, string theory admits of various 
higher rank antisymmetric tensor excitations as 
closed string modes which can also propagate as a bulk field. It was found that remarkably such fields are heavily suppressed on the brane in such warped
geometry model and thus offers a possible explanation of invisibility of these fields in current experiments \cite{ssg10,ssg11,ssg12,ssg13}.
Subsequently going beyond the stringy description, the implications of the presence of standard model fields in the bulk were also studied
in different variants of  warped braneworld models. All these models in general assumed the 3-brane hypersurface to be flat.   
These models were subsequently generalized to include non-flat branes \cite{ssg14,ssg15,ssg16} and also braneworld with larger number of extra dimensions 
\cite{ssg17,ssg18,ssg19,ssg20,ssg21,ssg22,ssg23,ssg24}.  
  
From a theoretical standpoint, warped geometry model has its underlying motivation
in the backdrop of string theory where the throat geometry (Klevanov- Strassler) \cite{klev}
solution exhibits warping character. However the Randall-Sundrum models captures the essence of such warped geometry models
in a simple way and also drew the attention in the context of 
AdS/CFT correspondence as the Randall-Sundrum model is defined on a ${\bf AdS_{5}}$ slice \cite{tony,rick,sund,anco,aga,hakg,brag,kaya,lotu}.

While the Randall-Sundrum model starts with Einstein's gravity in ${\bf AdS_{5}}$ manifold in five dimensional
space-time, there have been efforts to include the higher curvature effects in the nature of the warped geometry \cite{giov,andr,pen,mav,dot,sayan3,sayan6}.
Such corrections originate naturally in string theory where power expansion in terms of inverse string tension 
yields the higher order corrections to pure Einstein's gravity \cite{gari,gasp,aseem,turner,cogo,fara}.
Supergravity, as the low energy limit of heterotic string theory yields the Gauss-Bonnet ( GB ) term as the leading order correction \cite{sayan3,sayan6,boul} and therefore 
became an active area of interest as a modified theory of gravity.
Various cosmological implications of  GB correction have been studied extensively in the context of slow roll inflationary models
 \cite{sayan3,sayan6,lid,ish,sami1,sami2}, initial singularities \cite{whel,yaji,ben,kei,tret,kanti},
tensor perturbations \cite{glei,dota,peri,zho,rami} etc. 
It has been shown that the positivity or negativity of the GB couplings as well as it's magnitude can be strongly constrained from 
the WMAP9 \cite{wmap9} and PLANCK \cite{planck} data. 
In a different context like black hole Physics it has been shown that GB correction suppresses graviton emission \cite{kony,zkon,davbou,pari,kasi,alex,marteen,magmic}
 and therefore the black hole becomes more stable \cite{cve,roma,marb,past,rong,koko,chan,wil}.
Moreover the correction to black hole entropy due to GB term has also been an active area of interest \cite{sen1,qi,genr,sen2,alisa,sen3,james1,fin,ohta,tims}. 
Thus the Gauss-Bonnet gravity as a modified gravity theory has been studied exhaustively in different scenarios \cite{nijori,nijori1,luca,hide,motu,pres,pal,binpan,defeli,jili}
as a first step to include the higher curvature effects over Einstein's gravity. In addition to this string theory admits of higher loop corrections \cite{gasp,jeans,rich,nick,gava} which is
a further modification on Einstein-Gauss-Bonnet correction on Einstein's gravity.

In this work we investigate the role of the GB corrections and two loop higher genus contribution to the gravity action on the KK- modes and their 
interactions for different bulk fields which would enable one to
study various phenomenological implications of string loop corrected Gauss-Bonnet modified warped geometry model in one canvas.

Considering the GB correction as a small perturbative correction over the original Einstein's action, we first derive the modified warp factor and brane tensions
in a $Z_2$ orbifolded model.
We show that though the warp factor looks similar to RS model but the warp factor parameter which in RS model depends on the bulk cosmological
constant now have solutions in two branches which are functions  of the GB parameter $\alpha_{(5)}$, string loop correction parameter $A_1$ and also depends on
the extra dimensional coordinate. We determine a bound on the GB parameter so that the required Planck to TeV scale hierarchy can be achieved through the
warp factor in the light of recently discovered Higgs like boson at $125$ GeV. Moreover due to the presence
of GB as well as string loop corrections, here we show that the warping solution can be obtained for both de-Sitter and anti-de-Sitter bulk.
This feature is quite distinct from RS-scenario which is defined only on an ${\bf AdS_{5}}$ bulk in order to achieve a warped solution. 
After a detailed analysis of the character of the solutions of the warp factor and brane tensions we proceed to evaluate the zero mode and 
KK mode excitations of bulk graviton along with the 5-dimensional supergravity extension with bulk gravitino. In both the cases we find the profile of the bulk
wave functions. We then extend our calculations with bulk scalar as well as bulk gauge field by addressing both abelian and non-abelian cases including dilaton coupling.
Among other closed string modes, string theory admits of various higher rank antisymmetric tensor fields which are also possible candidates for bulk fields.
We study the KK-modes and the profiles of the bulk wave functions for various antisymmetric tensor fields including the possible dilaton and axion couplings. 
It is followed by a detailed  analysis of  bulk fermions where the profile of both left and right chiral modes are determined in presence of
the GB extended gravity model.  
We then evaluate various interactions among these bulk fields and determine the coupling parameters which is  crucial to understand the 
phenomenology of a string two loop corrected Einstein-Gauss-Bonnet warp geometry.
Finally we conclude by summarizing our results.      


\section{\bf Einstein Gauss-Bonnet warped geometry model with string loop correction in a 5-dimensional bulk spacetime}
\label{model}
We start our discussion with a warped model 
on the topological bulk manifold ${\bf {\cal M}_{5}:=dS_{5}/AdS_{5}\otimes S^{5}}$. Consider the $Ads_{5}$ slice in a  two brane framework described by the following action:
\be\begin{array}{llllll}\label{action}
   \displaystyle  S_{(5)}=S_{EH}+S_{GB}+S_{loop}+S_{Bulk}+S_{Brane},
   \end{array}\ee
where the contribution from the gravity sector is given by the {\it Einstein Hilbert}, {\it Gauss-Bonnet}  and {\it string two-loop correction} \cite{gasp} coming from the interaction with
dilatonic degrees of freedom via the Conformal Field Theory (CFT) in the bulk geometry such that,
\be\begin{array}{llllll}\label{eh}
   \displaystyle  S_{EH}=\frac{M^{3}_{(5)}}{2}\int d^{5}x \sqrt{-g_{(5)}}R_{(5)},
   \end{array}\ee

\be\begin{array}{llllll}\label{gb}
   \displaystyle  S_{GB}=\frac{\alpha_{(5)}M_{(5)}}{2}\int d^{5}x \sqrt{-g_{(5)}}\left[R^{ABCD(5)}R^{(5)}_{ABCD}-4R^{AB(5)}R^{(5)}_{AB}+R^{2}_{(5)}\right],
   \end{array}\ee

\be\begin{array}{llllll}\label{loop}
   \displaystyle  S_{loop}=-\frac{\alpha_{(5)}A_{1}M_{(5)}}{2}\int d^{5}x \sqrt{-g_{(5)}}e^{\theta_{1}\phi}\left[R^{ABCD(5)}R^{(5)}_{ABCD}-4R^{AB(5)}R^{(5)}_{AB}+R^{2}_{(5)}\right]
   \end{array}\ee
with $A,B,C,D=0,1,2,3,4(Extra~Dimension)$ and a conformal two-loop coupling constant $A_{1}$. \\
Other contributions come from bulk and two brane sector which are  given as:
\be\begin{array}{llllll}\label{bulk}
   \displaystyle  S_{Bulk}=\int d^{5}x \sqrt{-g_{(5)}}\left[{\cal L}^{field}_{Bulk}-2\Lambda_{(5)}e^{\theta_{2}\phi}\right],
   \end{array}\ee

\be\begin{array}{llllll}\label{brane}
   \displaystyle  S_{Brane}=\int d^{5}x \sum^{2}_{i=1}\sqrt{-g^{(i)}_{(5)}}\left[{\cal L}^{field}_{(i)}-T_{(i)}e^{\theta_{2}\phi}\right]\delta(y-y_{(i)}).
   \end{array}\ee
The ${\cal L}^{field}_{Bulk}$ represents the bulk field Lagrangian which may include different spin fields such as ${\cal U}(1)$ abelian gauge fields, ${\cal SU}({\cal N})$
 non-abelian gauge fields, spin $1/2$ fermions,
 dilaton , pure bulk scalar , rank-3 ({\it Kalb-Rammond}) and rank-4 antisymmetric tensor fields. Throughout 
the article we use $\alpha_{(5)}$ as Gauss-Bonnet coupling, $A_{1}$ as two-loop conformal coupling and $(\theta_{1},\theta_{2})$ for dilatonic coupling. 
In equation(\ref{brane}) the brane index $i=(1[hidden~brane],2[visible~brane])$
and ${\cal L}^{field}_{(i)}$ represents brane Lagrangian which contains brane fields. The bulk cosmological constant $\Lambda_{(5)}$ couples to
the dilatonic degrees of freedom.

The background metric describing slice of the ${\bf dS_{5}/AdS_{5}}$ warped geometry is given by \cite{lisa1},
\be\begin{array}{llllll}\label{brane}
   \displaystyle ds^{2}_{(5)}=g_{AB}dx^{A}dx^{B}=e^{-2A(y)}\eta_{\alpha\beta}dx^{\alpha}dx^{\beta}+r^{2}_{c}dy^{2},
   \end{array}\ee
where $r_{c}=e^{-B_{0}}(\sim {\cal O}(1))$ is the dimensionless quantity in the Planckian unit representing the compactification radius of extra dimension 
in a $\frac{S^{1}}{Z_{2}}$ orbifolding
and it is expressed in terms of the stabilized radion $B_{0}$. Most importantly the compactification radius
is assumed to be independent of four dimensional coordinates (by Poincare invariance) and extra dimensional coordinate (fifth dimension). 
Here the orbifold points are $y_{i}=[0,\pi]$
and pereodic boundary condition is imposed in the closed interval $-\pi\leq y\leq\pi$.
After orbifolding, the size of the extra dimensional interval is $\pi r_{c}$. Moreover in the above metric ansatz
$e^{-2A(y)}$ represents the warp factor and the Minkowski flat metric $\eta_{\alpha\beta}=(-1,+1,+1,+1)$. This will lead 
to dimensional reduction of the manifold ${\bf dS_{5}\rightarrow \left({\cal {\overline{M}}}^{1,3}\otimes\frac{S^{1}}{Z_{2}}\right)}$ 
or ${\bf AdS_{5}\rightarrow \left({\cal M}^{1,3}\otimes\frac{S^{1}}{Z_{2}}\right)}$ depending on the signature of the bulk cosmological constant.

\section{\bf Warp factor and brane tension}

Varying the action stated in equation(\ref{action}) and neglecting the back reaction of all the other brane/bulk fields except gravity,
the five dimensional Bulk Einstein's equation turns out to be
\be\begin{array}{lllll}\label{eneqn}
    \displaystyle \sqrt{-g_{(5)}}\left[G^{(5)}_{AB}+\frac{\alpha_{(5)}}{M^{2}_{(5)}}\left(1-A_{1}e^{\theta_{1}\phi}\right)H^{(5)}_{AB}\right]
=-\frac{e^{\theta_{2}\phi}}{M^{3}_{(5)}}\left[\Lambda_{(5)} \sqrt{-g_{(5)}}g^{(5)}_{AB}+\sum^{2}_{i=1}T_{(i)}\sqrt{-g^{(i)}_{(5)}}g^{(i)}_{\alpha\beta}\delta^{\alpha}_{A}\delta^{\beta}_{B}\delta(y-y_{(i)})\right]
   \end{array}\ee
where the five dimensional Einstein's tensor and the Gauss-Bonnet tensor \cite{sayan3,sayan6} is given by

\be\begin{array}{llll}\label{et}
    G^{(5)}_{AB}=\left[R^{(5)}_{AB}-\frac{1}{2}g^{(5)}_{AB}R_{(5)}\right],
   \end{array}\ee

\be\begin{array}{llll}\label{gbp}
  H^{(5)}_{AB}=2R^{(5)}_{ACDE}R_{B}^{CDE(5)}-4R_{ACBD}^{(5)}R^{CD(5)}
-4R_{AC}^{(5)}R_{B}^{C(5)}+2R^{(5)}R_{AB}^{(5)}\\ ~~~~~~~~~~~~~~~~~~~~~~~~~~~~~~~~~~~~~~~~~~~-\frac{1}{2}g^{(5)}_{AB}
\left(R^{ABCD(5)}R^{(5)}_{ABCD}-4R^{AB(5)}R^{(5)}_{AB}+R^{2}_{(5)}\right).
   \end{array}\ee
Similarly varying equation(\ref{action}) with respect to the dilaton field the gravidilaton equation of motion turns out to be
\be\begin{array}{llll}\label{jk11}
\displaystyle \frac{\theta_{2}}{M^{2}_{(5)}}\sum^{2}_{i=1}T_{(i)}\sqrt{-g^{(i)}_{(5)}}e^{\theta_{2}\phi}\delta(y-y_{(i)})
=\sqrt{-g_{(5)}}\left\{\alpha_{(5)}A_{1}\theta_{1}\left[R^{ABCD(5)}R^{(5)}_{ABCD}-4R^{AB(5)}R^{(5)}_{AB}+R^{2}_{(5)}\right]
\right.\\ \left.~~~~~~~~~~~~~~~~~~~~~~~~~~~~~~~~~~~~~~~~~~~~~~~~~~~~~~~~~~~~~~~~~~~~~~~~~~~~~~~~~~~~~~~~~~~~~~~~~~~~~~~~~~\displaystyle 
+2\frac{\Lambda_{(5)}}{M^{2}_{(5)}}\theta_{2}e^{\theta_{2}\phi}+\frac{\Box_{(5)} \phi}{M_{(5)}}\right\}
   \end{array}\ee
where the five dimensional D'Alembertian operator is defined as $\Box_{(5)}\phi=\frac{1}{\sqrt{-g_{(5)}}}\partial_{A}\left(\sqrt{-g_{(5)}}\partial^{A}\phi\right)$.
Now from the equation(\ref{eneqn}) $(A=\alpha,B=\beta)$ component of the Einstein's equation can be written as:
\be\begin{array}{llll}\label{gbf}
  \displaystyle  \frac{1}{r^{2}_{c}}\left\{6\left(\frac{dA(y)}{dy}\right)^{2}-3\frac{d^{2}A(y)}{dy^{2}}-\frac{8\alpha_{(5)}}{M^{2}_{(5)}}\left(1-A_{1}e^{\theta_{1}\phi}\right)
\left[\left(\frac{d^{2}A(y)}{dy^{2}}\right)^{2}-2\frac{d^{2}A(y)}{dy^{2}}\left(\frac{dA(y)}{dy}\right)^{2}+\left(\frac{dA(y)}{dy}\right)^{4}\right]\right\}\\
\displaystyle +\frac{4\alpha_{(5)}}{r^{4}_{c}M^{2}_{(5)}}\left(1-A_{1}e^{\theta_{1}\phi}\right)\left\{19\frac{d^{2}A(y)}{dy^{2}}\left(\frac{dA(y)}{dy}\right)^{2}
-5\left(\frac{d^{2}A(y)}{dy^{2}}\right)^{2}-14\left(\frac{dA(y)}{dy}\right)^{4}\right\}\\ \displaystyle  
=-\frac{e^{\theta_{2}\phi}}{M^{3}_{(5)}}\left[\Lambda_{(5)}+\frac{1}{r_{c}}
\left(T_{(1)}\delta(y)+T_{(2)}\delta(y-\pi)\right) \right]
   \end{array}\ee
and from $(A=4,B=4)$ component we get
\be\begin{array}{llll}\label{ffge}
   \displaystyle  \frac{1}{r^{2}_{c}}\left\{6\left(\frac{dA(y)}{dy}\right)^{2}-\frac{208\alpha_{(5)}}{M^{2}_{(5)}}\left(1-A_{1}e^{\theta_{1}\phi}\right)\left(\frac{dA(y)}{dy}\right)^{4}\right\}
+\frac{144\alpha_{(5)}}{r^{4}_{c}M^{2}_{(5)}}\left(1-A_{1}e^{\theta_{1}\phi}\right)\left(\frac{dA(y)}{dy}\right)^{4}=-\frac{\Lambda_{(5)}e^{\theta_{2}\phi}}{M^{3}_{(5)}}.
 \end{array}\ee

To solve equation(\ref{gbf}) and equation(\ref{ffge}) we assume that the dilaton is 
weakly coupled to gravity (weak coupling $\theta_{1}$) and the bulk cosmological constant (weak coupling $\theta_{2}$)  since 
the Gauss-Bonnet coupling is an outcome of perturbative correction to gravity at the quadratic order. In this context dilaton is function of extra dimension only in the 
bulk. Using this fact, the gravidilaton equation stated in equation(\ref{jk11}) is simplified to the following expression:
\be\begin{array}{lllll}\label{simple}
 \displaystyle \alpha_{(5)}A_{1}\theta_{1}\left[-\frac{16}{r^{4}_{c}}\left\{\left(\frac{d^{2}A(y)}{dy^{2}}\right)-4\left(\frac{dA(y)}{dy}\right)^{2}\right\}^{2}
+\frac{16}{r^{2}_{c}}\left\{\left(\frac{d^{2}A(y)}{dy^{2}}\right)-\left(\frac{dA(y)}{dy}\right)^{2}\right\}^{2}
+\frac{16}{r^{4}_{c}}\left\{2\left(\frac{d^{2}A(y)}{dy^{2}}\right)-5\left(\frac{dA(y)}{dy}\right)^{2}\right\}^{2}\right]\\
\displaystyle ~~~~~~~~~~~~~~~~~~~~~~~~~~~~~~~~~~~~~~~~~~~~~~~~~~~~~~~+\frac{1}{M_{(5)}}\frac{d}{dy}\left(e^{-4A(y)}\frac{d\phi}{dy}\right)
+\frac{\theta_{2}e^{\theta_{2}\phi}}{M^{2}_{(5)}}\left[2\Lambda_{(5)}-\frac{1}{r_{c}}
\left(T_{(1)}\delta(y)+T_{(2)}\delta(y-\pi)\right) \right]=0
   \end{array}\ee

Now including the well known ${\bf Z_{2}}$ orbifolding symmetry 
at the leading order of $\theta_{1}$, $\theta_{2}$ and $\alpha_{(5)}$  we get
\be\begin{array}{llll}\label{gradilatonic}
 \displaystyle  \phi(y)=\sum^{2}_{p=1}\left(\frac{|y|}{\theta^{\frac{5}{2}}_{p}}+\frac{1}{\theta_{p}} \right) 
   \end{array}\ee
with ${\bf \left\{\frac{\theta_{p}}{\theta_{q}}\rightarrow 1\forall (p,q)\right\}}$ and the corresponding warp factor turns out to be
\be\begin{array}{llll}\label{warp}
   \displaystyle A(y):= A_{\pm}(y)=k_{\pm}r_{c}|y|\end{array}\ee
where 
\be\begin{array}{llll}\label{wsol}
\displaystyle k_{\pm}=\sqrt{\frac{3M^{2}_{(5)}}{16\alpha_{(5)}\left(1-A_{1}e^{\theta_{1}\phi}\right)}
\left\{1\pm \frac{r_{c}}{3M^{\frac{5}{2}}_{(5)}}\sqrt{\left[\frac{9M^{5}_{(5)}}{r^{2}_{c}}+\left(208
-\frac{144}{r^{2}_{c}}\right)\alpha_{(5)}\left(1-A_{1}e^{\theta_{1}\phi}\right)\Lambda_{(5)}e^{\theta_{2}\phi}\right]}\right\}}\end{array}\ee

along with a stringent constraint

\be\begin{array}{lllll}\label{cont}
   \displaystyle \alpha_{(5)}\Lambda_{(5)}\geq -\left[\frac{9M^{5}_{(5)}}{r^{2}_{c}\left(208
-\frac{144}{r^{2}_{c}}\right)\left(1-A_{1}e^{\theta_{1}\phi}\right)e^{\theta_{2}\phi}}\right]. 
   \end{array}\ee

It may be observed that though the warp factor looks similar to RS warp factor \cite{lisa1} but the parameter $k$ is now defined over two different
branches $k_{+}$ and $k_{-}$ and unlike RS scenario it is dependent on the extra dimensional coordinate $y$.
It can be easily observed that after taking $(\theta_{1},\theta_{2},A_{1},\alpha_{(5)})\rightarrow 0$ limit $k_{-}$ branch asymptotically reaches to Randall-Sundrum limit.
On the contrary the $k_{+}$ branch asymptotically diverges.

The brane tensions for the visible and hidden brane turn out to be
\be\begin{array}{lllll}\label{tension1}
\displaystyle   T^{\mp}_{\bf hid}:=T_{(1)}=\mp\left\{\Lambda_{(5)}r_{c}+6 k^{2}_{\pm}M^{3}_{(5)}r_{c}e^{-\theta_{2}\phi}\left[1-\frac{4\alpha_{(5)}
\left(1-A_{1}e^{\theta_{1}\phi}\right)}{3M^{2}_{(5)}}k^{2}_{\pm}\left(r^{2}_{c}+7\right)\right]\right\},\\
 \displaystyle   T^{\pm}_{\bf vis}:=T_{(2)}=\pm\left\{\Lambda_{(5)}r_{c}+ 6 k^{2}_{\pm}M^{3}_{(5)}r_{c}e^{-\theta_{2}\phi}\left[1-\frac{4\alpha_{(5)}
\left(1-A_{1}e^{\theta_{1}\phi}\right)}{3M^{2}_{(5)}}k^{2}_{\pm}\left(r^{2}_{c}+7\right)\right]\right\}.
 
\end{array}\ee

Furthermore the modified four dimensional effective Planck mass in presence of Gauss-Bonnet perturbative coupling 
is given by
\be\begin{array}{lllll}\label{planckmass}
 \displaystyle    M_{PL}:=M_{(4)}=\sqrt{M^{3}_{(5)}r_{c}\int^{+\pi}_{-\pi}dy~~ e^{-2k_{\pm}r_{c}|y|}}\\
~~~~~~~~~~~~~~~~~~\displaystyle =\frac{M^{\frac{3}{2}}}{\sqrt{k_{\pm}}}\sqrt{\left[1-e^{-2k_{\pm}r_{c}\pi}\right]}.
 \end{array}\ee

\begin{figure}[htb]
\centering
\subfigure[]{
    \includegraphics[width=8.5cm,height=7cm] {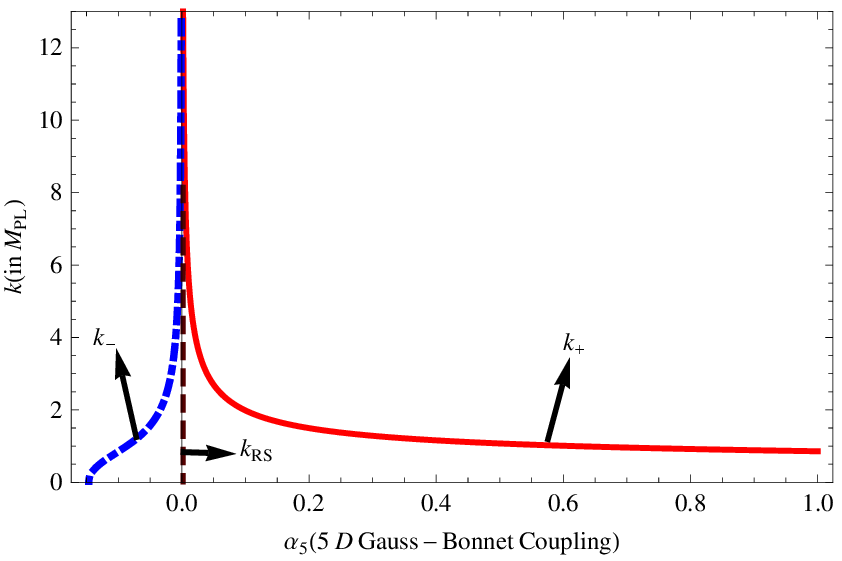}
    \label{fig:subfig1}
}
\subfigure[]{
    \includegraphics[width=8.5cm,height=7cm] {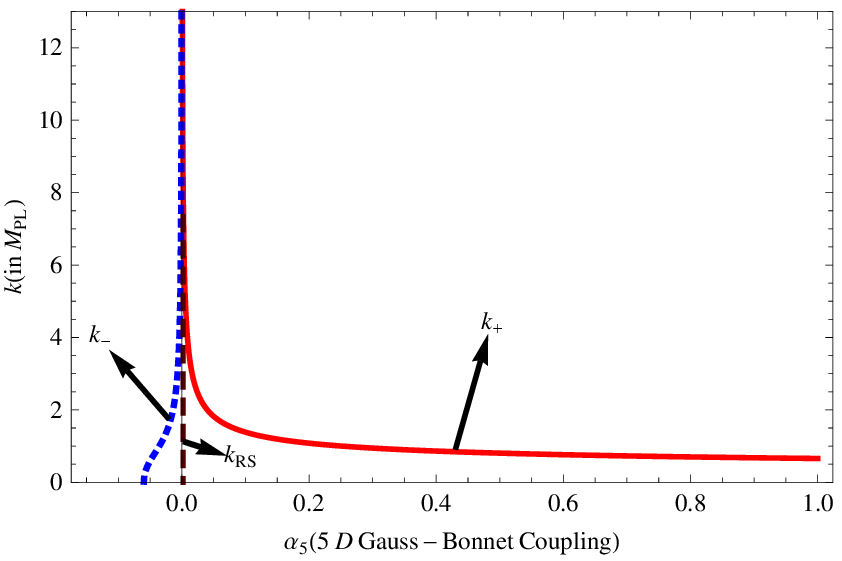}
    \label{fig:subfig2}
}
\subfigure[]{
    \includegraphics[width=8.5cm,height=7cm] {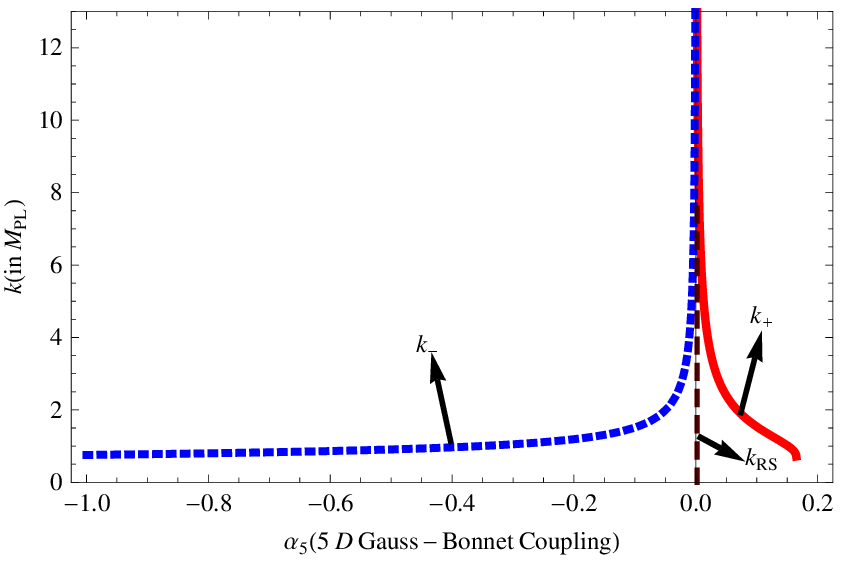}
    \label{fig:subfig3}
}
\subfigure[]{
    \includegraphics[width=8.5cm,height=7cm] {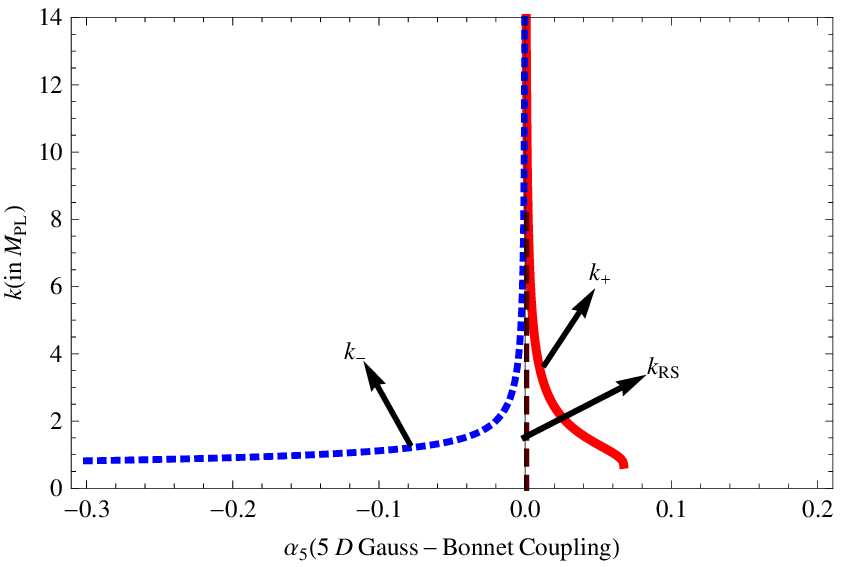}
    \label{fig:subfig4}
}
\caption[Optional caption for list of figures]{Variation
 of $k_{\pm}$  
vs Gauss-Bonnet coupling $\alpha_{(5)}$ for \subref{fig:subfig1} $\Lambda_{(5)}>0$ and $A_{1}>0$, \subref{fig:subfig2} $\Lambda_{(5)}>0$ and $A_{1}<0$, 
\subref{fig:subfig3} $\Lambda_{(5)}<0$ and $A_{1}>0$ and
\subref{fig:subfig4} $\Lambda_{(5)}<0$ and $A_{1}<0$. In this context $B_{0}=0.002$, $r_{c}=0.996\sim 1$, $|A_{1}|=0.04$, $\theta_{1}=0.05$ and $\theta_{2}=0.04$.}
\label{fig:subfigureExample51}
\end{figure}

\begin{figure}[htb]
\centering
\subfigure[]{
    \includegraphics[width=8.5cm,height=7cm] {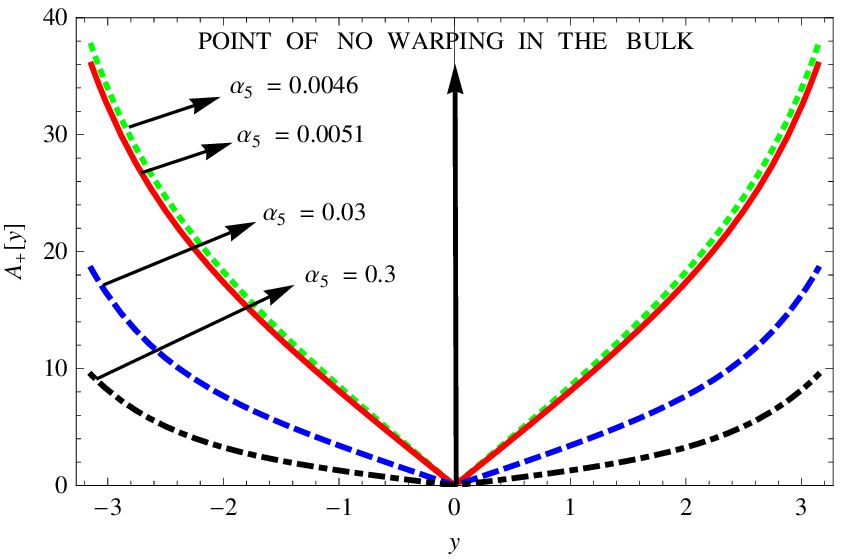}
    \label{fig:subfig5}
}
\subfigure[]{
    \includegraphics[width=8.5cm,height=7cm] {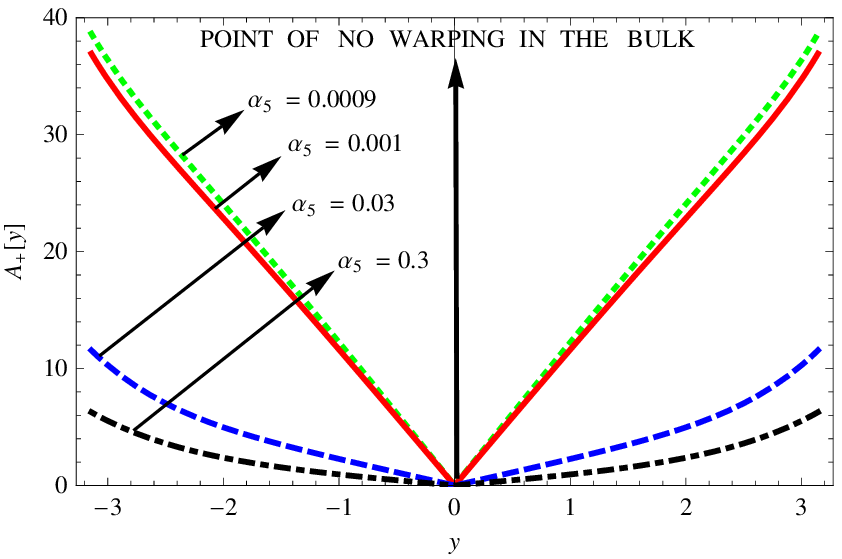}
    \label{fig:subfig6}
}
\subfigure[]{
    \includegraphics[width=8.5cm,height=7cm] {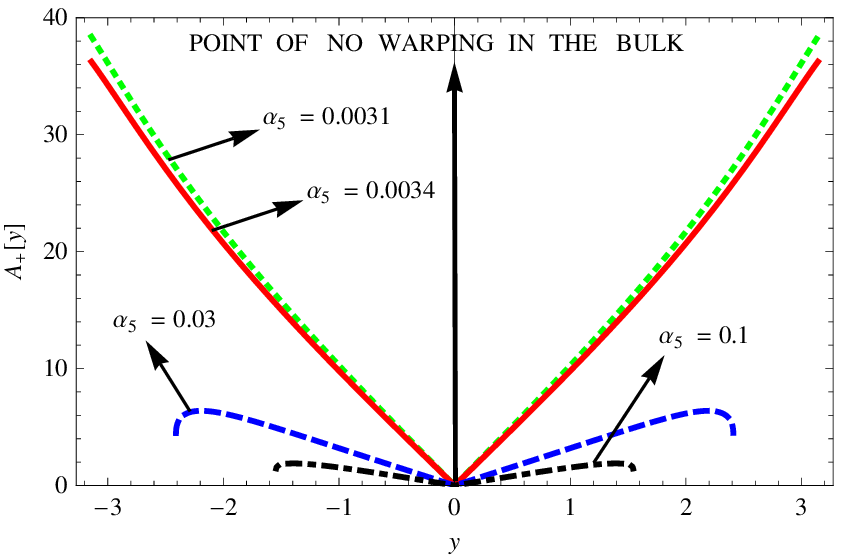}
    \label{fig:subfig7}
}
\subfigure[]{
    \includegraphics[width=8.5cm,height=7cm] {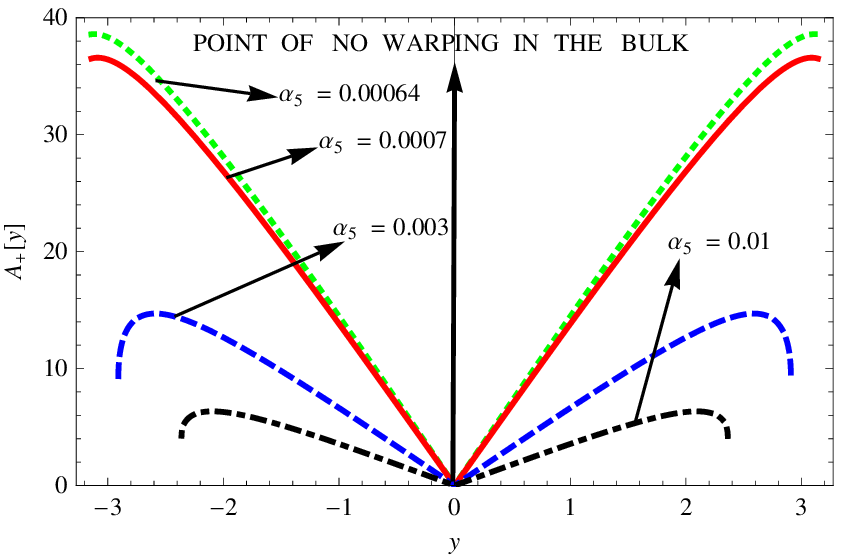}
    \label{fig:subfig8}
}
\caption[Optional caption for list of figures]{Variation
 of warp function $A_{+}$  
vs extra dimensional coordinate $y$ for \subref{fig:subfig5} $\Lambda_{(5)}>0$ and $A_{1}>0$, \subref{fig:subfig6} $\Lambda_{(5)}>0$ and $A_{1}<0$, 
\subref{fig:subfig7} $\Lambda_{(5)}<0$ and $A_{1}>0$ and
\subref{fig:subfig8} $\Lambda_{(5)}<0$ and $A_{1}<0$. In this context $B_{0}=0.002$, $r_{c}=0.996\sim 1$, $|A_{1}|=0.04$, $\theta_{1}=0.05$ and $\theta_{2}=0.04$ with three 
different sets of Gauss-Bonnet coupling coefficient $\alpha_{(5)}$. The region between $\textcolor{red}{red}$ and $\textcolor{green}{green}$ colored line corresponds to the recently discovered
Higgs like boson at 125 GeV.}
\label{fig:subfigureExample52}
\end{figure}
\begin{figure}[htb]
\centering
\subfigure[]{
    \includegraphics[width=8.5cm,height=7cm] {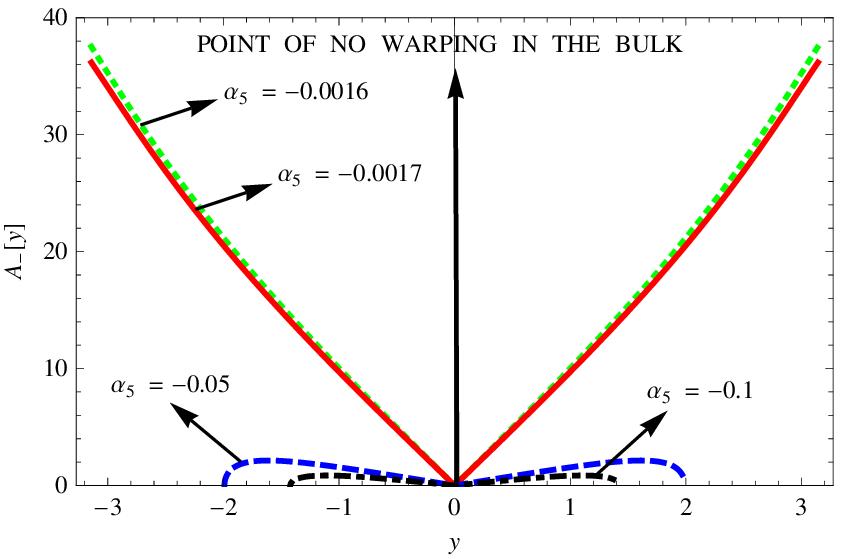}
    \label{fig:subfig9}
}
\subfigure[]{
    \includegraphics[width=8.5cm,height=7cm] {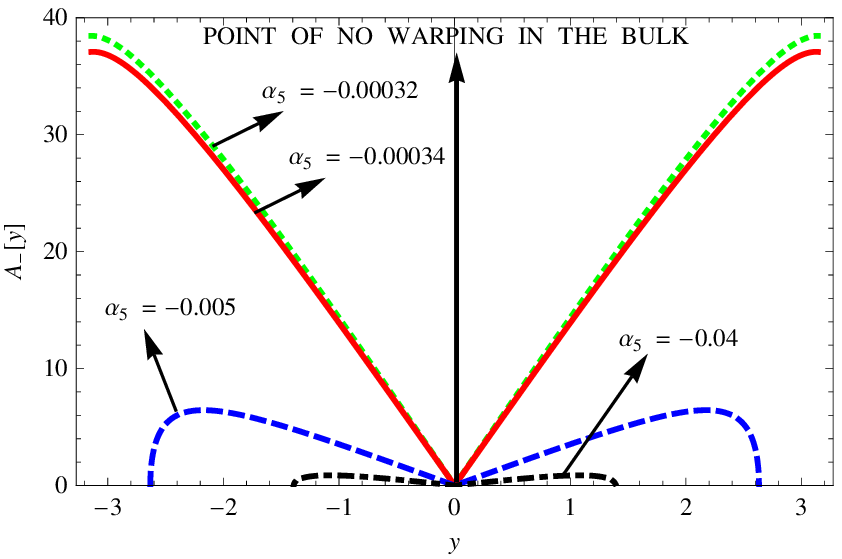}
    \label{fig:subfig10}
}
\subfigure[]{
    \includegraphics[width=8.5cm,height=7cm] {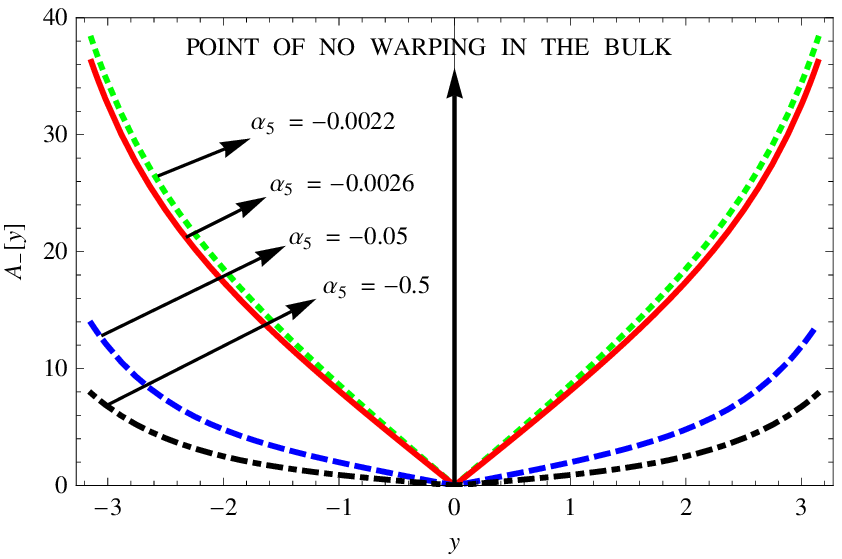}
    \label{fig:subfig11}
}
\subfigure[]{
    \includegraphics[width=8.5cm,height=7cm] {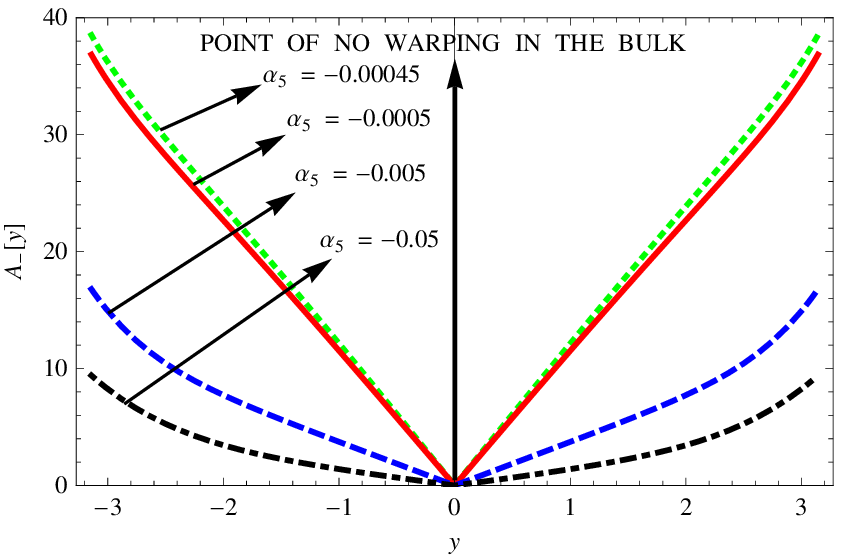}
    \label{fig:subfig12}
}
\caption[Optional caption for list of figures]{Variation
 of warp function $A_{-}$  
vs extra dimensional coordinate $y$ for\subref{fig:subfig9} $\Lambda_{(5)}>0$ and $A_{1}>0$, \subref{fig:subfig10} $\Lambda_{(5)}>0$ and $A_{1}<0$, 
\subref{fig:subfig11} $\Lambda_{(5)}<0$ and $A_{1}>0$ and
\subref{fig:subfig12} $\Lambda_{(5)}<0$ and $A_{1}<0$. In this context $B_{0}=0.002$, $r_{c}=0.996\sim 1$, $|A_{1}|=0.04$, $\theta_{1}=0.05$ and $\theta_{2}=0.04$
 with four 
different sets of Gauss-Bonnet coupling coefficient $\alpha_{(5)}$. The region between $\textcolor{red}{red}$ and $\textcolor{green}{green}$ colored line corresponds to the recently discovered
Higgs like boson at 125 GeV.}
\label{fig:subfigureExample53}
\end{figure}

\begin{figure}[htb]
\centering
\subfigure[]{
    \includegraphics[width=8.5cm,height=7cm] {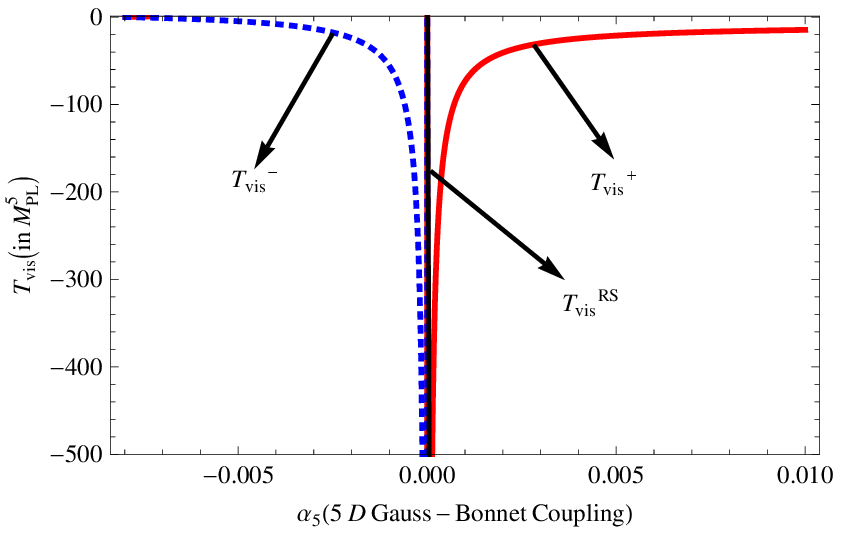}
    \label{fig:subfig1a}
}
\subfigure[]{
    \includegraphics[width=8.5cm,height=7cm] {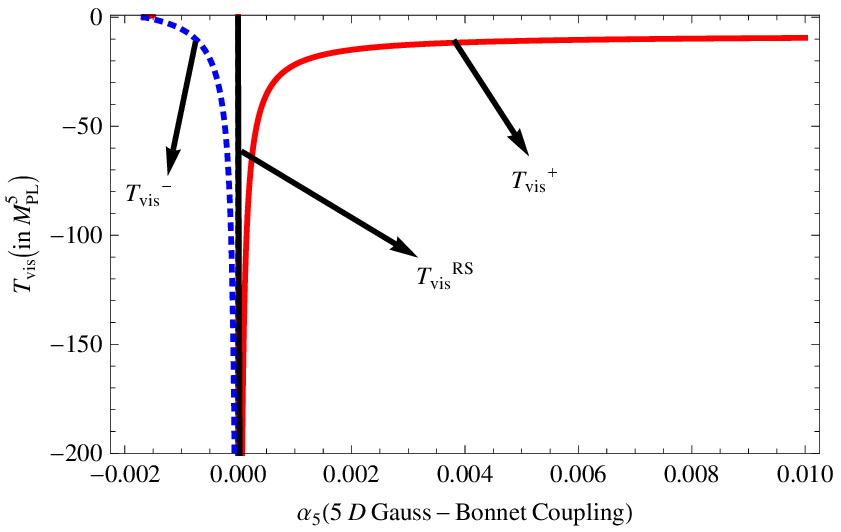}
    \label{fig:subfig2a}
}
\subfigure[]{
    \includegraphics[width=8.5cm,height=7cm] {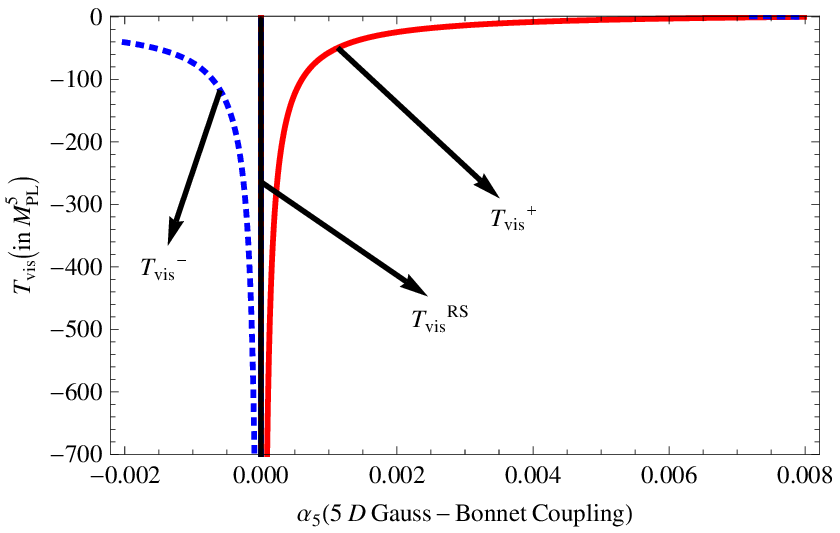}
    \label{fig:subfig3a}
}
\subfigure[]{
    \includegraphics[width=8.5cm,height=7cm] {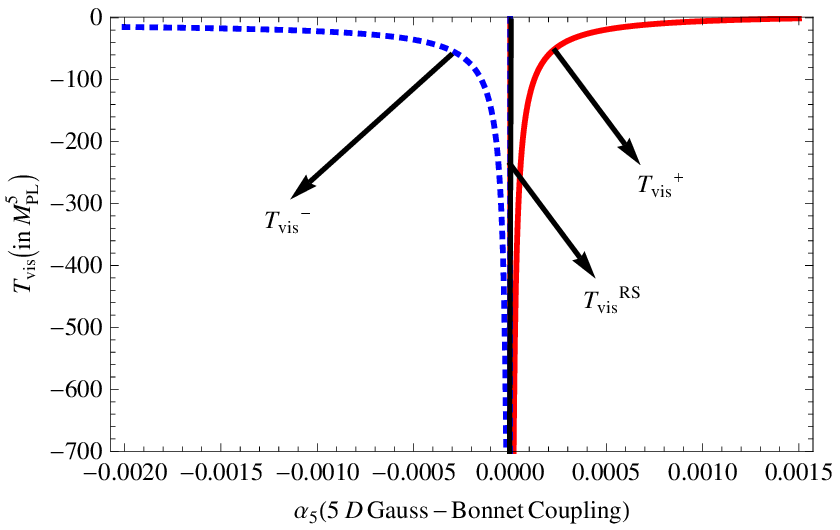}
    \label{fig:subfig4a}
}
\caption[Optional caption for list of figures]{Variation
 of visible brane tension $T_{vis}$  
vs Gauss-Bonnet coupling $\alpha_{(5)}$ for \subref{fig:subfig1a} $\Lambda_{(5)}>0$ and $A_{1}>0$, \subref{fig:subfig2a} $\Lambda_{(5)}>0$ and $A_{1}<0$, 
\subref{fig:subfig3a} $\Lambda_{(5)}<0$ and $A_{1}>0$ and
\subref{fig:subfig4a} $\Lambda_{(5)}<0$ and $A_{1}<0$. In this context we use 
$B_{0}=0.002$, $r_{c}=0.996\sim 1$, $|A_{1}|=0.04$, $\theta_{1}=0.05$ and $\theta_{2}=0.04$.}
\label{fig:subfigureExample51a}
\end{figure}

\begin{table}[htb]
\tiny
\begin{tabular}{|c|c|c|c|c|c|c|c|c|c|c|c|c|c|c|c|c|}
 \hline $\alpha_{(5)}(>0)$ & $\alpha_{(5)}(<0)$& $\Lambda_{(5)}$ & $A_{1}$ &
$\theta_{1}$&$\theta_{2}$&$r_{c}$&$k_{+}r_{c}$&$k_{-}r_{c}$& $A_{+}(\pi)$&$A_{-}(\pi)$
 \\
(for $k_{+}$ branch)&(for $k_{+}$ branch) & (in $M^{5}_{PL}$) &  &
(in $M^{-1}_{PL}$)&(in $M^{-1}_{PL}$)&(in $M^{-1}_{PL}$)& & & &
 \\
 \hline
0.00460~-~0.00510&(-0.00160)~-~(-0.00170) &1 & 0.04&0.05 &0.04 &0.996 &11.46~-~12.10 &11.52~-~12.10 &36.0~-~38.0 &36.2~-~38.0\\
\hline
0.00090~-~0.00100&(-0.00032)~-~(-0.00034) & 1&-0.04 &0.05 &0.04 &0.996 &11.68~-~12.38 &11.68~-~12.19 &36.7~-~38.9  &36.7~-~38.3\\
 \hline
0.00310~-~0.00340&(-0.00220)~-~(-0.00260) & -1& 0.04&0.05 &0.04 &0.996 &11.49~-~12.19 &11.49~-~12.16 &36.1~-~38.3  &36.1~-~38.2\\
 \hline
0.00064~-~0.00070&(-0.00045)~-~(-0.00050) &-1 &-0.04 &0.05 &0.04 &0.996 &11.52~-~12.13 &11.71~-~12.38 &36.2~-~38.1  &36.8~-~38.9\\
\hline
\end{tabular}
\caption{Allowed parameter space for $k_{+}$ and $k_{-}$ branch to produce Planck to TeV scale warping. }\label{tab100}
\end{table}

In figure(\ref{fig:subfigureExample51}) we have plotted the behavior of characteristic parameter $k_{\pm}$ with respect to Gauss-Bonnet coupling $\alpha_{(5)}$ for all 
possible signatures of five dimensional bulk cosmological constant $\Lambda_{(5)}$ and two-loop conformal coupling coefficient $A_{1}$. For $\Lambda_{(5)}>0$ 
the $k_{-}$ solution touches the $\alpha_{(5)}$ axis at zero for $\alpha_{(5)}<0$ in 
figure(\ref{fig:subfigureExample51}\subref{fig:subfig1}) and figure(\ref{fig:subfigureExample51}\subref{fig:subfig2}) which is physically redundant. The only features that are 
accepted $\Lambda_{(5)}<0$ situation where
the $k_{-}$ solution is asymptotic in nature for $\alpha_{(5)}<0$ are clearly exhibited in figure(\ref{fig:subfigureExample51}\subref{fig:subfig3})
 and figure(\ref{fig:subfigureExample51}\subref{fig:subfig4}).  It is evident from the 
figure(\ref{fig:subfigureExample51}\subref{fig:subfig1}) and figure(\ref{fig:subfigureExample51}\subref{fig:subfig2}) is that 
the $k_{+}$ solution shows the asymptotic behavior for $\Lambda_{(5)}>0$, $\alpha_{(5)}>0$ and non-zero for rest of the two situations. It is important to mention 
here that for all situations in figure(\ref{fig:subfigureExample51}) $\alpha_{(5)}\rightarrow 0$ shows the well known Randal-Sundrum feature 
for $k_{-}$ branch ($k_{-}\rightarrow k_{RS}$). On the other hand in the same limit $k_{+}$ branch asymptotically diverges.
The overall parameter space satisfies
the criteria $k_{\pm}r_{c}\simeq 12$  and $A_{\pm}(\pm\pi)\simeq 36$, which is a necessary requirement  to solve the well known gauge hierarchy problem in the two brane set up
are explicitly mentioned in table(\ref{tab100}).

It is interesting to observe from the figures(\ref{fig:subfigureExample51}\subref{fig:subfig1}-\ref{fig:subfigureExample51}\subref{fig:subfig4}) that both  $k_{\pm}$ decreases with increase in the GB parameter $\alpha_{(5)}$ which would cause a fall in the warping through the
warp factor unless the value of the modulus $r_c$ is changed accordingly. So if one wants to resolve the gauge hierarchy problem then a little hierarchy will enter through
$r_c$ ( which is now greater than the RS value ) due to the non-vanishing value of the GB coupling $\alpha_{(5)}$. We shall see it's implications in the subsequent sections.

In figure(\ref{fig:subfigureExample52}) explicitly shows the behavior of positive warp function $A_{+}(y)$ with respect to the coordinate characterizing the 
extra dimension $y$ for all 
possible signatures of five dimensional bulk cosmological constant $\Lambda_{(5)}$ and two-loop conformal coupling coefficient $A_{1}$. In this context we 
use three positive fixed values of the Gauss-Bonnet coupling $\alpha_{(5)}$ to explain the behavior of $A_{+}(y)$. It is important to mention here that 
that the point $y=0$ satisfies the no warping condition which is evident from equation(\ref{warp}).
If we tune the numerical value of Gauss-Bonnet
coupling to a small value then $A_{+}(\pm\pi)\simeq 36$ which can easily solve the well known hierarchy problem. As we increase the strength of the 
coupling then we see that $A_{+}(\pm\pi)<36$ and it is no longer possible to address the gauge hierarchy problem due to 
Gauss-Bonnet correction. This feature directly sets the constraint on Gauss-Bonnet coupling strength appearing as a perturbative correction on  
Einstein-Hillbert term in the gravity action.

Similarly figure(\ref{fig:subfigureExample53}) explicitly shows the behavior of negative warp function $A_{-}(y)$ with respect to the coordinate characterizing the 
extra dimension $y$ for all 
possible signatures of five dimensional bulk cosmological constant $\Lambda_{(5)}$ and two-loop conformal coupling coefficient $A_{1}$. 
The graphical behavior of negative warp function is significantly different from the positive one. But it is clear from the 
above diagrams that both the positive and negative bulk cosmological constant 
can solve the hierarchy problem in the perturbative regime. In this connection the most significant 
result comes from figure(\ref{fig:subfigureExample52}) and figure(\ref{fig:subfigureExample53})
 the tiny parameter space between the \textcolor{red}{red} and \textcolor{green}{green} colored curve
corresponds to the recently observed Higgs like scalar at 125 GeV.

The graphical behavior of visible brane tension with respect to the Gauss- Bonnet Coupling 
for all signatures of five dimensional bulk cosmological constant $\Lambda_{(5)}$ and two-loop conformal coupling co-efficient $A_{1}$
 is explicitly shown in figure(\ref{fig:subfigureExample51a}). In the $\alpha_{(5)}\rightarrow 0$ limit 
visible brane tension asymptotically follows Randall-Sundrum feature. It is important to mention here that 
in figure(\ref{fig:subfigureExample51a}) we have maintained the restriction 
for $k_{+}$ branch $\alpha_{(5)}>0$ and for $k_{-}$ branch $\alpha_{(5)}<0$ which is strictly valid throughout our article. Consequently the visible brane tension is 
always negative for negative Gauss-Bonnet coupling followed by the $k_{+}$ branch and positive signature of 
the Gauss-Bonnet coupling followed by the $k_{-}$ branch.



\section{\bf Analysis of Bulk Kaluza-Klien Spectrum and their coupling for different bulk fields in presence of Gauss-Bonnet Coupling}
\label{dreduc}
In this section we elaborately discuss the technical details of the dimensional reduction technique of the bulk fields appearing in the bulk action 
via Kaluza-Klien spectrum analysis in presence of perturbative Gauss-Bonnet coupling in modified Randall-Sundrum scenario.


\subsection{\bf Bulk Graviton Field}
\label{graviton}

In this context we are interested to find out the Kaluza-Klien spectrum of spin-2 bulk graviton field.
To explore the characteristic features of bulk graviton we rescale the four dimensional counterpart of the five dimensional Randall-Sundrum metric stated in equation(\ref{brane}) 
in presence of
perturbative Gauss-Bonnet coupling $\alpha_{(5)}$. This induces the tensor perturbation in the gravity sector via the fluctuation through graviton degrees of freedom.
Such spin-2 field content is the 
essential ingredient in the context of phenomenology of extra dimension studied from 
Randall-Sundrum two brane scenario. This picture is constructed out of the underlying assumption that gravity is 
the only candidate which propagates in the bulk via the bulk graviton. In this section we will concentrate solely on the graviton degrees of freedom. After rescaling
the old four dimensional counterpart of the five dimensional metric can be recast as 
\be\begin{array}{llll}\label{new}
    \displaystyle g_{\alpha\beta}=e^{-2A_{\pm}(y)}\left[\eta_{\alpha\beta}+{\cal K}_{(5)}{\bf h}_{\alpha\beta}(x,y)\right]
   \end{array}\ee
where ${\cal K}_{(5)}:=\frac{2}{M^{\frac{3}{2}}_{(5)}}$ represents the expansion parameter for tensor perturbation.
Here in the context of perturbative graviton field theory we also use the fact that such expansion 
parameter is much smaller than unity.
Consequently the total resulting metric for the tensor perturbation is given by
\be\begin{array}{llll}\label{totmetricd}
    \displaystyle ds^{2}_{(5);new}=e^{-2A_{\pm}(y)}\left[\eta_{\alpha\beta}+{\cal K}_{(5)}{\bf h}_{\alpha\beta}(x,y)\right]dx^{\alpha}dx^{\beta}
+r^{2}_{c}dy^{2}.
   \end{array}\ee
Plugging the new metric stated in equation(\ref{totmetricd}) in the Einstein-Hilbert action via five dimensional Ricci scalar 
the five dimensional perturbative action for graviton can be written as
\be\begin{array}{llll}\label{abshg1g}
 \tiny  \displaystyle S_{{EH}}=\frac{1}{2}\int d^{4}x\int^{+\pi}_{-\pi}dy~r_{c}e^{-4A_{\pm}(y)}\left\{1+{\cal K}_{(5)}~Tr({\bf h}_{\alpha\beta}(x,y))
+\frac{{\cal K}^{2}_{(5)}}{2}
\left[\left(Tr({\bf h}_{\alpha\beta}(x,y))\right)^{2}-Tr({\bf h}^{2}_{\alpha\beta}(x,y))\right]\right.\\ \left.~~~~~~~~~~\displaystyle
+{\cal K}^{3}_{(5)}{\bf h}_{\alpha}^{[\alpha}(x,y){\bf h}_{\beta}^{\beta}(x,y)
{\bf h}_{\gamma}^{\gamma]}(x,y)\right\}\displaystyle\left[\frac{1}{r^{2}_{c}}\left(-\frac{1}{2}\overrightarrow{D}_{y}\left\{e^{2A_{\pm}(y)}\left(\eta^{\alpha\beta}
+{\cal K}_{(5)}{\bf h}^{\alpha\beta}(x,y)\right)\left[\eta_{\beta\alpha}\overrightarrow{D}_{y}\left(e^{-4A_{\pm}(y)}\right)\right.\right.\right.\right.\\
\left.\left.\left.\left.~~~~~~~~~~\displaystyle +{\cal K}_{(5)}
\overrightarrow{D}_{y}\left(e^{-4A_{\pm}(y)}{\bf h}_{\beta\alpha}(x,y)\right)\right]\right\}-\displaystyle\frac{1}{4}e^{4A_{\pm}(y)}\left(\eta^{\alpha\lambda}
+{\cal K}_{(5)}{\bf h}^{\alpha\lambda}(x,y)\right)\left(\eta^{\beta\rho}
+{\cal K}_{(5)}{\bf h}^{\beta\rho}(x,y)\right)\right.\right.\\ \left.\left.
~~\displaystyle\times\left\{\eta_{\lambda\beta}\overrightarrow{D}_{y}\left(e^{-2A_{\pm}(y)}\right)+{\cal K}_{(5)}
\overrightarrow{D}_{y}\left(e^{-2A_{\pm}(y)}{\bf h}_{\lambda\beta}(x,y)\right)\right\}
\left\{\eta_{\rho\alpha}\overrightarrow{D}_{y}\left(e^{-2A_{\pm}(y)}\right)+{\cal K}_{(5)}
\overrightarrow{D}_{y}\left(e^{-2A_{\pm}(y)}{\bf h}_{\rho\alpha}(x,y)\right)\right\}\right)\right.\\ \left.\displaystyle~~~~~~~
+\displaystyle e^{2A_{\pm}(y)}\left(\eta^{\alpha\beta}
+{\cal K}_{(5)}{\bf h}^{\alpha\beta}(x,y)\right)\left\{-\frac{1}{2r^{2}_{c}}\left\{\eta_{\beta\alpha}\overrightarrow{D}^{2}_{y}
\left(e^{-2A_{\pm}(y)}\right)+{\cal K}_{(5)}
\overrightarrow{D}^{2}_{y}\left(e^{-2A_{\pm}(y)}{\bf h}_{\beta\alpha}(x,y)\right)\right\}
\right.\right.\\ \left.\left.\displaystyle~~~~+\frac{{\cal K}_{(5)}}{2}\partial_{\gamma}\left\{\left(\eta^{\gamma\lambda}
+{\cal K}_{(5)}{\bf h}^{\gamma\lambda}(x,y)\right)\left[\partial_{\beta}{\bf h}_{\alpha\lambda}(x,y)+\partial_{\alpha}{\bf h}_{\lambda\beta}(x,y)-
\partial_{\lambda}{\bf h}_{\alpha\beta}(x,y)\right]\right\}-\frac{{\cal K}_{(5)}}{2}\partial_{\beta}\left\{\left(\eta^{\gamma\lambda}
+{\cal K}_{(5)}{\bf h}^{\gamma\lambda}(x,y)\right)\right.\right.\right.\\ \left.\left.\left.\displaystyle~~~~\times\displaystyle\left[\partial_{\gamma}{\bf h}_{\alpha\lambda}(x,y)+\partial_{\alpha}{\bf h}_{\lambda\gamma}(x,y)-
\partial_{\lambda}{\bf h}_{\alpha\gamma}(x,y)\right]\right\}+\displaystyle\frac{e^{2A_{\pm}(y)}}{4r^{2}_{c}}\left(\eta^{\lambda\alpha^{'}}
+{\cal K}_{(5)}{\bf h}^{\lambda\alpha^{'}}(x,y)\right)\right.\right.\\ \left.\left.
~~\displaystyle\times\left\{\eta_{\alpha\lambda}\overrightarrow{D}_{y}\left(e^{-2A_{\pm}(y)}\right)+{\cal K}_{(5)}
\overrightarrow{D}_{y}\left(e^{-2A_{\pm}(y)}{\bf h}_{\alpha\lambda}(x,y)\right)\right\}
\left\{\eta_{\beta\alpha^{'}}\overrightarrow{D}_{y}\left(e^{-2A_{\pm}(y)}\right)+{\cal K}_{(5)}
\overrightarrow{D}_{y}\left(e^{-2A_{\pm}(y)}{\bf h}_{\beta\alpha^{'}}(x,y)\right)\right\}\right.\right.\\ \left.\left.~~~~
-\displaystyle\frac{e^{2A_{\pm}(y)}}{4r^{2}_{c}}\left(\eta^{\lambda\delta}
+{\cal K}_{(5)}{\bf h}^{\lambda\delta}(x,y)\right)\left\{\eta_{\beta\lambda}\overrightarrow{D}_{y}\left(e^{-2A_{\pm}(y)}\right)+{\cal K}_{(5)}
\overrightarrow{D}_{y}\left(e^{-2A_{\pm}(y)}{\bf h}_{\beta\lambda}(x,y)\right)\right\}
\left\{\eta_{\delta\alpha}\overrightarrow{D}_{y}\left(e^{-2A_{\pm}(y)}\right)\right.\right.\right.\\ \left.\left.\left.\displaystyle~~~~+{\cal K}_{(5)}
\overrightarrow{D}_{y}\left(e^{-2A_{\pm}(y)}{\bf h}_{\delta\alpha}(x,y)\right)\right\}
-\displaystyle\frac{e^{2A_{\pm}(y)}}{4r^{2}_{c}}\left(\eta^{\lambda\delta}
+{\cal K}_{(5)}{\bf h}^{\lambda\delta}(x,y)\right)\left\{\eta_{\delta\beta}\overrightarrow{D}_{y}\left(e^{-2A_{\pm}(y)}\right)
\right.\right.\right.\\ \left.\left.\left.\displaystyle~~~~+{\cal K}_{(5)}
\overrightarrow{D}_{y}\left(e^{-2A_{\pm}(y)}{\bf h}_{\delta\beta}(x,y)\right)\right\}
\left\{\eta_{\lambda\alpha}\overrightarrow{D}_{y}\left(e^{-2A_{\pm}(y)}\right)+{\cal K}_{(5)}
\overrightarrow{D}_{y}\left(e^{-2A_{\pm}(y)}{\bf h}_{\lambda\alpha}(x,y)\right)\right\}-\displaystyle\frac{{\cal K}^{2}_{(5)}}{4}\left(\eta^{\lambda\delta}
+{\cal K}_{(5)}{\bf h}^{\lambda\delta}(x,y)\right)\right.\right.\\ \left.\left.\displaystyle\times\left(\eta^{\eta\rho}
+{\cal K}_{(5)}{\bf h}^{\eta\rho}(x,y)\right)\left[\partial_{\beta}{\bf h}_{\eta\delta}(x,y)+\partial_{\eta}{\bf h}_{\delta\beta}(x,y)-
\partial_{\delta}{\bf h}_{\eta\beta}(x,y)\right]\left[\partial_{\lambda}{\bf h}_{\alpha\rho}(x,y)+\partial_{\alpha}{\bf h}_{\rho\lambda}(x,y)-
\partial_{\rho}{\bf h}_{\lambda\alpha}(x,y)\right]\right.\right.\\ \left.\left.
\displaystyle~~~~+\displaystyle\frac{{\cal K}^{2}_{(5)}}{4}\left(\eta^{\lambda\delta}
+{\cal K}_{(5)}{\bf h}^{\lambda\delta}(x,y)\right)\left(\eta^{\eta\rho}
+{\cal K}_{(5)}{\bf h}^{\eta\rho}(x,y)\right)\left[\partial_{\beta}{\bf h}_{\eta\delta}(x,y)+\partial_{\eta}{\bf h}_{\delta\beta}(x,y)-
\partial_{\delta}{\bf h}_{\eta\beta}(x,y)\right]\right.\right.\\\left.\left.~~~~~~~~~~~~~~~~~~~~~~~~~~~~~~~~~~~~~~~~~~~~~~~~~~~~~~~~~~~~~~~~~~~~~~~~~~~~~~~~~~~~~
\times\displaystyle\left[\partial_{\lambda}{\bf h}_{\alpha\rho}(x,y)+\partial_{\alpha}{\bf h}_{\rho\lambda}(x,y)-
\partial_{\rho}{\bf h}_{\lambda\alpha}(x,y)\right]\displaystyle\right\}\right]
   \end{array}\ee

where we introduce a new symbol $\overrightarrow{{\cal D}_{y}}:=\frac{d}{dy}$. \\
Let the Kaluza-Klien expansion
of the spin-2 graviton field is given by
\be\begin{array}{lllll}\label{KK41g}
   \displaystyle {\bf h}_{\alpha\beta}(x,y)=\sum^{\infty}_{n=0}{\bf h}^{(n)}_{\alpha\beta}(x)~\frac{\chi^{(n)}_{\pm;\bf G}(y)}{\sqrt{r_{c}}}.
   \end{array}\ee
Now plugging equation(\ref{KK41g}) in equation(\ref{abshg1g}) and including the transverse and traceless criteria of the graviton given by
\be\begin{array}{llll}\label{trtrlass}
   \displaystyle {\bf h}^{\mu~(n)}_{~\mu}=0=\eta^{\alpha\beta}{\bf h}^{(n)}_{\alpha\beta},\\
\displaystyle  \partial^{\mu}{\bf h}^{(n)}_{\mu\nu}=0=\eta^{\alpha\beta}\partial_{\alpha}{\bf h}^{(n)}_{\beta\gamma} 
   \end{array}\ee
the leading order contribution to the effective four dimensional action reduces to the following expression:

\be\begin{array}{llll}\label{as1}
\displaystyle {\bf S}^{EH}\simeq
\frac{M^{3}_{(5)}{\cal K}^{2}_{(5)}}{2}\int d^{4}x\sum^{\infty}_{n=0}
{\bf h}^{\alpha\beta~(n)}(x){\bf h}^{(n)}_{\alpha\beta}(x)\left(m^{{\bf G}}_{n}\right)^{2}_{\pm}\end{array}\ee

In this context we impose the following orthonormalization condition of extra dimension dependent wave functions
\be\begin{array}{llll}\label{no121g}
 \displaystyle   \int^{+\pi}_{-\pi}dy~e^{-2A_{\pm}(y)}~\chi^{(m)}_{\pm;\bf G}(y)~\chi^{(n)}_{\pm;\bf G}(y)=\delta^{mn}
   \end{array}\ee
The mass term of the graviton field is defined through the following differential equation as
\be\begin{array}{llll}\label{dif181g}
  \displaystyle   -\frac{1}{r^{2}_{c}}\overrightarrow{{\cal D}_{y}}\left(e^{-4A_{\pm}(y)}\overrightarrow{{\cal D}_{y}}\chi^{(n)}_{\pm;\bf G}(y)\right)=
e^{-2A_{\pm}(y)}\left(m^{{\bf G}}_{n}\right)^{2}_{\pm}\chi^{(n)}_{\pm;\bf G}(y).
   \end{array}\ee
Introducing two new quantities $z^{\pm;{\bf G}}_{n}:=\frac{\left(m^{{\bf G}}_{n}\right)_{\pm}}{k_{\pm}}e^{A_{\pm}(y)}$ and 
$f^{n}_{\pm;\bf G}:=e^{-2A_{\pm}(y)}\chi^{(n)}_{\pm;\bf G}$ the
equation(\ref{dif181g}) can be recast in terms of Bessel differential equation of order two as
\be\begin{array}{llll}\label{dif281g}
    \displaystyle  \left[\left(z^{\pm;{\bf G}}_{n}\right)^{2}\overrightarrow{{\cal D}^{2}}_{z^{\pm;{\bf G}}_{n}}
+z^{\pm;{\bf G}}_{n}\overrightarrow{{\cal D}}_{z^{\pm;{\bf G}}_{n}}+\left\{\left(z^{\pm;{\bf G}}_{n}\right)^{2}-4\right\}\right]f^{n}_{\pm;\bf G}
=0
   \end{array}\ee

The analytical solution of this equation turns out to be
\be\begin{array}{llll}\label{sol1xc1dg}
\displaystyle    \chi^{(n)}_{\pm;\bf G}(y)=\frac{e^{2A_{\pm}(y)}}{{\cal N}^{\pm;\bf G}_{(n)}}\left[{\cal J}_{2}
(z^{\pm;{\bf G}}_{n})+\alpha^{\pm;\bf G}_{n}{\cal Y}_{2}(z^{\pm;{\bf G}}_{n})\right].
   \end{array}\ee
Here ${\cal N}^{\pm;\bf G}_{(n)}$ is the normalization constant of the extra dimension dependent wave function and $\alpha^{\pm;\bf G}_{n}$ is the
integration constant to be determined from the orthonormalization condition and the continuity conditions at the orbifold fixed point.
Self-adjointness and hermiticity of the differential operator appearing in equation(\ref{dif281g})
demands that $\overrightarrow{{\cal D}_{y}}\chi^{(n)}_{\pm;\bf G}(y)$ is
continuous at the orbifold fixed points $y_{i}=0,\pi$. Consequently we have
\be\begin{array}{llll}\label{cond1zx1dg}
 \displaystyle  \overrightarrow{{\cal D}_{y}}\chi^{(n)}_{\pm;\bf G}|_{y_{i}=0}=0~~\implies \alpha^{\pm;\bf G}_{n}=
\frac{\left[\frac{\left(m^{{\bf G}}_{n}\right)_{\pm}}{k_{\pm}}{\cal J}^{'}_{2}\left(\frac{\left(m^{{\bf G}}_{n}\right)_{\pm}}{k_{\pm}}\right)
+{\cal J}_{2}\left(\frac{\left(m^{{\bf G}}_{n}\right)_{\pm}}{k_{\pm}}\right)\right]}{\left[2{\cal Y}_{2}\left(\frac{\left(m^{{\bf G}}_{n}\right)_{\pm}}{k_{\pm}}\right)
+\frac{\left(m^{{\bf G}}_{n}\right)_{\pm}}{k_{\pm}}{\cal Y}^{'}_{2}\left(\frac{\left(m^{{\bf G}}_{n}\right)_{\pm}}{k_{\pm}}\right)\right]}.
   \end{array}\ee

\be\begin{array}{llll}\label{cond2zx1dg}
 \displaystyle  \overrightarrow{{\cal D}_{y}}\chi^{(n)}_{\pm;{\bf G}}|_{y_{i}=\pi}=0~~\implies \alpha^{\pm;{\bf G}}_{n}=
\frac{\left[{\cal J}_{2}\left(x^{\pm;{\bf G}}_{n}\right)
+x^{\pm;{\bf G}}_{n}{\cal J}^{'}_{2}\left(x^{\pm;{\bf G}}_{n}\right)\right]}{\left[x^{\pm;{\bf G}}_{n}{\cal Y}^{'}_{2}\left(x^{\pm;{\bf G}}_{n}\right)
+{\cal Y}_{2}\left(x^{\pm;{\bf G}}_{n}\right)\right]}
   \end{array}\ee
where $z^{\pm;{\bf G}}_{n}(\pi):=x^{\pm;{\bf G}}_{n}=\frac{\left(m^{{\bf G}}_{n}\right)_{\pm}}{k_{\pm}}e^{k_{\pm}r_{c}\pi}$.
For $e^{k_{\pm}r_{c}\pi}\gg 1,~\frac{\left(m^{{\bf G}}_{n}\right)_{\pm}}{k_{\pm}}\ll 1$ the mass spectrum for the graviton field
is expected to be of the order of TeV scale i.e.
\be\begin{array}{llll}\label{approxcv1dg}
\displaystyle    \alpha^{\pm;{\bf G}}_{n}\simeq-\frac{\pi}{4}\left(x^{\pm;{\bf G}}_{n}\right)^{2}e^{-2k_{\pm}r_{c}\pi}. 
\end{array}\ee

Now using equation(\ref{approxcv1dg}) and equation(\ref{cond1zx1dg}) we get
\be\begin{array}{llllll}\label{rootzxg}
 \displaystyle    \frac{\pi}{4}\left(x^{\pm;{\bf G}}_{n}\right)^{2}e^{-2k_{\pm}r_{c}\pi}
= \frac{\left[{\cal J}_{2}\left(x^{\pm;{\bf G}}_{n}\right)
+x^{\pm;{\bf G}}_{n}{\cal J}^{'}_{2}\left(x^{\pm;{\bf G}}_{n}\right)\right]}{\left[{\cal Y}_{2}\left(x^{\pm;{\bf G}}_{n}\right)
+x^{\pm;{\bf G}}_{n}{\cal Y}^{'}_{2}\left(x^{\pm;{\bf G}}_{n}\right)
\right]} ~~~~~~~~~~~
\displaystyle \Rightarrow {\cal J}_{1}\left(x^{\pm;{\bf G}}_{n}\right)\simeq -\frac{\pi}{4}\left(x^{\pm;{\bf G}}_{n}\right)^{2}e^{-2k_{\pm}r_{c}\pi}
{\cal Y}_{1}\left(x^{\pm;{\bf G}}_{n}\right)\approx 0
   \end{array}\ee
which is a transcendental equation of $x^{\pm;{\bf G}}_{n}$ and the roots of this equation gives the graviton field mass spectrum 
$\left(m^{{\bf G}}_{n}\right)_{\pm}$ in presence of perturbative Gauss-Bonnet coupling $\alpha_{(5)}$.
This leads to approximately the various Kaluza-Klien mode masses for the graviton field as,
\be\begin{array}{llll}\label{massasdgrav}
   \displaystyle \left(m^{{\bf G}}_{n}\right)_{\pm}\approx
 \left(n+\frac{1}{2}\mp\frac{1}{4}\right)\pi k_{\pm}e^{-k_{\pm}r_{c}\pi}.
   \end{array}\ee
\begin{figure}[ht]
\centering
\subfigure[]{
    \includegraphics[width=8.5cm,height=7cm] {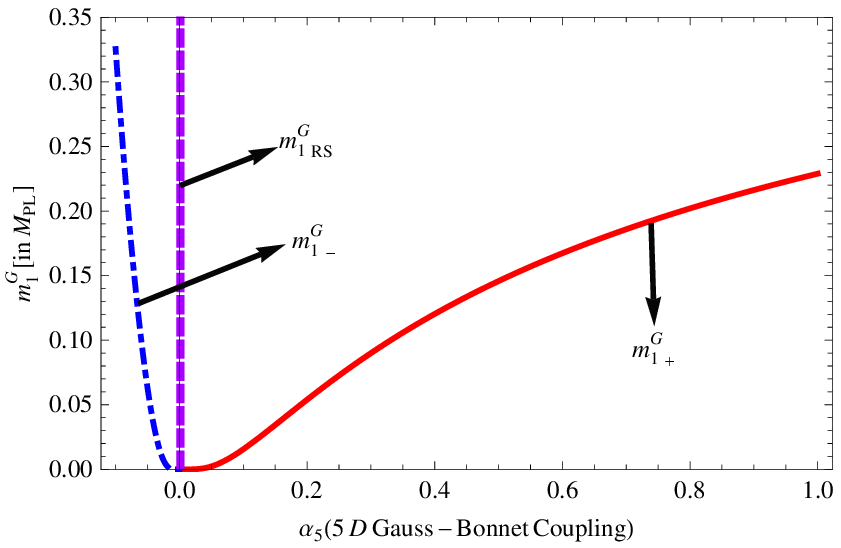}
    \label{fig:subfig1a}
}
\subfigure[]{
    \includegraphics[width=8.5cm,height=7cm] {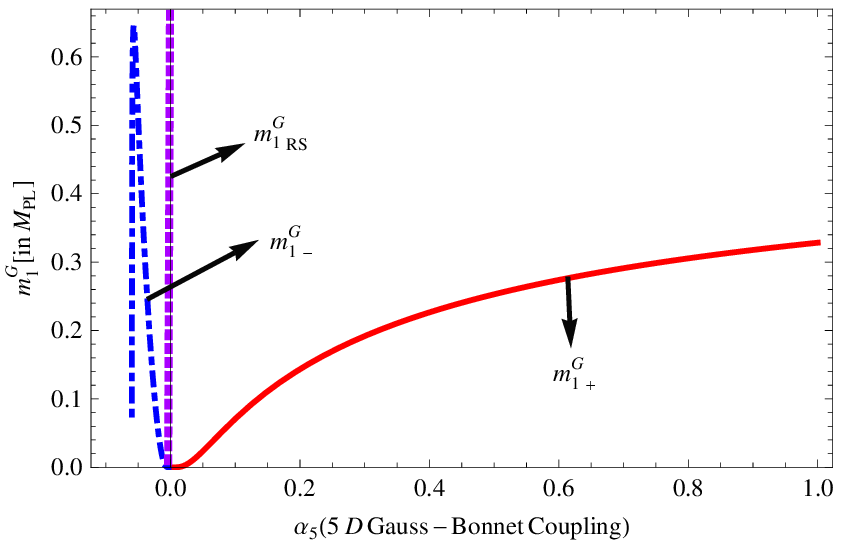}
    \label{fig:subfig2a}
}
\subfigure[]{
    \includegraphics[width=8.5cm,height=7cm] {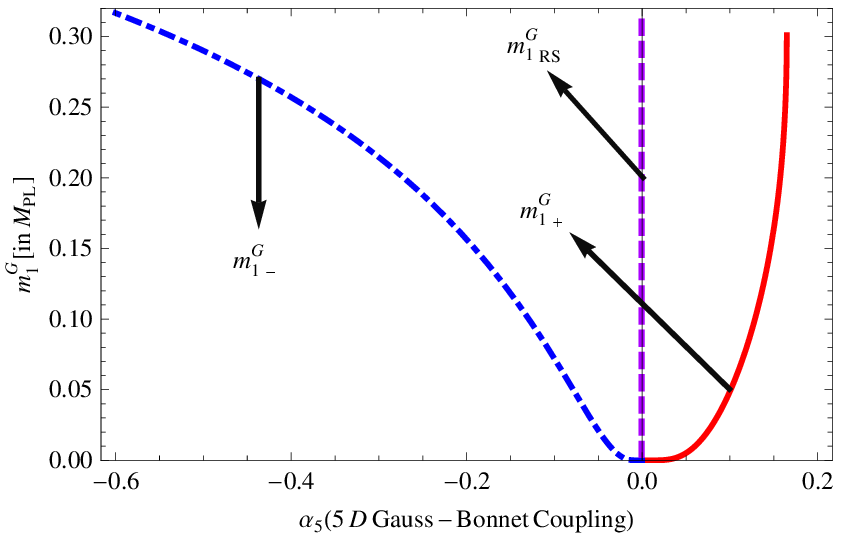}
    \label{fig:subfig3a}
}
\subfigure[]{
    \includegraphics[width=8.5cm,height=7cm] {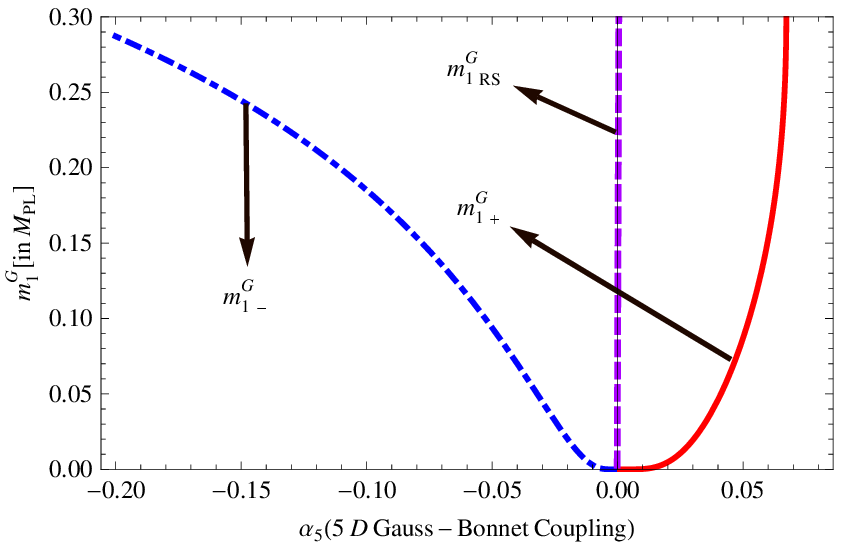}
    \label{fig:subfig4a}
}
\caption[Optional caption for list of figures]{Variation
 of Kaluza Klien graviton mass $(m^{G}_{1})_{\pm}$  
vs Gauss-Bonnet coupling $\alpha_{(5)}$ in the first excited state 
for \subref{fig:subfig1a} $\Lambda_{(5)}>0$ and $A_{1}>0$, \subref{fig:subfig2a} $\Lambda_{(5)}>0$ and $A_{1}<0$, 
\subref{fig:subfig3a} $\Lambda_{(5)}<0$ and $A_{1}>0$ and
\subref{fig:subfig4a} $\Lambda_{(5)}<0$ and $A_{1}<0$. In this context we use 
$B_{0}=0.002$, $r_{c}=0.996\sim 1$, $|A_{1}|=0.04$, $\theta_{1}=0.05$ and $\theta_{2}=0.04$.}
\label{fig:subfigureExample51a}
\end{figure}

This shows that for both $A_{+}$ and $A_{-}$, unless the Gauss-Bonnet coupling $\alpha_{(5)}$ is 
sufficiently small i.e. much much smaller than 0.005, the warp factor suppression will not be sufficient to
produce warping of the oder of $10^{-16}$, and hence, then graviton KK-modes will be much higher than the TeV
scale and beyond the scope of detection in LHC. The graphical behavior of KK mass spectrum of graviton mode in the 
first excited state with respect to the Gauss-Bonnet coupling $\alpha_{(5)}$ is explicitly shown in figure(\ref{fig:subfigureExample51a}) for all possible signatures
of cosmological constant $\Lambda_{(5)}$ and the conformal factor $A_{1}$ appearing in the string loop correction for the two branches of warping solution.

Now using equation(\ref{no121g}) the normalization constant for $n\neq 0$ mode reduces to the following expression
\be\begin{array}{llll}\label{vcvc1g}
    \displaystyle {\cal N}^{\pm;{\bf G}}_{(n)}=\frac{e^{k_{\pm}r_{c}\pi}}{
\sqrt{k_{\pm}r_{c}}}\sqrt{\left\{\left[{\cal J}_{2}\left(x^{\pm;{\bf G}}_{n}\right)+\alpha^{\pm;{\bf G}}_{n}{\cal Y}_{2}\left(x^{\pm;{\bf G}}_{n}\right)
\right]^{2}-e^{-2k_{\pm}r_{c}}\left[{\cal J}_{2}\left(x^{\pm;{\bf G}}_{n}e^{-k_{\pm}r_{c}\pi}
\right)+\alpha^{\pm;{\bf G}}_{n}{\cal Y}_{2}\left(x^{\pm;{\bf G}}_{n}e^{-k_{\pm}r_{c}\pi}\right)
\right]^{2}\right\}}. 
   \end{array}\ee
For $e^{k_{\pm}r_{c}\pi}\gg 1,~\frac{\left(m^{{\bf G}}_{n}\right)_{\pm}}{k_{\pm}}\ll 1$ the integration constant $\alpha^{\pm;{\bf G}}_{n}\ll 1$.
Consequently ${\cal Y}_{2}(z^{\pm;{\bf G}}_{n})$ is neglected compared to ${\cal J}_{2}(z^{\pm;{\bf G}}_{n})$ in equation(\ref{sol1xc1dg}) 
and then the normalization constant for $n\neq 0$ mode
turns out to be
\be\begin{array}{llll}\label{vcvc1gg}
    \displaystyle {\cal N}^{\pm;{\bf G}}_{(n)}=\frac{e^{k_{\pm}r_{c}\pi}}{
\sqrt{k_{\pm}r_{c}}}{\cal J}_{2}\left(x^{\pm;{\bf G}}_{n}\right). 
   \end{array}\ee
Consequently the extra dimensional dependent wave function for $n\neq 0$ is, 
\be\begin{array}{llll}\label{sol1xc1g}
\displaystyle    \chi^{(n)}_{\pm;\bf G}(y)=\frac{\sqrt{k_{\pm}r_{c}}~e^{2A_{\pm}(y)}}{e^{k_{\pm}r_{c}\pi}}\frac{{\cal J}_{2}
(z^{\pm;{\bf G}}_{n})}{{\cal J}_{2}
(x^{\pm;{\bf G}}_{n})}.
   \end{array}\ee

For massless $n=0$ mode the solution of the equation(\ref{dif181g}) is
\be\begin{array}{lllll}\label{jkloss1dg}
   \displaystyle  \chi^{(0)}_{\pm;{\bf G}}=\frac{C_{1}}{4k_{\pm}r_{c}}e^{4A_{\pm}(y)}+C_{2}.
   \end{array}\ee
Here $C_{1}$ and $C_{2}$ are arbitrary integration constants. Now applying the boundary condition through the continuity
of the wave function we get $C_{1}=0$. As a result the zero mode solution turns out to be $ \chi^{(0)}_{\pm;{\cal \phi}}=C_{2}$.
Now applying the normalization condition the ground state massless zero mode wave function turns out to be
\be\begin{array}{llll}\label{mlsxg}
   \displaystyle  \chi^{(0)}_{\pm;{\bf G}}=C_{2}=\sqrt{k_{\pm}r_{c}}.
   \end{array}\ee
The ground state obtained for graviton for our set up is exactly same as the massless graviton obtained in the context of Randall-Sundrum scenario.

\subsection{\bf Gravitino Field}
\label{gravitino}
We now address the supersymmetric version of the above model in a 5-dimensional supergravity framework then 
the onshell supergravity multiplet consists of the vierbein ($V^{\alpha}_{M}$), the graviphoton degrees of freedom ($B_{M}$) and 
two simplectic -Majorana gravitinos $\left({\bf \Psi}^{(j)}_{\bf sG}\right)_{P}$ with $i=1,2$.
In a ${\bf dS_{(5)}/AdS_{(5)}}$ slice the five dimensional action for the spin~$\frac{3}{2}$ supersymmetric gravitino field in the context of ${\cal N}$ = 1
supergravity can be written as \cite{nem,tony3,poko},
\be\begin{array}{llll}\label{abshg1gr}
   \displaystyle S_{{\bf \Psi_{sG}}}=-\frac{i M^{3}_{(5)}}{2}\int d^{5}x~Det({\cal V})~\sum^{2}_{i=1}\sum^{2}_{j=1}\left[
\left(\bar{\bf \Psi}^{(i)}_{\bf sG}\right)_{M}{\bf \Upsilon}^{MNP}\overleftrightarrow{{\bf D}}^{\bf sG}_{N}
\left({\bf \Psi}^{(j)}_{\bf sG}\right)_{P}\delta_{ij}-\frac{3}{2}\overrightarrow{D}_{y}A_{\pm}(y)\left(\bar{\bf \Psi}^{(i)}_{\bf sG}\right)_{M}{\bf \Upsilon}^{MN}
\Sigma^{ij}_{3}\left({\bf \Psi}^{(j)}_{\bf sG}\right)_{P}\right]
   \end{array}\ee
where the
index $i$ and $j$ label the fundamental representation of the ${\cal SU}(2)_{R}$ automorphism group of the ${\cal N}$ = 1
supersymmetry algebra in five dimensions. In this context 
\be\begin{array}{llll}\label{perm}
  \displaystyle {\bf \Upsilon}^{MNP}:=\frac{1}{3!}\gamma^{[M}\otimes\gamma^{N}\otimes\gamma^{P]}=\sum_{p=permutaion}
\frac{(-1)^{p}}{3!}\gamma^{M}\otimes\gamma^{N}\otimes\gamma^{P},\\
 \displaystyle {\bf \Upsilon}^{MN}:=\frac{1}{2!}\gamma^{[M}\otimes\gamma^{N]}=\sum_{p=permutaion}
\frac{(-1)^{p}}{2!}\gamma^{M}\otimes\gamma^{N}
   \end{array}\ee
are the antisymmetrized tensor product of five dimensional gamma matrices. 
Without loosing any physical information here we choose a physical gauge in which the graviphoton degrees of freedom vanishes.
The gravitino supersymmetry transformation is
given by
\be\begin{array}{llll}\label{abse1grgr}
   \displaystyle \delta\left({\bf \Psi}^{(i)}_{\bf sG}\right)_{P}(x,y)=\sum^{2}_{j=1}\left(
\overrightarrow{{\bf D}}^{\bf sG}_{N}\delta^{ij}+\frac{1}{2}\overrightarrow{D}_{y}A_{\pm}(y)\gamma_{N}\Sigma^{ij}_{3}\right){\bf \vartheta}^{j}(x,y)
   \end{array}\ee
where $\Sigma_{3}=diag(1,-1)$ and ${\bf \vartheta}^{i}$ is the symplectic-Majorana spinor which represents a parameter of supergravity
transformation. We define the ${\bf Z}_{2}$ transformation of the symplectic -Majorana spinor as 
\be\begin{array}{llll}\label{cxgr}
   \displaystyle  {\bf \vartheta}^{i}(x,\bf y)\xrightarrow{{\bf Z}_{2}}{\bf \vartheta}^{i}(x,\bf-y):
=\sum^{2}_{j=1}\Sigma^{ij}_{3}\gamma_{5}{\bf \vartheta}^{j}(\bf y)
   \end{array}\ee
for which local supersymmetry is intact due to 
$\delta\left({\bf \Psi}^{(i)}_{\bf sG}\right)_{P}=0$
subject
to the Killing condition

\be\begin{array}{llll}\label{kill}
   \displaystyle \overrightarrow{{\bf D}}^{\bf sG}_{N}{\bf \vartheta}^{i}(x,y)=-\frac{1}{2}\overrightarrow{D}_{y}A_{\pm}(y)\sum^{2}_{j=1}\gamma_{N}\Sigma^{ij}_{3}{\bf \vartheta}^{j}(x,y)
   \end{array}\ee
which is always valid in non compact ${\bf AdS_{(5)}}$ bulk space. But after imposing ${\bf \frac{S^{(1)}}{Z_{2}}}$
orbifold symmetry the surface term satisfies an extra condition $\gamma_{5}{\bf \vartheta}^{i}=\sum^{2}_{j=1}\sigma^{ij}_{3}{\bf \vartheta}^{j}$ which implies that after
orbifold compactification we have ${\cal N}$=1 supergravity theory instead of ${\cal N}$=2 supergravity.

The Kaluza-Klien expansion
of the five dimensional gravitino $\left({\bf \Psi}^{(j)}_{\bf sG}\right)_{P}$ and five dimensional
supergravity parameter (${\bf \vartheta}^{i}$) are given by
\be\begin{array}{lllll}\label{KK41gr1}
   \displaystyle \left({\bf \Psi}_{{\bf L,R};\bf sG}\right)_{\mu}(x,y)=\sum^{\infty}_{n=0}\left({\bf \Psi}^{(n)}_{{\bf L,R};\bf sG}\right)_{\mu}(x)
~\frac{\chi^{(n);{\bf L,R}}_{\pm;\bf sG}(y)}{\sqrt{r_{c}}}
   \end{array}\ee

\be\begin{array}{lllll}\label{KK41gr2}
   \displaystyle \left({\bf \Psi}_{{\bf L,R};\bf sG}\right)_{4}(x,y)=\sum^{\infty}_{n=0}\left({\bf \Psi}^{(n)}_{{\bf L,R};\bf sG}\right)_{4}(x)
~\frac{\chi^{(n);{\bf 4~L,R}}_{\pm;\bf sG}(y)}{\sqrt{r_{c}}}
   \end{array}\ee

 \be\begin{array}{lllll}\label{KK41gr3}
   \displaystyle {\bf \vartheta}_{{\bf L,R}}(x,y)=\sum^{\infty}_{n=0}{\bf \vartheta}^{(n)}_{{\bf L,R}}(x)
~\frac{\chi^{(n);{\bf L,R}}_{\pm;\bf sG}(y)}{\sqrt{r_{c}}}
   \end{array}\ee
where henceforth we have omitted the ${\cal SU}(2)_{R}$ index $i$ since we only consider $i=1$. The remaining $i=2$ component is 
obtained by imposing symplectic-Majorana criteria. Here ${\bf L,R}$ stands for left and right chiral five dimensional gravitino
which is responsible for chiral flipping along with a overall signature under the action of the chiral matrix $\gamma_{5}$ i.e. 
$\gamma_{5}\left({\bf \Psi}_{{\bf L,R};\bf sG}\right)=\gamma_{5}{\cal P}_{\bf L,R}\left({\bf \Psi}_{\bf sG}\right)=\pm\left({\bf \Psi}_{{\bf R,L};\bf sG}\right)$.
Throughout our analysis including the contribution from Gauss-Bonnet coupling we use $\overrightarrow{D}_{y}A_{\pm}(y)=k_{\pm}r_{c}sgn(y)$
and $\overrightarrow{D}^{2}_{y}A_{\pm}(y)=2k_{\pm}r_{c}\overrightarrow{D}_{y}sgn(y)=2k_{\pm}r_{c}\left[\delta(y)-\delta(y-\pi)\right]$.
Consequently the supergravity transformation for $i=1$ turns out to be
\be\begin{array}{lll}\label{sugtr1}
   \displaystyle \delta\left({\bf \Psi}_{\bf sG}\right)_{\mu}(x,y)= \left(\overrightarrow{\partial}_{\mu}+k_{\pm}r_{c}sgn(y)\gamma_{\mu}{\cal P}_{\bf L}\right){\bf \vartheta}(x,y),\end{array}\ee
\be\begin{array}{lll}\label{sugtr2}\displaystyle \delta\left({\bf \Psi}_{\bf sG}\right)_{4}(x,y)= \left(\overrightarrow{\partial}_{4}+\frac{1}{2}k_{\pm}r_{c}sgn(y)\gamma_{5}\right)
{\bf \vartheta}(x,y).
   \end{array}\ee

After substituting the Kaluza-Klien expansion stated in equation(\ref{KK41gr1}) and equation(\ref{KK41gr3}) in equation(\ref{sugtr1}) the 
supergravity transformation for nth gravitino mode reduces to the following expressions:
\be\begin{array}{lll}\label{sugtr3}
   \displaystyle \delta\left({\bf \Psi}^{(n)}_{{\bf L};\bf sG}\right)_{\mu}(x)
=\sum^{\infty}_{k=0}\left(\delta^{nk}\overrightarrow{\partial}_{\mu}{\bf \vartheta}^{(k)}_{\bf R}(x)+\tilde{\gamma}_{\mu}{\bf d}^{(nk)}{\bf \vartheta}^{(k)}_{\bf L}(x)\right),\\
\displaystyle \delta\left({\bf \Psi}^{(n)}_{{\bf R};\bf sG}\right)_{\mu}(x)
=\sum^{\infty}_{k=0}\delta^{nk}\overrightarrow{\partial}_{\mu}{\bf \vartheta}^{(k)}_{\bf R}(x).
   \end{array}\ee

where $\tilde{\gamma}_{\mu}$ is the four dimensional Minkowski gamma matrix. The expansion coefficients appearing as 
an outcome of dimensional reduction takes the following form:
\be\begin{array}{llll}\label{excof}
    \displaystyle {\bf d}^{(nk)}:=k_{\pm}r_{c}\int^{+\pi}_{-\pi}dy~sgn(y)e^{-2A_{\pm}(y)}\chi^{(n);{\bf L}}_{\pm;{\bf sG}}(y)\chi^{(k);{\bf R}}_{\pm;{\bf sG}}(y).
   \end{array}\ee
Consequently we have 
\be\begin{array}{llll}\label{jklabel}
 \displaystyle   \chi^{(n);{\bf 4~L,R}}_{\pm;{\bf sG}}(y)=\frac{1}{\left(m^{{\bf sG}}_{n}\right)_{\pm}}\left(\pm \overrightarrow{\partial}_{4}+\frac{k_{\pm}r_{c}}{2}sgn(y)\right)
  \chi^{(n);{\bf L,R}}_{\pm;{\bf sG}}(y)
   \end{array}\ee
and the ${\cal N}$=1 supergravity transformation for the fifth component of the gravitino field in presence of Gauss-Bonnet coupling 
can be recast in terms of the Kaluza-Klien modes as
\be\begin{array}{llll}\label{wfnun}
\displaystyle  \delta\left({\bf \Psi}^{(n)}_{{\bf L,R};\bf sG}\right)_{4}(x):=\pm \left(m^{{\bf sG}}_{n}\right)_{\pm}{\bf\vartheta}_{\bf L,R}(x).   
   \end{array}\ee
This directly shows that under ${\cal N}$=1 supergravity transformation 
the nth Kaluza-Klien mode of the fifth component of the gravitino transform as a Goldstino realized in term 
of the parameter for the supergravity transformation ${\bf\vartheta}_{\bf L,R}$. This is usually known as {\it superHiggs mechanism}. Considering all these facts
the four dimensional Kaluza-Klien gravitino can be redefined in terms of physical degrees of freedom as
\be\begin{array}{llll}\label{newfnc}
  \displaystyle \widehat{{\bf \Psi}^{(n)}_{\mu~{\bf L};\bf sG}}:=\left\{\left(m^{{\bf sG}}_{n}\right)_{\pm}\left({\bf \Psi}^{(n)}_{\mu~{\bf L};\bf sG}
+\tilde{\gamma}_{\mu}\sum^{\infty}_{k=0}\frac{{\bf d}^{(nk)}}{\left(m^{{\bf sG}}_{k}\right)_{\pm}}{\bf \Psi}^{(k)}_{4~{\bf R};\bf sG}\right)
-\partial_{\mu}{\bf \Psi}^{(n)}_{4~{\bf L};\bf sG}\right\},\\
\displaystyle \widehat{{\bf \Psi}^{(n)}_{\mu~{\bf R};\bf sG}}:=\left\{\left(m^{{\bf sG}}_{n}\right)_{\pm}{\bf \Psi}^{(n)}_{\mu~{\bf L};\bf sG}
-\partial_{\mu}{\bf \Psi}^{(n)}_{4~{\bf L};\bf sG}\right\}.
   \end{array}\ee
which are invariant under ${\cal N}$=1 supergravity transformations. To find out the Kaluza-Klien spectrum for the bulk gravitino field we start with
the master equation for the bulk five dimensional gravitino field, commonly known as {\it Rarita-Schwinger} equation which can be written as
\be\begin{array}{llll}\label{bulktino}
    \displaystyle \left[{\bf\Upsilon}^{MNP}\overrightarrow{{\bf D}}^{sG}_{N}-\frac{3}{2}k_{\pm}sgn(y){\bf\Upsilon}^{MP}\right]
\left({\bf \Psi}^{(i)}_{\bf sG}\right)_{P}=0,\\
\displaystyle \left(\bar{\bf \Psi}^{(i)}_{\bf sG}\right)_{M}\left[{\bf\Upsilon}^{MNP}\overleftarrow{{\bf D}}^{sG}_{N}-\frac{3}{2}k_{\pm}sgn(y){\bf\Upsilon}^{MP}\right]
=0.
   \end{array}\ee
After dimensional reduction the four dimensional effective action for rescaled four dimensional gravitino in the context of ${\cal N}$=1 supergravity 
can be written as:
\be\begin{array}{llll}\label{abshg1gr1}
   \displaystyle S_{{\bf \Psi_{sG}}}=-\frac{i M^{3}_{(5)}}{2}\int d^{4}x\sum^{\infty}_{n=0}\sum^{2}_{i=1}\sum^{2}_{j=1}\left[
\widehat{\bar{\bf \Psi}^{(n);(i)}_{\alpha~{\bf L,R};\bf sG}}(x){\gamma}^{\alpha\beta\sigma}\overleftrightarrow
{\partial}_{\beta}
\widehat{{\bf \Psi}^{(n);(j)}_{\sigma~{\bf L,R};\bf sG}}(x)\delta_{ij}-\left(m^{{\bf sG}}_{n}\right)_{\pm}
\widehat{\bar{\bf \Psi}^{(n);(i)}_{\alpha~{\bf L,R};\bf sG}}(x){\gamma}^{\alpha\sigma}\widehat{{\bf \Psi}^{(n);(j)}_{\sigma~{\bf L,R};\bf sG}}(x)\right]
   \end{array}\ee
and the effective {\it Rarita-Schwinger} equation in four dimension turns out to be  
\be\begin{array}{llll}\label{bulktinoeff}
    \displaystyle \left[{\gamma}^{\alpha\beta\sigma}\overrightarrow
{\partial}_{\beta}-\left(m^{{\bf sG}}_{n}\right)_{\pm}{\gamma}^{\alpha\sigma}\right]\widehat{{\bf \Psi}^{(n)}_{\sigma~{\bf L,R};\bf sG}}(x)=0,\\
\displaystyle \widehat{\bar{\bf \Psi}^{(n)}_{\alpha~{\bf L,R};\bf sG}}(x)\left[{\gamma}^{\alpha\beta\sigma}\overleftarrow
{\partial}_{\beta}-\left(m^{{\bf sG}}_{n}\right)_{\pm}{\gamma}^{\alpha\sigma}\right]=0
   \end{array}\ee
 
The extra dimension dependent Kaluza-Klien wave function for gravitino field is determined from the following two differential equations 
\be\begin{array}{lll}\label{fmassgr1}
    \displaystyle \left(\overrightarrow{\cal D}_{y}+\frac{1}{2}k_{\pm}r_{c}sgn(y)\right)\chi^{(n);{\bf L}}_{\pm;\bf sG}(y)
=\left(m^{{\bf sG}}_{n}\right)_{\pm}\chi^{(n);{\bf R}}_{\pm;\bf sG}(y)
   \end{array}\ee

\be\begin{array}{lll}\label{fmassgr2}
    \displaystyle \left(\overrightarrow{\cal D}_{y}-\frac{5}{2}k_{\pm}r_{c}sgn(y)\right)\chi^{(n);{\bf R}}_{\pm;\bf sG}(y)
=-\left(m^{{\bf sG}}_{n}\right)_{\pm}\chi^{(n);{\bf L}}_{\pm;\bf sG}(y)
   \end{array}\ee

subject to the following boundary conditions:
\be\begin{array}{lll}\label{bvchjbcgr}
   \displaystyle\chi^{(n);{\bf L}}_{\pm;\bf sG}(y_{i})\chi^{(n);{\bf R}}_{\pm;\bf sG}(y_{i})=0,\\
 \displaystyle\chi^{(n);{\bf L}}_{\pm;\bf sG}(y_{i})=0,~~~\chi^{(n);{\bf R}}_{\pm;\bf sG}(y_{i})=0
  \end{array}\ee
where at $y_{i}=0,\pi$ the ${\bf Z_{2}}$ orbifold symmetry is imposed. This follows from the fact that 
left-handed or all right-handed fermionic wave functions are ${\bf Z_{2}}$ odd. In this context the gravitino differential operator 
$\left(\overrightarrow{\cal D}_{y}+\frac{1}{2}k_{\pm}r_{c}sgn(y)\right)$ and $\left(\overrightarrow{\cal D}_{y}-\frac{5}{2}k_{\pm}r_{c}sgn(y)\right)$ 
are hermitian and the mass eigen values are real. Consequently $\chi^{(n);{\bf L, R}}_{\pm;\bf sG}(y)$
is chosen to be real.  Additionally we impose the following orthonormalization condition
\be\begin{array}{llll}\label{orstring}
    \displaystyle \int^{+\pi}_{-\pi}dy~e^{-A_{\pm}(y)}\chi^{(n);{\bf L, R}}_{\pm;\bf sG}(y)\chi^{(m);{\bf L, R}}_{\pm;\bf sG}(y)=\delta^{nm}.
   \end{array}\ee

Now introducing two new variables $z^{\pm;{\bf L},{\bf R}}_{n}:=\frac{\left(m^{\bf sG}_{n}\right)_{\pm}}{k_{\pm}}e^{A_{\pm}(y)}$ 
and $\hat{\bf g}^{(n)}_{{\bf L},{\bf R};\bf sG}:=e^{-A_{\pm}(y)}\chi^{(n);{\bf L,R}}_{\pm;\bf sG}$
equation(\ref{fmassgr1}) can be recast in terms of Bessel differential equation of order two as
\be\begin{array}{llll}\label{dif4sg1}
    \displaystyle \left[\left(z^{\pm;{\bf L}}_{n}\right)^{2}\overrightarrow{{\cal D}^{2}}_{z^{\pm;{\bf L}}_{n}}
+z^{\pm;{\bf L}}_{n}\overrightarrow{{\cal D}}_{z^{\pm;{\bf L}}_{n}}
+\left\{\left(z^{\pm;{\bf L}}_{n}\right)^{2}-4\right\}\right]\hat{\bf g}^{(n)}_{{\bf L}}(z^{\pm;{\bf L}}_{n})
=0
   \end{array}\ee

and the analytical solution for left chiral $n\neq 0$ gravitino Kaluza-Klien modes turn out to be
\be\begin{array}{llll}\label{sol6sgr1}
\displaystyle    \chi^{(n);{\bf L}}_{\pm;\bf sG}(z^{\pm;{\bf L}}_{n})=\frac{e^{\frac{3}{2}A_{\pm}(y)}}{{\cal N}^{\pm;{\bf L}}_{(n);\bf sG}}
\left[{\cal J}_{2}
(z^{\pm;{\bf L}}_{n})+\beta^{\pm;{\bf L}}_{n}{\cal Y}_{2}(z^{\pm;{\bf L}}_{n})\right].
   \end{array}\ee

Substituting equation(\ref{fmassgr1}) and equation(\ref{fmassgr2}) 
the analytical solution for the right chiral $n\neq 0$ gravitino Kaluza-Klien modes takes the following form:
\be\begin{array}{llll}\label{sol6sgr2}
\displaystyle    \chi^{(n);{\bf R}}_{\pm;\bf sG}(z^{\pm;{\bf R}}_{n})=\frac{r_{c}e^{\frac{3}{2}A_{\pm}(y)}}{{\cal N}^{\pm;{\bf R}}_{(n);\bf sG}}
\left[{\cal J}_{1}
(z^{\pm;{\bf R}}_{n})+\beta^{\pm;{\bf R}}_{n}{\cal Y}_{1}(z^{\pm;{\bf R}}_{n})\right].
   \end{array}\ee

Here ${\cal N}^{\pm;{\bf L},{\bf R}}_{(n)}$ be the normalization constant of the extra dimension dependent wave function and $\beta^{\pm;{\bf L},{\bf R}}_{n}$ is the
integration constant determined from the orthonormalization condition and the continuity conditions at the orbifold fixed points.
Now applying the boundary condition on equation(\ref{sol6sgr1}) and equation(\ref{sol6sgr2}) we get
\be\begin{array}{llll}\label{cond16l}
 \displaystyle  \chi^{(n);{\bf L}}_{\pm;\bf sG}|_{y_{i}=0}=0~~\implies \beta^{\pm;{\bf L}}_{n}=
-\frac{{\cal J}_{2}\left(\frac{\left(m^{{\bf sG}}_{n}\right)_{\pm}}{k_{\pm}}\right)}
{{\cal Y}_{2}\left(\frac{\left(m^{{\bf sG}}_{n}\right)_{\pm}}{k_{\pm}}\right)}.
   \end{array}\ee

\be\begin{array}{llll}\label{cond21l}
 \displaystyle  \chi^{(n);{\bf L}}_{\pm;\bf sG}|_{y_{i}=\pi}=0~~\implies \beta^{\pm;{\bf L}}_{n}=
-\frac{{\cal J}_{2}\left(x^{\pm;{\bf L}}_{n}\right)
}{{\cal Y}_{2}\left(x^{\pm;{\bf L}}_{n}\right)}
   \end{array}\ee

\be\begin{array}{llll}\label{cond16r}
 \displaystyle  \chi^{(n);{\bf R}}_{\pm;\bf sG}|_{y_{i}=0}=0~~\implies \beta^{\pm;{\bf R}}_{n}=
-\frac{{\cal J}_{1}\left(\frac{\left(m^{{\bf sG}}_{n}\right)_{\pm}}{k_{\pm}}\right)}
{{\cal Y}_{1}\left(\frac{\left(m^{{\bf sG}}_{n}\right)_{\pm}}{k_{\pm}}\right)}.
   \end{array}\ee

\be\begin{array}{llll}\label{cond21r}
 \displaystyle  \chi^{(n);{\bf R}}_{\pm;\bf sG}|_{y_{i}=\pi}=0~~\implies \beta^{\pm;{\bf L}}_{n}=
-\frac{{\cal J}_{1}\left(x^{\pm;{\bf L}}_{n}\right)
}{{\cal Y}_{1}\left(x^{\pm;{\bf L}}_{n}\right)}
   \end{array}\ee

where $z^{\pm;{\bf L},{\bf R}}_{n}(\pi):=x^{\pm;{\bf L},{\bf R}}_{n}=\frac{\left(m^{{\bf sG}}_{n}\right)_{\pm}}{k_{\pm}}e^{k_{\pm}r_{c}\pi}$.
Now using equation(\ref{cond16l}-\ref{cond21r}) we get
\be\begin{array}{llllll}\label{root1r}
 \displaystyle   \frac{{\cal J}_{2}\left(x^{\pm;{\bf L}}_{n}e^{-k_{\pm}r_{c}\pi}\right)}
{{\cal Y}_{2}\left(x^{\pm;{\bf L}}_{n}e^{-k_{\pm}r_{c}\pi}\right)}
= \frac{{\cal J}_{2}\left(x^{\pm;{\bf L}}_{n}\right)
}{{\cal Y}_{2}\left(x^{\pm;{\bf L}}_{n}\right)} 
   \end{array}\ee

\be\begin{array}{llllll}\label{root2r}
 \displaystyle   \frac{{\cal J}_{1}\left(x^{\pm;{\bf R}}_{n}e^{-k_{\pm}r_{c}\pi}\right)}
{{\cal Y}_{1}\left(x^{\pm;{\bf R}}_{n}e^{-k_{\pm}r_{c}\pi}\right)}
= \frac{{\cal J}_{1}\left(x^{\pm;{\bf R}}_{n}\right)
}{{\cal Y}_{1}\left(x^{\pm;{\bf R}}_{n}\right)} 
   \end{array}\ee

which is an transcendental equation of $x^{\pm;{\bf L},{\bf R}}_{n}$ and the roots of this equation give the left and right chiral 
fermionic field mass spectrum $\left(m^{{\bf L},{\bf R}}_{n}\right)_{\pm}$
 in presence of perturbative 
Gauss-Bonnet coupling $\alpha_{(5)}$.
This approximately leads to the various Kaluza-Klien mode masses for the gravitino field as,
\be\begin{array}{llll}\label{massasd}
   \displaystyle  \left(m^{{\bf sG}}_{n}\right)_{\pm}\approx \left(n+\frac{1}{4}\right)\pi k_{\pm}e^{-k_{\pm}r_{c}\pi}.
   \end{array}\ee
The gravitino mass spectrum exhibits similar feature as graviton mode.
 Now using equation(\ref{orstring}) the normalization constant for $n\neq 0$ mode reduces to the following expression
\be\begin{array}{llll}\label{vcvcasar1}
    \displaystyle {\cal N}^{\pm;{\bf L}}_{(n)}=\frac{e^{k_{\pm}r_{c}\pi}}{
\sqrt{k_{\pm}r_{c}}}\left\{\left[{\cal J}_{2}\left(x^{\pm;{\bf L}}_{n}\right)
+\beta^{\pm;{\bf L}}_{n}{\cal Y}_{2}\left(x^{\pm;{\bf L}}_{n}\right)
\right]^{2}
\displaystyle-e^{-2k_{\pm}r_{c}}\left[{\cal J}_{2}\left(x^{\pm;{\bf L}}_{n}e^{-k_{\pm}r_{c}\pi}
\right)+\beta^{\pm;{\bf L}}_{n}{\cal Y}_{2}\left(x^{\pm;{\bf L}}_{n}e^{-k_{\pm}r_{c}\pi}\right)
\right]^{2}\right\}^{\frac{1}{2}}\end{array}\ee

\be\begin{array}{llll}\label{vcvcasar2}
    \displaystyle {\cal N}^{\pm;{\bf R}}_{(n)}=\frac{e^{k_{\pm}r_{c}\pi}}{
\sqrt{k_{\pm}r_{c}}}\left\{\left[{\cal J}_{1}\left(x^{\pm;{\bf R}}_{n}\right)
+\beta^{\pm;{\bf L}}_{n}{\cal Y}_{1}\left(x^{\pm;{\bf R}}_{n}\right)
\right]^{2}
\displaystyle-e^{-2k_{\pm}r_{c}}\left[{\cal J}_{1}\left(x^{\pm;{\bf R}}_{n}e^{-k_{\pm}r_{c}\pi}
\right)+\beta^{\pm;{\bf L}}_{n}{\cal Y}_{1}\left(x^{\pm;{\bf R}}_{n}e^{-k_{\pm}r_{c}\pi}\right)
\right]^{2}\right\}^{\frac{1}{2}}.\end{array}\ee
 For $e^{k_{\pm}r_{c}\pi}\gg 1,~\frac{\left(m^{{\bf sG}}_{n}\right)_{\pm}}{k_{\pm}}\ll 1$ the integration constant $\beta^{\pm;{\bf L},{\bf R}}_{n}\ll 1$.
Consequently ${\cal Y}_{2}(z^{\pm;{\bf L}}_{n})$ and  ${\cal Y}_{1}(z^{\pm;{\bf R}}_{n})$
 are neglected compared to ${\cal J}_{2}(z^{\pm;{\bf L}}_{n})$ and ${\cal J}_{1}(z^{\pm;{\bf R}}_{n})$ in equation(\ref{sol6sgr1}) and equation(\ref{sol6sgr2}).
Hence the normalization constant for $n\neq 0$ mode
turns out to be
\be\begin{array}{llll}\label{vcvczx11r}
    \displaystyle {\cal N}^{\pm;{\bf L}}_{(n)}=\frac{e^{k_{\pm}r_{c}\pi}}{
\sqrt{k_{\pm}r_{c}}}{\cal J}_{2}\left(x^{\pm;{\bf L}}_{n}\right),
   \end{array}\ee

\be\begin{array}{llll}\label{vcvczx12r}
    \displaystyle {\cal N}^{\pm;{\bf R}}_{(n)}=\frac{e^{k_{\pm}r_{c}\pi}}{
\sqrt{k_{\pm}r_{c}}}{\cal J}_{1}\left(x^{\pm;{\bf R}}_{n}\right). 
   \end{array}\ee
Consequently the extra dimension dependent wave function for $n\neq 0$ mode turns out to be 

\be\begin{array}{llll}\label{sol61r1z}
\displaystyle    \chi^{(n);{\bf L}}_{\pm;\bf sG}(z^{\pm;{\bf L}}_{n})=\frac{\sqrt{k_{\pm}r_{c}}e^{\frac{3}{2}A_{\pm}(y)}}{e^{k_{\pm}r_{c}\pi}
}\frac{{\cal J}_{2}
(z^{\pm;{\bf L}}_{n})}{{\cal J}_{2}\left(x^{\pm;{\bf L}}_{n}\right)},
   \end{array}\ee

\be\begin{array}{llll}\label{sol61r2z}
\displaystyle    \chi^{(n);{\bf R}}_{\pm;\bf sG}(z^{\pm;{\bf R}}_{n})=\frac{\sqrt{k_{\pm}r_{c}}e^{\frac{3}{2}A_{\pm}(y)}}{e^{k_{\pm}r_{c}\pi}
}\frac{{\cal J}_{1}
(z^{\pm;{\bf R}}_{n})}{{\cal J}_{1}\left(x^{\pm;{\bf R}}_{n}\right)}.
   \end{array}\ee

For massless $n=0$ mode the solution of the equation(\ref{fmass}) turns out to be
\be\begin{array}{lllll}\label{j1}
   \displaystyle \chi^{(0);{\bf L}_{\pm;\bf sG}}(y)=\frac{e^{-\frac{1}{2}A_{\pm}(y)}}{ {\cal N}^{\pm;{\bf L}}_{(0)}},\\
\displaystyle \chi^{(0);{\bf R}_{\pm;\bf sG}}(y)=\frac{e^{\frac{5}{2}A_{\pm}(y)}}{ {\cal N}^{\pm;{\bf R}}_{(0)}}.
   \end{array}\ee
Here ${\cal N}^{\pm;{\bf L},{\bf R}}_{(0)}$ normalization constant for zero mode. Now applying the normalization condition 
we get ${\cal N}^{\pm;{\bf L}}_{(0)}=\sqrt{\frac{\left(1-e^{-2k_{\pm}r_{c}\pi}\right)}{k_{\pm}r_{c}}}$ and
 ${\cal N}^{\pm;{\bf L}}_{(0)}=\sqrt{\frac{\left(e^{4k_{\pm}r_{c}\pi}-1\right)}{2k_{\pm}r_{c}}}$. Consequently the ground state massless zero
 mode wave function for gravitino species turns out to be
\be\begin{array}{llll}\label{mlkl}
   \displaystyle \chi^{(0);{\bf L}_{\pm;\bf sG}}(y)=\sqrt{\frac{\left(1-e^{-2k_{\pm}r_{c}\pi}\right)}{k_{\pm}r_{c}}}e^{-\frac{1}{2}A_{\pm}(y)},\\
\displaystyle \chi^{(0);{\bf R}_{\pm;\bf sG}}(y)=\sqrt{\frac{\left(e^{4k_{\pm}r_{c}\pi}-1\right)}{2k_{\pm}r_{c}}}e^{\frac{5}{2}A_{\pm}(y)}.
\end{array}\ee
For each of the left and right chiral mode shows two fold characteristics due to the presence of two branches ($k_{\pm}$) in the context
of Gauss-Bonnet coupling induced string phenomenology. In the asymptotic limit $\alpha_{(5)}\rightarrow 0$ the $k_{-}$ branch will reproduce the 
well known Randall-Sundrum behavior for both left and right chiral mode. Also the $k_{+}$ branch gives us completely new informations about 
the warped phenomenology in presence of Gauss-Bonnet coupling.
Most significantly the ground state obtained for gravitino
in the brane is fixed, but in the bulk it goes with the extra dimensional coordinate $y$.

\subsection{\bf Bulk Scalar Field}
\label{bscal}

The five dimensional action for bulk scalar field can be written as
\be\begin{array}{llll}\label{abshg1bs}
   \displaystyle S_{{\cal {\bf \Phi}}}=\frac{1}{2}\int d^{5}x\sqrt{-g_{(5)}}~\left[g^{AB}\left(\overrightarrow{\partial}_{A}{\bf \Phi}(x,y)\right)\left(
\overrightarrow{\partial}_{B}{\bf \Phi}(x,y)\right)
-m^{2}_{{\bf \Phi}}{\bf\Phi}^{2}(x,y)\right]\\
\displaystyle ~~~~=\frac{1}{2}\int d^{4}x\int^{+\pi}_{-\pi}r_{c}\left[e^{-2A_{\pm}(y)}\eta^{\mu\nu}\left(\overrightarrow{\partial}_{\mu}{\bf \Phi}(x,y)\right)
\left(\overrightarrow{\partial}_{\nu}{\bf \Phi}(x,y)\right)+\frac{1}{r^{2}_{c}}{\bf \Phi}(x,y)\overrightarrow{\partial}_{y}\left\{e^{-4A_{\pm}(y)}
\left(\overrightarrow{\partial}_{y}{\bf \Phi}(x,y)\right)\right\}\right.\\ \left.~~~~~~~~~~~~~~~~~~~~~~~~~~~~~~~~~~~~~~~~~
~~~~~~~~~~~~~~~~~~~~~~~~~~~~~~~~~~~~~~~~~~~~~~~~~~~~~~~~~~~~~~~~~~~~\displaystyle-m^{2}_{{\bf \Phi}}e^{-4A_{\pm}(y)}{\bf\Phi}^{2}(x,y)\right].
   \end{array}\ee

 We choose the Kaluza-Klien expansion
of the bulk scalar as,
\be\begin{array}{lllll}\label{KK41bs}
   \displaystyle {\bf \Phi}(x,y)=\sum^{\infty}_{n=0}{\bf \Phi}^{(n)}(x)~\frac{\chi^{(n)}_{\pm;\bf \Phi}(y)}{\sqrt{r_{c}}}. 
   \end{array}\ee


\begin{figure}[htb]
\centering
\subfigure[]{
    \includegraphics[width=8.5cm,height=7cm] {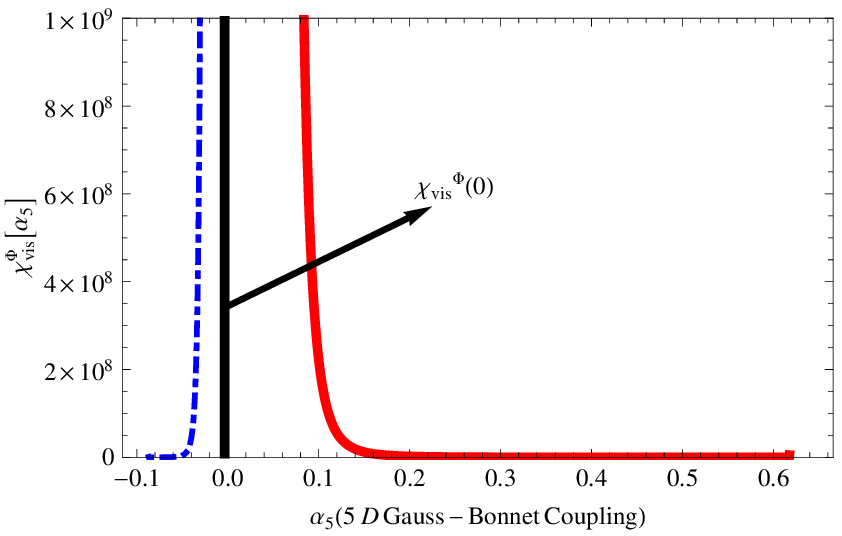}
    \label{fig:subfig4500}
}
\subfigure[]{
    \includegraphics[width=8.5cm,height=7cm] {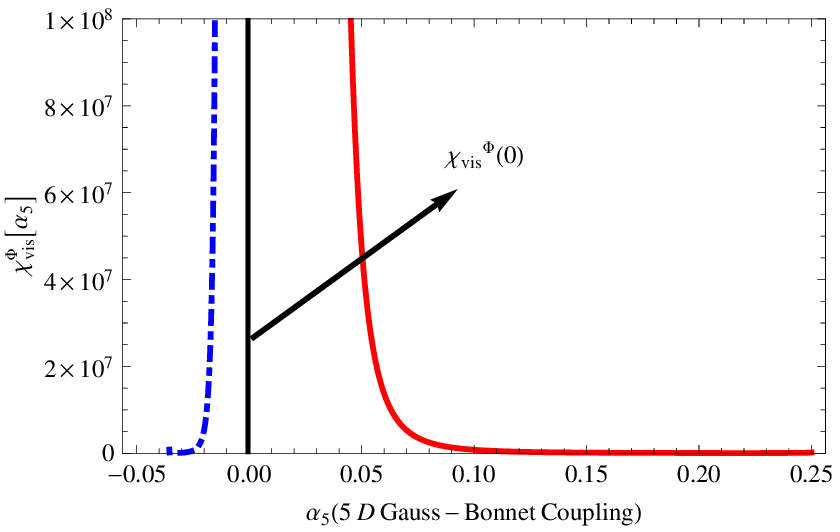}
    \label{fig:subfig4600}
}
\subfigure[]{
    \includegraphics[width=8.5cm,height=7cm] {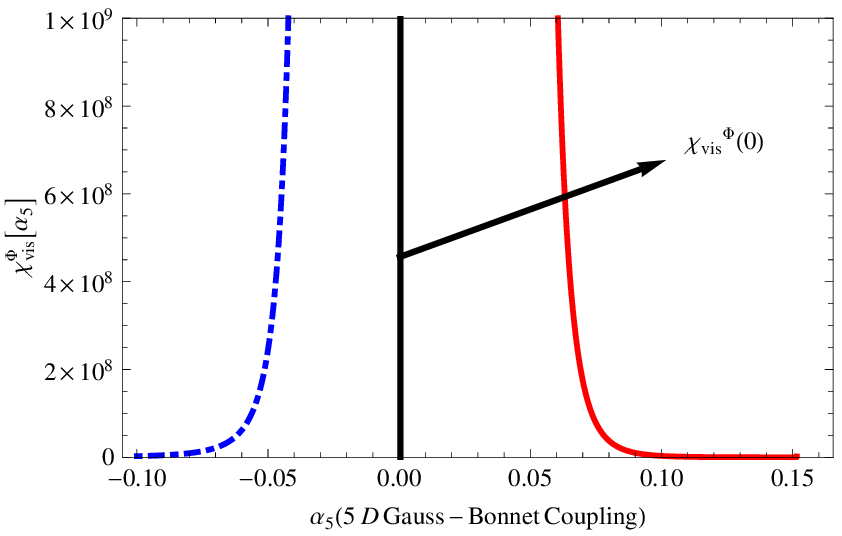}
    \label{fig:subfig4700}
}
\subfigure[]{
    \includegraphics[width=8.5cm,height=7cm] {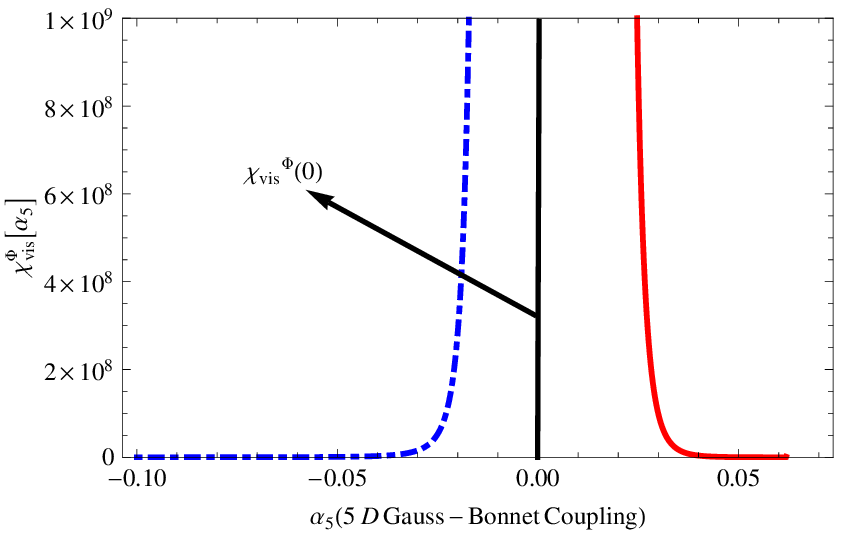}
    \label{fig:subfig4800}
}
\caption[Optional caption for list of figures]{Variation
 of $\chi^{(1)}_{\pm;\bf \Phi}(\pi)(=\chi^{(1)}_{\pm;vis})$ 
vs Gauss-Bonnet coupling $\alpha_{(5)}$ for \subref{fig:subfig4500} $\Lambda_{(5)}>0$ and $A_{1}>0$, \subref{fig:subfig4600} $\Lambda_{(5)}>0$ and $A_{1}<0$, 
\subref{fig:subfig4700} $\Lambda_{(5)}<0$ and $A_{1}>0$ and
\subref{fig:subfig4800} $\Lambda_{(5)}<0$ and $A_{1}<0$. In this context $B_{0}=0.002$, $r_{c}=0.996\sim 1$, $|A_{1}|=0.04$, $\theta_{1}=0.05$ and $\theta_{2}=0.04$.}
\label{fig:subfigureExample58500}
\end{figure}

\begin{figure}[htb]
\centering
\subfigure[]{
    \includegraphics[width=8.5cm,height=7cm] {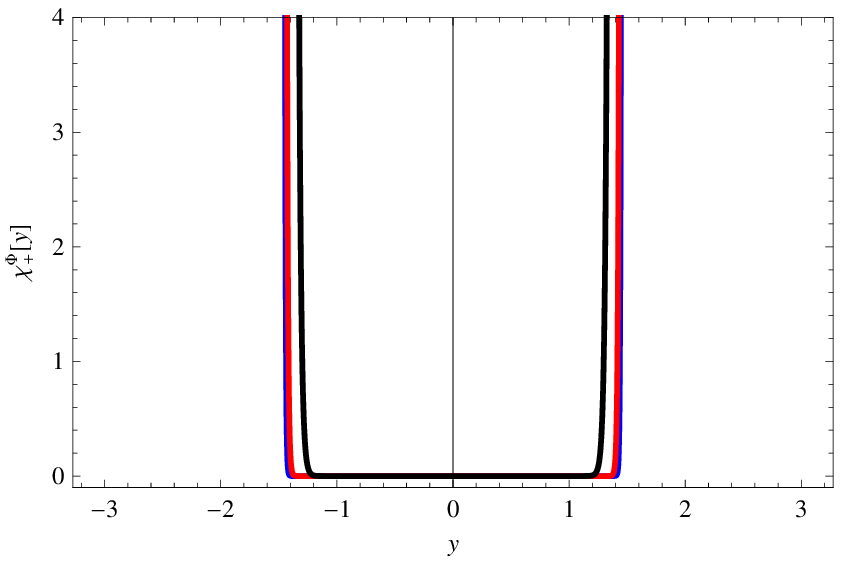}
    \label{fig:subfig5100}
}
\subfigure[]{
    \includegraphics[width=8.5cm,height=7cm] {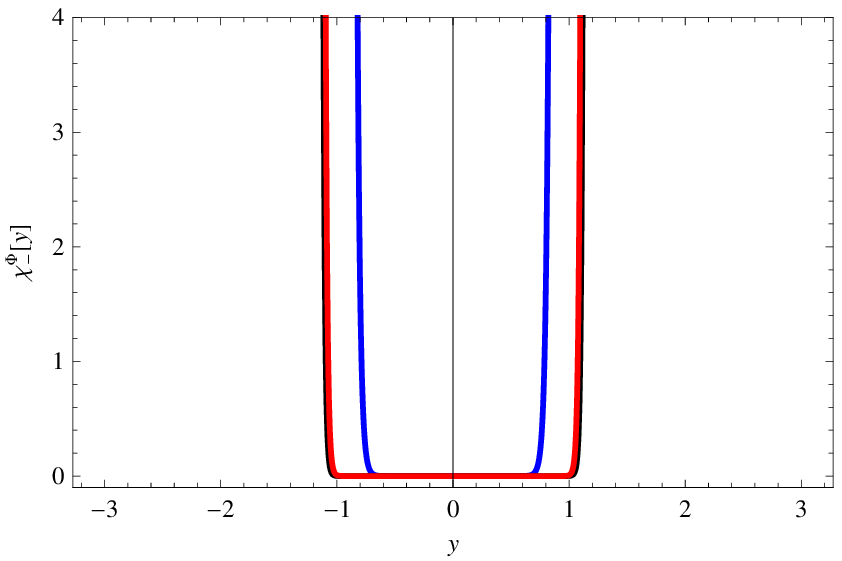}
    \label{fig:subfig5200}
}
\caption[Optional caption for list of figures]{Variation
 of $\chi^{(1)}_{+;\bf \Phi}(y)$ and $\chi^{(1)}_{-;\bf \Phi}(y)$
vs extra dimensional coordinate $y$ for  
\subref{fig:subfig5100} $\Lambda_{(5)}<0$ and $A_{1}<0$ and \subref{fig:subfig5200} $\Lambda_{(5)}>0$ and $A_{1}<0$ respectively.
 In this context we use $B_{0}=0.002$, $r_{c}=0.996\sim 1$, $|A_{1}|=0.04$, $\theta_{1}=0.05$ and $\theta_{2}=0.04$
three distinct values of Gauss-Bonnet Coupling $\alpha_{(5)}$.}
\label{fig:subfigureExample59500}
\end{figure}


Plugging equation(\ref{KK41bs}) in equation(\ref{abshg1bs}) the effective four dimensional action reduces to the following form:
\be\begin{array}{llll}\label{abs2321bs}
   \displaystyle S_{{\bf \Phi}}=\int d^{4}x\sum^{\infty}_{n=0}\left[\eta^{\mu\nu}\left(\overrightarrow{\partial}_{\mu}{\bf \Phi}^{(n)}(x)\right)
\left(\overrightarrow{\partial}_{\nu}{\bf \Phi}^{(n)}(x)\right)-
\left(m^{{\bf \Phi}}_{n}\right)^{2}_{\pm}\left({\bf \Phi}^{(n)}(x)\right)^{2}\right]\end{array}\ee

subject to the orthonormalization condition of extra dimension dependent wave functions
\be\begin{array}{llll}\label{no121bs}
 \displaystyle   \int^{+\pi}_{-\pi}dy~e^{-2A_{\pm}(y)}~\chi^{(m)}_{\pm;\bf \Phi}(y)~\chi^{(n)}_{\pm;\bf \Phi}(y)=\delta^{mn}
   \end{array}\ee
and the mass term of the bulk scalar field is defined through the following differential equation as
\be\begin{array}{llll}\label{dif181bs}
  \displaystyle   -\frac{1}{r^{2}_{c}}\overrightarrow{{\cal D}_{y}}\left(e^{-4A_{\pm}(y)}\overrightarrow{{\cal D}_{y}}\chi^{(n)}_{\pm;\bf \Phi}(y)\right)
+m^{2}_{{\bf \Phi}}e^{-4A_{\pm}(y)}\chi^{(n)}_{\pm;\bf \Phi}(y)=
e^{-2A_{\pm}(y)}\left(m^{{\bf \Phi}}_{n}\right)^{2}_{\pm}\chi^{(n)}_{\pm;\bf \Phi}(y).
   \end{array}\ee

Now introducing a new variable $z^{\pm;{\bf \Phi}}_{n}:=\frac{\left(m^{{\bf \Phi}}_{n}\right)_{\pm}}{k_{\pm}}e^{A_{\pm}(y)}$ 
equation(\ref{dif181bs}) can be recast in terms of Bessel differential equation of order $\nu^{\bf \Phi}_{\pm}:=\sqrt{4+\frac{m^{2}_{{\bf \Phi}}}{k^{2}_{\pm}}}$ as
\be\begin{array}{llll}\label{dif281bs}
    \displaystyle  \left[\left(z^{\pm;{\bf \Phi}}_{n}\right)^{2}\overrightarrow{{\cal D}^{2}}_{z^{\pm;{\bf \Phi}}_{n}}
+z^{\pm;{\bf \Phi}}_{n}\overrightarrow{{\cal D}}_{z^{\pm;{\bf \Phi}}_{n}}+\left\{\left(z^{\pm;{\bf \Phi}}_{n}\right)^{2}-\left(\nu^{\bf \Phi}_{\pm}\right)^{2}\right\}\right]\chi^{(n)}_{\pm;\bf \Phi}
=0
   \end{array}\ee
and the analytical solution turns out to be
\be\begin{array}{llll}\label{sol1xc1bs}
\displaystyle    \chi^{(n)}_{\pm;\cal Z}(y)=\frac{e^{2A_{\pm}(y)}}{{\cal N}^{\pm;\bf \Phi}_{(n)}}\left[{\cal J}_{\nu^{\bf \Phi}_{\pm}}
(z^{\pm;{\bf \Phi}}_{n})+\alpha^{\pm;\bf \Phi}_{n}{\cal Y}_{\nu^{\bf \Phi}_{\pm}}(z^{\pm;{\bf \Phi}}_{n})\right].
   \end{array}\ee
Here ${\cal N}^{\pm;\bf \Phi}_{(n)}$ be the normalization constant of the extra dimension dependent wave function and $\alpha^{\pm;\bf \Phi}_{n}$ is the
integration constant determined from the orthonormalization condition and the continuity conditions at the orbifold fixed point.
Self-adjointness and hermiticity of the differential operator appearing in equation(\ref{dif281bs})
demands that $\overrightarrow{{\cal D}_{y}}\chi^{(n)}_{\pm;\bf \Phi}(y)$ is
 continuous at the orbifold fixed points $y_{i}=0,\pi$. Consequently we have
\be\begin{array}{llll}\label{cond1zx1bs}
 \displaystyle  \overrightarrow{{\cal D}_{y}}\chi^{(n)}_{\pm;\bf \Phi}|_{y_{i}=0}=0~~\implies \alpha^{\pm;\bf\Phi}_{n}=
\frac{\left[\frac{\left(m^{{\bf \Phi}}_{n}\right)_{\pm}}{k_{\pm}}{\cal J}^{'}_{\nu^{\bf \Phi}_{\pm}}\left(\frac{\left(m^{{\bf\Phi}}_{n}\right)_{\pm}}{k_{\pm}}\right)
+2{\cal J}_{\nu^{\bf \Phi}_{\pm}}\left(\frac{\left(m^{{\bf\Phi}}_{n}\right)_{\pm}}{k_{\pm}}\right)\right]}{\left[2{\cal Y}_{\nu^{\bf \Phi}_{\pm}}\left(\frac{\left(m^{{\bf\Phi}}_{n}\right)_{\pm}}{k_{\pm}}\right)
+\frac{\left(m^{{\bf\Phi}}_{n}\right)_{\pm}}{k_{\pm}}{\cal Y}^{'}_{\nu^{\bf \Phi}_{\pm}}\left(\frac{\left(m^{{\bf\Phi}}_{n}\right)_{\pm}}{k_{\pm}}\right)\right]}.
   \end{array}\ee

\be\begin{array}{llll}\label{cond2zx1bs}
 \displaystyle  \overrightarrow{{\cal D}_{y}}\chi^{(n)}_{\pm;{\bf \Phi}}|_{y_{i}=\pi}=0~~\implies \alpha^{\pm;{\bf\Phi}}_{n}=
\frac{\left[2{\cal J}_{\nu^{\bf \Phi}_{\pm}}\left(x^{\pm;{\bf\Phi}}_{n}\right)
+x^{\pm;{\bf\Phi}}_{n}{\cal J}^{'}_{\nu^{\bf \Phi}_{\pm}}\left(x^{\pm;{\bf\Phi}}_{n}\right)\right]}{\left[x^{\pm;{\bf\Phi}}_{n}{\cal Y}^{'}_{\nu^{\bf \Phi}_{\pm}}\left(x^{\pm;{\bf\Phi}}_{n}\right)
+2{\cal Y}_{\nu^{\bf \Phi}_{\pm}}\left(x^{\pm;{\bf\Phi}}_{n}\right)\right]}
   \end{array}\ee
where $z^{\pm;{\bf \Phi}}_{n}(\pi):=x^{\pm;{\bf\Phi}}_{n}=\frac{\left(m^{{\bf\Phi}}_{n}\right)_{\pm}}{k_{\pm}}e^{k_{\pm}r_{c}\pi}$.
For $e^{k_{\pm}r_{c}\pi}\gg 1,~\frac{\left(m^{{\bf\Phi}}_{n}\right)_{\pm}}{k_{\pm}}\ll 1$ the mass spectrum for the bulk scalar field is expected to be of the order of TeV scale i.e.

\be\begin{array}{llllll}\label{rootzxbs}
 \displaystyle   \frac{\left[x^{\pm;{\bf\Phi}}_{n}e^{-k_{\pm}r_{c}\pi}{\cal J}^{'}_{\nu^{\bf \Phi}_{\pm}}\left(x^{\pm;{\bf\Phi}}_{n}e^{-k_{\pm}r_{c}\pi}\right)
+2{\cal J}_{\nu^{\bf \Phi}_{\pm}}\left(x^{\pm;{\bf\Phi}}_{n}e^{-k_{\pm}r_{c}\pi}\right)\right]}{\left[2{\cal Y}_{\nu^{\bf \Phi}_{\pm}}\left(x^{\pm;{\bf\Phi}}_{n}e^{-k_{\pm}r_{c}\pi}\right)
+x^{\pm;{\bf\Phi}}_{n}e^{-k_{\pm}r_{c}\pi}{\cal Y}^{'}_{\nu^{\bf \Phi}_{\pm}}\left(\frac{\left(m^{{\bf\Phi}}_{n}\right)_{\pm}}{k_{\pm}}\right)\right]}
= \frac{\left[2{\cal J}_{\nu^{\bf \Phi}_{\pm}}\left(x^{\pm;{\bf\Phi}}_{n}\right)
+x^{\pm;{\bf\Phi}}_{n}{\cal J}^{'}_{\nu^{\bf \Phi}_{\pm}}\left(x^{\pm;{\bf\Phi}}_{n}\right)\right]}{\left[x^{\pm;{\bf\Phi}}_{n}{\cal Y}^{'}_{\nu^{\bf \Phi}_{\pm}}\left(x^{\pm;{\bf\Phi}}_{n}\right)
+2{\cal Y}_{\nu^{\bf \Phi}_{\pm}}\left(x^{\pm;{\bf\Phi}}_{n}\right)\right]} \\
\\
\displaystyle \Rightarrow {\cal J}_{\nu^{\bf \Phi}_{\pm}}\left(x^{\pm;{\cal Z}}_{n}\right)+x^{\pm;{\bf \Phi}}_{n}
{\cal J}^{'}_{\nu^{\bf \Phi}_{\pm}}\left(x^{\pm;{\bf \Phi}}_{n}\right)\approx 0
   \end{array}\ee
which is an transcendental equation of $x^{\pm;{\bf\Phi}}_{n}$ and the roots of this equation gives the scalar field mass spectrum 
$\left(m^{{\bf\Phi}}_{n}\right)_{\pm}$ in presence of perturbative 
Gauss-Bonnet coupling $\alpha_{(5)}$. This leads to approximately
\be\begin{array}{llll}\label{massasdphi}
   \displaystyle  \left(m^{{\bf \Phi}}_{n}\right)_{\pm}\approx \left(n+\frac{1}{2}\nu^{\bf \Phi}_{\pm}-\frac{3}{4}\right)\pi k_{\pm}e^{-k_{\pm}r_{c}\pi}.
   \end{array}\ee

Now using equation(\ref{no121bs}) the normalization constant for $n\neq 0$ mode reduces to the following expression
\be\begin{array}{llll}\label{vcvc1bs}
    \displaystyle {\cal N}^{\pm;{\bf \Phi}}_{(n)}=\frac{e^{k_{\pm}r_{c}\pi}}{
\sqrt{k_{\pm}r_{c}}}\sqrt{\left\{\left[{\cal J}_{\nu^{\bf \Phi}_{\pm}}\left(x^{\pm;{\bf \Phi}}_{n}\right)+\alpha^{\pm;{\bf \Phi}}_{n}{\cal Y}_{\nu^{\bf \Phi}_{\pm}}\left(x^{\pm;{\bf \Phi}}_{n}\right)
\right]^{2}-e^{-2k_{\pm}r_{c}}\left[{\cal J}_{\nu^{\bf \Phi}_{\pm}}\left(x^{\pm;{\bf \Phi}}_{n}e^{-k_{\pm}r_{c}\pi}
\right)+\alpha^{\pm;{\bf \Phi}}_{n}{\cal Y}_{\nu^{\bf \Phi}_{\pm}}\left(x^{\pm;{\bf \Phi}}_{n}e^{-k_{\pm}r_{c}\pi}\right)
\right]^{2}\right\}}. 
   \end{array}\ee
For $e^{k_{\pm}r_{c}\pi}\gg 1,~\frac{\left(m^{{\cal Z}}_{n}\right)_{\pm}}{k_{\pm}}\ll 1$ the inegration constant $\alpha^{\pm;{\cal Z}}_{n}\ll 1$.
Consequently ${\cal Y}_{\nu^{\bf \Phi}_{\pm}}(z^{\pm;{\bf \Phi}}_{n})$ is neglected compared to ${\cal J}_{\nu^{\bf \Phi}_{\pm}}(z^{\pm;{\bf \Phi}}_{n})$ in equation(\ref{sol1xc1bs}) 
and then the normalization constant for $n\neq 0$ mode
turns out to be
\be\begin{array}{llll}\label{vcvc1bsbs}
    \displaystyle {\cal N}^{\pm;{\bf \Phi}}_{(n)}=\frac{e^{k_{\pm}r_{c}\pi}}{
\sqrt{k_{\pm}r_{c}}}{\cal J}_{\nu^{\bf \Phi}_{\pm}}\left(x^{\pm;{\bf \Phi}}_{n}\right)\sqrt{1+\frac{4-\left(\nu^{\bf \Phi}_{\pm}\right)^{2}}{\left(x^{\pm;{\bf \Phi}}_{n}\right)^{2}}}. 
   \end{array}\ee
Consequently the extra dimensional dependent wave function for $n\neq 0$ turns out to be 
\be\begin{array}{llll}\label{sol1xc1bs}
\displaystyle    \chi^{(n)}_{\pm;\bf \Phi}(y)=\frac{\sqrt{k_{\pm}r_{c}}~e^{2A_{\pm}(y)}}{\left(\sqrt{1+\frac{4-\left(\nu^{\bf \Phi}_{\pm}\right)^{2}}{\left(x^{\pm;{\bf \Phi}}_{n}\right)^{2}}}
\right)e^{k_{\pm}r_{c}\pi}}\frac{{\cal J}_{\nu^{\bf \Phi}_{\pm}}
(z^{\pm;{\bf \Phi}}_{n})}{{\cal J}_{\nu^{\bf \Phi}_{\pm}}
(x^{\pm;{\bf \Phi}}_{n})}.
   \end{array}\ee

In figure(\ref{fig:subfigureExample58500}) we have clearly depicted the behavior of the extra dimension
dependent first excited state for all possible signatures of Gauss-Bonnet coupling for $k_{+}$ and $k_{-}$ branch for scalar field
different from dilatonic degrees of freedom.
The asymptotic behavior in the $\alpha_{(5)}\rightarrow 0$ is different for two existing physical branches in 
the context of Gauss-Bonnet coupling induced string phenomenology. Most significantly in this free asymptotic limit 
$k_{-}$ branch reproduces the well known Randall-Sundrum feature.
Additionally in figure(\ref{fig:subfigureExample59500}) we have plotted the graphical behavior 
of the extra dimension dependent wave function for $k_{+}$ and $k_{-}$ branch corresponding the first excited 
state for two signatures of bulk cosmological constant $\Lambda_{(5)}$ and negative two -loop conformal coupling $A_{1}$
for three distinct values of Gauss-Bonnet coupling $\alpha_{(5)}$. The behavior of the scalar wave function for the Kaluza Klien 
first excited state explicitly shows that the confinement of the scalar degrees of freedom in the ${\bf dS_{5}}/{\bf AdS_{5}}$
bulk topological space is larger compared to the dilatonic degrees of freedom. If if include the possibility of self interaction via pure
quartic coupling or through derivative coupling which we have elaborately discussed later induces the appearance of ${\cal SU}(2)$ 
Higgs doublet in the bulk topological space.

For massless $n=0$ mode the solution of the equation(\ref{dif181bs}) turns out to be
\be\begin{array}{lllll}\label{jkloss1bs}
   \displaystyle  \chi^{(0)}_{\pm;{\bf \Phi}}=\frac{C_{1}}{4k_{\pm}r_{c}}e^{4A_{\pm}(y)}+C_{2}.
   \end{array}\ee
Here $C_{1}$ and $C_{2}$ are arbitrary integration constants. Now applying the boundary condition through the continuity
of the wave function we get $C_{1}=0$. As a result the zero mode solution turns out to be $ \chi^{(0)}_{\pm;{\bf \Phi}}=C_{2}$.
Now applying the normalization condition the ground state massless zero mode wave function turns out to be
\be\begin{array}{llll}\label{mlsxbs}
   \displaystyle  \chi^{(0)}_{\pm;{\bf \Phi}}=C_{2}=\sqrt{\frac{k_{\pm}r_{c}}{1-e^{-2k_{\pm}r_{c}\pi}}}\approx \frac{1}{\sqrt{2\pi}}.
   \end{array}\ee
The ground state obtained for bulk scalar field for our set up is exactly same as it is obtained in the context of Randall-Sundrum scenario.
Additionally the ground sate obtained for the graviton and bulk scalar field is exactly identical.

\subsection{\bf ${\cal U}(1)$ Abelian Gauge Field }
\label{gaf}
The five dimensional action for the pure ${\cal U}(1)$ abelian gauge theory can be written as
\be\begin{array}{llll}\label{abs}
   \displaystyle S_{{\cal A}}=-\frac{1}{4}\int d^{5}x\sqrt{-g_{(5)}}~{\cal F}_{MN}(x,y){\cal F}^{MN}(x,y)
   \end{array}\ee
where the five dimensional rank-2 antisymmetric ${\cal U}(1)$ abelian gauge field strength tensor is given by
\be\begin{array}{llll}\label{abse}
   \displaystyle {\cal F}_{MN}:=\overrightarrow{\partial}_{[M}{\cal A}_{N]} (x,y)
   \end{array}\ee
with ${\cal A}_{M}:=\left({\cal A}_{\alpha},{\cal A}_{4}\right)$. This leads to the five dimensional Maxwell's equation
\be\begin{array}{llll}\label{max}
 \displaystyle   \frac{1}{\sqrt{-g_{(5)}}}\overrightarrow{\partial}_{N}\left(\sqrt{-g_{(5)}}{\cal F}^{MN}(x,y)\right)=0.
   \end{array}\ee
Equation(\ref{abs}) does not involve the affine connection terms 
due to the antisymmetry of the ${\cal U}(1)$ abelian gauge field strength tensor. To find out effective four dimensional action 
through the Kaluza-Klien spectrum we assume that ${\cal A}_{\alpha}$ (satisfies Neumann boundary condition) 
and ${\cal A}_{4}$ (satisfies Dirichlet boundary condition) are ${\bf Z_{2}}$ even and odd respectively.
Depending on this crucial choice of the ${\bf Z_{2}}$ parity the gauge-fermion interactions are preserved. It also ensures that ${\cal A}_{4}$
does not have zero mode in the four dimensional effective theory. This leads to
 \be\begin{array}{llll}\label{abs1}
   \displaystyle S_{{\cal A}}=-\frac{1}{4}\int d^{5}x\left[\eta^{\mu\kappa}\eta^{\nu\lambda}
{\cal F}_{\kappa\lambda}(x,y){\cal F}_{\mu\nu}(x,y)+2e^{-2A_{\pm}(y)}\eta^{\nu\lambda}\overrightarrow{{\cal D}_{y}}{\cal A}_{\nu}(x,y)\overrightarrow{{\cal D}_{y}}{\cal A}_{\lambda}(x,y)\right]
   \end{array}\ee
where we introduce a new symbol $\overrightarrow{{\cal D}_{y}}:=\frac{d}{dy}$. In equation(\ref{abs1}) we use the gauge degrees of freedom
\be\begin{array}{lll}\label{ggga}
    {\cal A}_{4}=0~~~~~\implies~~~~~{\cal A}_{4}(x,y_{(i)})=\overrightarrow{\partial}_{4}{\cal A}(x,y_{(i)})=\overrightarrow{{\cal D}_{y}}{\cal A}(x,y_{(i)})=0
   \end{array}\ee

where at the orbifolding ${\bf Z_{2}}$ symmetry is imposed on $y_{(i)}$. This is consistent with the gauge invariant equation
$\oint d^{4}x {\cal A}_{4}=0$ which is the outcome of previous parity assignment.
Consequently the theory on the 3-brane is completely free
from ${\cal A}_{4}$ and the gauge invariance of the effective four dimensional gauge theory in intact. Let the Kaluza-Klien expansion
of the ${\cal A}_{\mu}(x,y)$ gauge field is given by
\be\begin{array}{lllll}\label{KK1}
   \displaystyle {\cal A}_{\mu}(x,y)=\sum^{\infty}_{n=0}{\cal A}^{(n)}_{\mu}(x)~\frac{\chi^{(n)}_{\pm;\cal A}(y)}{\sqrt{r_{c}}}. 
   \end{array}\ee
 Additionally the Gauss law constraint is given by
\be\begin{array}{lll}\label{gl}
   \displaystyle \overrightarrow{\partial}_{4}\left(\overrightarrow{\partial}_{\mu}{\cal A}^{\mu}(x,y)\right)=0~\implies~\eta^{\mu\nu}\sum^{\infty}_{n=0}
\overrightarrow{\partial}_{\mu}{\cal A}^{(n)}_{\nu}(x)\overrightarrow{{\cal D}_{y}}\chi^{(n)}_{\pm;\cal A}(y)=0.
   \end{array}\ee an outcome 
But this implies for $n=0$ we have $\overrightarrow{{\cal D}_{y}}\chi^{(0)}_{\pm;\cal A}=0$ and due the four dimensional ${\cal U}(1)$ gauge invariance 
this condition is not imposed on the zero mode ${\cal A}^{(0)}_{\mu}$. On the other hand 
\be\begin{array}{lll}\label{gl1}
   \displaystyle \eta^{\mu\nu}
\overrightarrow{\partial}_{\mu}{\cal A}^{(n)}_{\nu}(x)=0~~~~~~\forall ~n\neq 0
   \end{array}\ee
due to $\overrightarrow{{\cal D}_{y}}\chi^{(n)}_{\pm;\cal A}(y)\neq 0$. This is very important criteria 
satisfied by the massive vector particles in four dimensional flat Minkowski space.
Now plugging equation(\ref{KK1}) in equation(\ref{abs1}) the effective four dimensional action reduces to the following form:
\be\begin{array}{llll}\label{abs2}
   \displaystyle S_{{\cal A}}=-\int d^{4}x\sum^{\infty}_{n=0}\left[\frac{1}{4}\eta^{\mu\kappa}\eta^{\nu\lambda}
{\cal F}^{(n)}_{\kappa\lambda}(x){\cal F}^{(n)}_{\mu\nu}(x)+\frac{1}{2}\left(m^{{\cal A}}_{n}\right)^{2}_{\pm}
\eta^{\nu\lambda}{\cal A}^{(n)}_{\nu}(x){\cal A}^{(n)}_{\lambda}(x)\right]\end{array}\ee

where the effective four dimensional ${\cal U}(1)$ abelian gauge field strength is defined as ${\cal F}^{(n)}_{\mu\nu}(x):=\overrightarrow{\partial}_{[\mu}{\cal A}^{(n)}_{\nu]}(x)$.
In this context we impose the following orthonormalization condition of extra dimension dependent wave functions
\be\begin{array}{llll}\label{no1}
 \displaystyle   \int^{+\pi}_{-\pi}dy~\chi^{(m)}_{\pm;\cal A}(y)~\chi^{(n)}_{\pm;\cal A}(y)=\delta^{mn}
   \end{array}\ee
and the mass term of the gauge field is defined through the following differential equation as
\be\begin{array}{llll}\label{dif1}
  \displaystyle   -\frac{1}{r^{2}_{c}}\overrightarrow{{\cal D}_{y}}\left(e^{-2A_{\pm}(y)}\overrightarrow{{\cal D}_{y}}\chi^{(n)}_{\pm;\cal A}(y)\right)=\left(m^{{\cal A}}_{n}\right)^{2}_{\pm}\chi^{(n)}_{\pm;\cal A}(y).
   \end{array}\ee

Now introducing two new variables $z^{\pm;{\cal A}}_{n}:=\frac{\left(m^{{\cal A}}_{n}\right)_{\pm}}{k_{\pm}}e^{A_{\pm}(y)}$ and $f^{\pm;{\cal A}}_{n}:=e^{-A_{\pm}(y)}\chi^{(n)}_{\cal A}(y)$
equation(\ref{dif1}) can be recast in terms of Bessel differential equation of order $1$ as
\be\begin{array}{llll}\label{dif2}
    \displaystyle  \left[\left(z^{\pm;{\cal A}}_{n}\right)^{2}\overrightarrow{{\cal D}^{2}}_{z^{\pm;{\cal A}}_{n}}
+z^{\pm;{\cal A}}_{n}\overrightarrow{{\cal D}}_{z^{\pm;{\cal A}}_{n}}+\left\{\left(z^{\pm;{\cal A}}_{n}\right)^{2}-1\right\}\right]f^{\pm}_{n}
=0
   \end{array}\ee
and the analytical solution turns out to be
\be\begin{array}{llll}\label{sol1}
\displaystyle    \chi^{(n)}_{\pm;\cal A}(y)=\frac{e^{{\cal A}_{\pm}(y)}}{{\cal N}^{\pm;\cal A}_{(n)}}\left[{\cal J}_{1}
(z^{\pm;{\cal A}}_{n})+\alpha^{\pm;\cal A}_{n}{\cal Y}_{1}(z^{\pm;{\cal A}}_{n})\right].
   \end{array}\ee
Here ${\cal N}^{\pm;\cal A}_{(n)}$ be the normalization constant of the extra dimension dependent wave function and $\alpha^{\pm;\cal A}_{n}$ is the
integration constant determined from the orthonormalization condition and the continuity conditions at the orbifold fixed point.
Self-adjointness and hermiticity of the differential operator appearing in equation(\ref{dif2})
demands that $\overrightarrow{{\cal D}_{y}}\chi^{(n)}_{\pm;\cal A}(y)$ is
 continious at the orbifold fixed points $y_{i}=0,\pi$. Consequently we have
\be\begin{array}{llll}\label{cond1}
 \displaystyle  \overrightarrow{{\cal D}_{y}}\chi^{(n)}_{\pm;\cal A}|_{y_{i}=0}=0~~\implies \alpha^{\pm;\cal A}_{n}=
-\frac{\left[{\cal J}_{1}\left(\frac{\left(m^{{\cal A}}_{n}\right)_{\pm}}{k_{\pm}}\right)
+\frac{\left(m^{{\cal A}}_{n}\right)_{\pm}}{k_{\pm}}{\cal J}^{'}_{1}\left(\frac{\left(m^{{\cal A}}_{n}\right)_{\pm}}{k_{\pm}}\right)\right]}{\left[{\cal Y}_{1}\left(\frac{\left(m^{{\cal A}}_{n}\right)_{\pm}}{k_{\pm}}\right)
+\frac{\left(m^{{\cal A}}_{n}\right)_{\pm}}{k_{\pm}}{\cal Y}^{'}_{1}\left(\frac{\left(m^{{\cal A}}_{n}\right)_{\pm}}{k_{\pm}}\right)\right]}.
   \end{array}\ee

\be\begin{array}{llll}\label{cond2}
 \displaystyle  \overrightarrow{{\cal D}_{y}}\chi^{(n)}_{\pm;\cal A}|_{y_{i}=\pi}=0~~\implies \alpha^{\pm;\cal A}_{n}=
-\frac{\left[{\cal J}_{1}\left(x^{\pm;A}_{n}\right)
+x^{\pm;A}_{n}{\cal J}^{'}_{1}\left(x^{\pm;A}_{n}\right)\right]}{\left[{\cal Y}_{1}\left(x^{\pm;A}_{n}\right)
+x^{\pm;A}_{n}{\cal Y}^{'}_{1}\left(x^{\pm;A}_{n}\right)\right]}
   \end{array}\ee
where $z^{\pm;A}_{n}(\pi):=x^{\pm;A}_{n}=\frac{\left(m^{{\cal A}}_{n}\right)_{\pm}}{k_{\pm}}e^{k_{\pm}r_{c}\pi}$.
For $e^{k_{\pm}r_{c}\pi}\gg 1,~\frac{\left(m^{{\cal A}}_{n}\right)_{\pm}}{k_{\pm}}\ll 1$ the mass spectrum for the gauge fields is expected to be of the order of TeV scale i.e.
\be\begin{array}{llll}\label{approx}
 \displaystyle    \alpha^{\pm;\cal A}_{n}\simeq -\frac{\pi}{2\left[\ln\left(\frac{x^{\pm;A}_{n}}{2}\right)-k_{\pm}r_{c}\pi+\gamma+\frac{1}{2}\right]} 
   \end{array}\ee
where $\gamma=0.5772=-\psi(1)$ is the Euler-Mascheroni constant. In general $\psi(n+1)$ is defined through the well known Gamma function as
\be\begin{array}{llll}\label{gama}
  \displaystyle  \Gamma(\epsilon-n)=\frac{(-1)^{n}}{n!}\left[\frac{1}{\epsilon}+\psi(n+1)+\frac{\epsilon}{2}\left\{\frac{\pi^{2}}{3}+\psi^{2}(n+1)-\psi^{'}(n+1)
\right\}+{\cal O}(\epsilon^{2})\right]
   \end{array}\ee
where
\be\begin{array}{llll}\label{si}
    \displaystyle \psi(n+1)=\sum^{n}_{m=1}\frac{1}{m}-\gamma,
   \end{array}\ee
\be\begin{array}{lllll}\label{ders}
  \displaystyle  \psi^{'}(n+1)=\frac{\pi^{2}}{6}-\sum^{n}_{m=1}\frac{1}{m^{2}}.
   \end{array}\ee

Now using equation(\ref{approx}) and equation(\ref{cond1}) we get
\be\begin{array}{llllll}\label{root}
 \displaystyle    \frac{\pi}{2\left[\ln\left(\frac{x^{\pm;A}_{n}}{2}\right)-k_{\pm}r_{c}\pi+\gamma+\frac{1}{2}\right]}
=  \frac{\left[{\cal J}_{1}\left(x^{\pm;A}_{n}e^{-k_{\pm}r_{c}\pi}\right)
+\frac{x^{\pm;A}_{n}}{2}e^{-k_{\pm}r_{c}\pi}\left\{{\cal J}_{0}\left(x^{\pm;A}_{n}e^{-k_{\pm}r_{c}\pi}\right)-{\cal J}_{2}\left(x^{\pm;A}_{n}e^{-k_{\pm}r_{c}\pi}\right)\right\}\right]}{\left[{\cal Y}_{1}\left(x^{\pm;A}_{n}e^{-k_{\pm}r_{c}\pi}\right)
+\frac{x^{\pm;A}_{n}}{2}e^{-k_{\pm}r_{c}\pi}\left\{{\cal Y}_{0}\left(x^{\pm;A}_{n}e^{-k_{\pm}r_{c}\pi}\right)-{\cal Y}_{2}\left(x^{\pm;A}_{n}e^{-k_{\pm}r_{c}\pi}\right)\right\}\right]} 
   \end{array}\ee
which is an transcendental equation of $x^{\pm;A}_{n}$ and the roots of this equation gives the gauge field mass spectrum $\left(m^{{\cal A}}_{n}\right)_{\pm}$ in presence of perturbative 
Gauss-Bonnet coupling $\alpha_{(5)}$.
This leads to approximately
\be\begin{array}{llll}\label{massasdgag}
   \displaystyle \left(m^{{\cal A}}_{n}\right)_{\pm}\approx
 \left(n\mp\frac{1}{4}\right)\pi k_{\pm}e^{-k_{\pm}r_{c}\pi}.
   \end{array}\ee
Once again as in case of graviton, Gauss-Bonnet coupling $\alpha_{(5)}$ has to have an upper bound to detect their signature in TeV scale experiment.
\begin{figure}[ht]
\centering
\subfigure[]{
    \includegraphics[width=8.5cm,height=7cm] {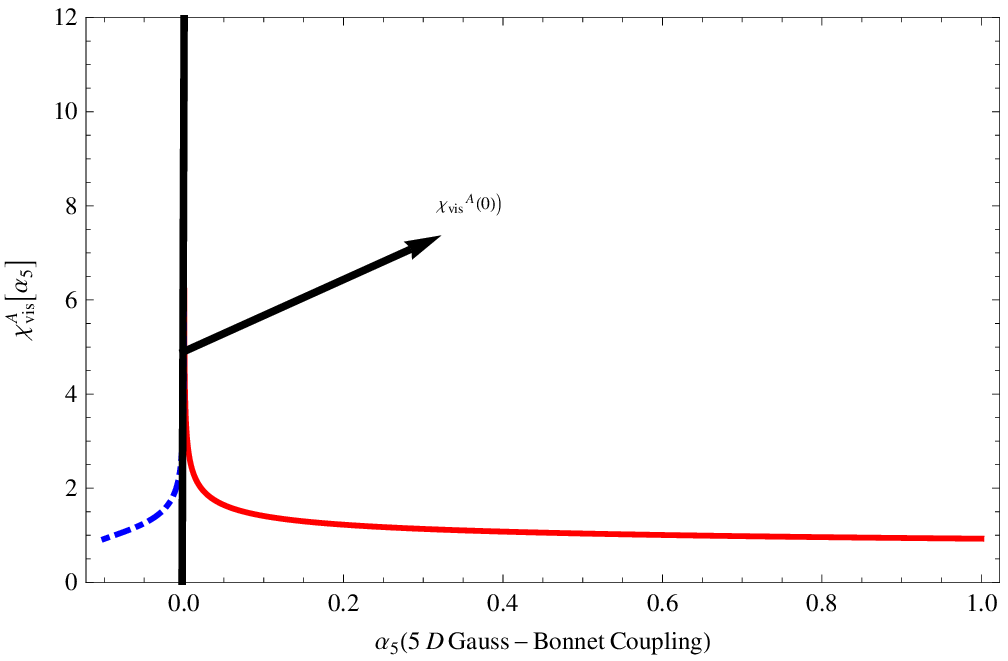}
    \label{fig:subfig12}
}
\subfigure[]{
    \includegraphics[width=8.5cm,height=7cm] {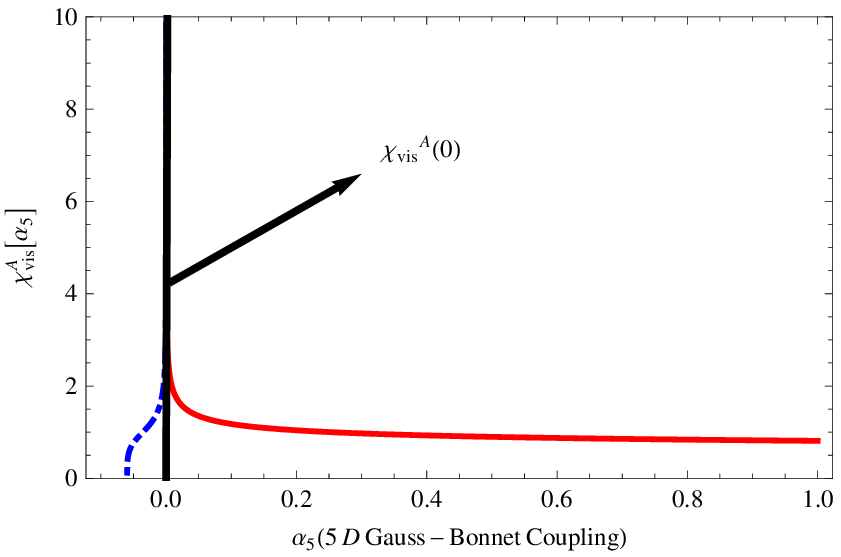}
    \label{fig:subfig13}
}
\subfigure[]{
    \includegraphics[width=8.5cm,height=7cm] {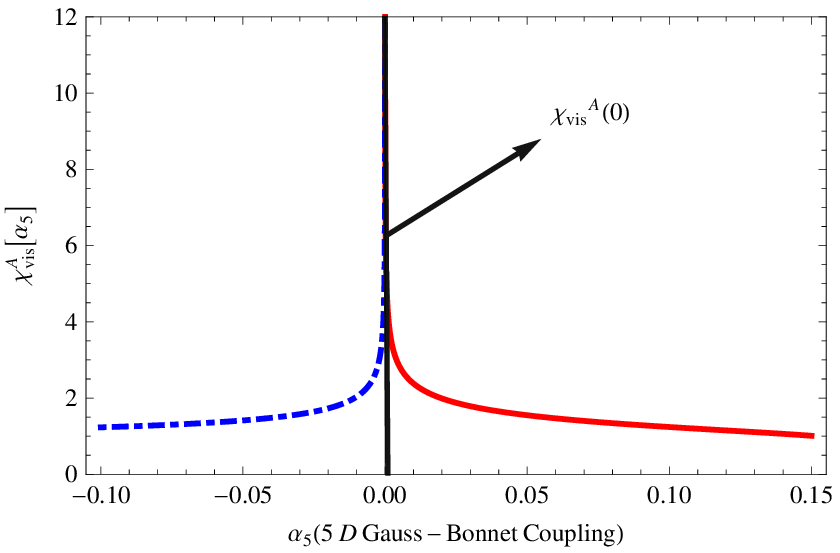}
    \label{fig:subfig14}
}
\subfigure[]{
    \includegraphics[width=8.5cm,height=7cm] {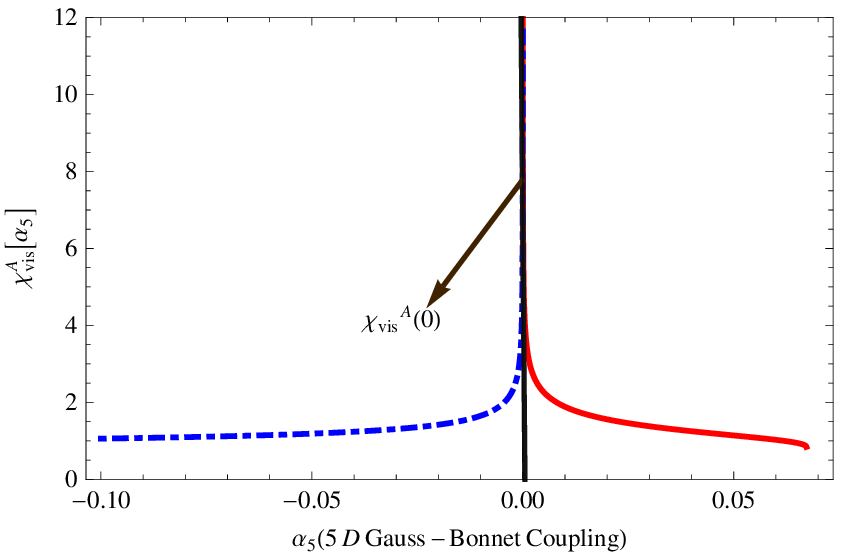}
    \label{fig:subfig15}
}
\caption[Optional caption for list of figures]{Variation
 of $\chi^{(1)}_{\pm;\cal A}(\pi)(=\chi^{(1)}_{\pm;vis})$ 
vs Gauss-Bonnet coupling $\alpha_{(5)}$ for \subref{fig:subfig12} $\Lambda_{(5)}>0$ and $A_{1}>0$, \subref{fig:subfig13} $\Lambda_{(5)}>0$ and $A_{1}<0$, 
\subref{fig:subfig14} $\Lambda_{(5)}<0$ and $A_{1}>0$ and
\subref{fig:subfig15} $\Lambda_{(5)}<0$ and $A_{1}<0$. In this context $B_{0}=0.002$, $r_{c}=0.996\sim 1$, $|A_{1}|=0.04$, $\theta_{1}=0.05$ and $\theta_{2}=0.04$.}
\label{fig:subfigureExample505}
\end{figure}

\begin{figure}[ht]
\centering
\subfigure[]{
    \includegraphics[width=8.5cm,height=7cm] {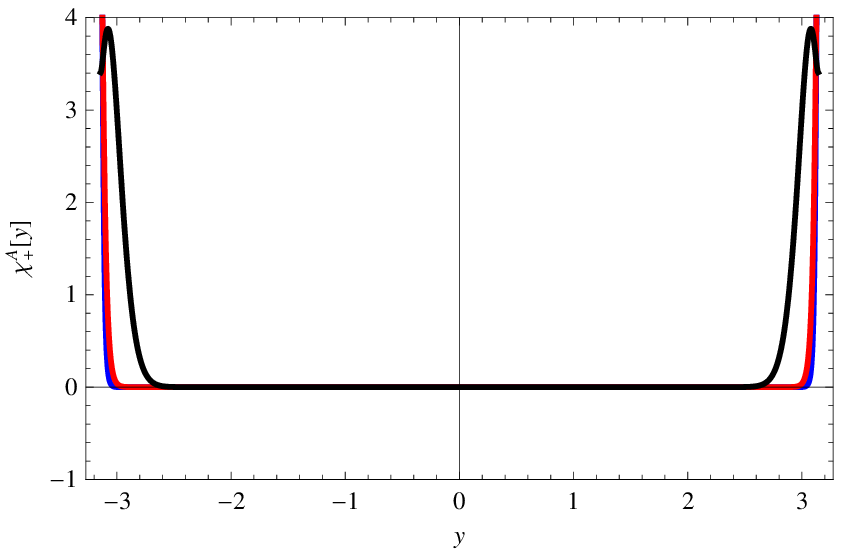}
    \label{fig:subfig19}
}
\subfigure[]{
    \includegraphics[width=8.5cm,height=7cm] {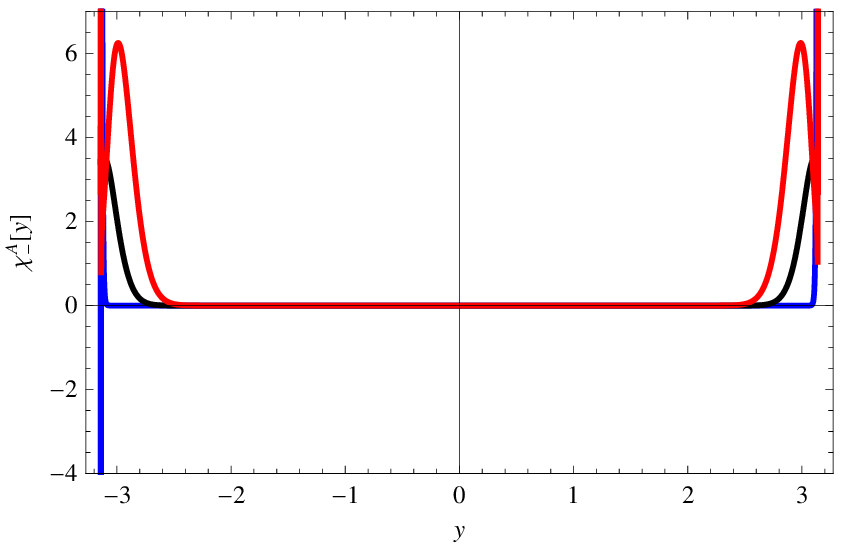}
    \label{fig:subfig20}
}
\caption[Optional caption for list of figures]{Variation
 of $\chi^{(1)}_{+;\cal A}(y)$ and $\chi^{(1)}_{-;\cal A}(y)$
vs extra dimensional coordinate $y$ for \subref{fig:subfig19} $\Lambda_{(5)}<0$ and $A_{1}<0$
and \subref{fig:subfig20} $\Lambda_{(5)}>0$ and $A_{1}<0$ respectively.
 In this context we use $B_{0}=0.002$, $r_{c}=0.996\sim 1$, $|A_{1}|=0.04$, $\theta_{1}=0.05$ and $\theta_{2}=0.04$
three distinct values of Gauss-Bonnet Coupling $\alpha_{(5)}$.}
\label{fig:subfigureExample515}
\end{figure}


 Now using equation(\ref{no1}) the normalization constant for $n\neq 0$ mode reduces to the following expression
\be\begin{array}{llll}\label{vcvc}
    \displaystyle {\cal N}^{\pm;\cal A}_{(n)}=\frac{e^{k_{\pm}r_{c}\pi}}{
\sqrt{k_{\pm}r_{c}}}\sqrt{\left\{\left[{\cal J}_{1}\left(x^{\pm;A}_{n}\right)+\alpha^{\pm;\cal A}_{n}{\cal Y}_{1}\left(x^{\pm;A}_{n}\right)
\right]^{2}-e^{-2k_{\pm}r_{c}}\left[{\cal J}_{1}\left(x^{\pm;A}_{n}e^{-k_{\pm}r_{c}\pi}
\right)+\alpha^{\pm;\cal A}_{n}{\cal Y}_{1}\left(x^{\pm;A}_{n}e^{-k_{\pm}r_{c}\pi}\right)
\right]^{2}\right\}}. 
   \end{array}\ee
For $e^{k_{\pm}r_{c}\pi}\gg 1,~\frac{\left(m^{{\cal A}}_{n}\right)_{\pm}}{k_{\pm}}\ll 1$ the integration constant $\alpha^{\pm;\cal A}_{n}\ll 1$.
Consequently ${\cal Y}_{1}(z^{\pm;{\cal A}}_{n})$ is neglected compared to ${\cal J}_{1}(z^{\pm;{\cal A}}_{n})$ in equation(\ref{sol1}) 
and then the normalization constant for $n\neq 0$ mode
turns out to be
\be\begin{array}{llll}\label{vcvc}
    \displaystyle {\cal N}^{\pm;\cal A}_{(n)}=\frac{e^{k_{\pm}r_{c}\pi}}{
\sqrt{k_{\pm}r_{c}}}{\cal J}_{1}\left(x^{\pm;A}_{n}\right). 
   \end{array}\ee
Consequently the extra dimension dependent wave function for $n\neq 0$ modes 
turns out to be 
 \be\begin{array}{lllll}\label{wavnewdf}\displaystyle
\chi^{(n)}_{\pm;\cal A}(y)=
\frac{e^{{\cal A}_{\pm}(y)}\sqrt{k_{\pm}r_{c}}}{e^{k_{\pm}r_{c}\pi}}
\frac{{\cal J}_{1}(z^{\pm;{\cal A}}_{n})}{{\cal J}_{1}(x^{\pm;{\cal A}}_{n})}.
\end{array}\ee
In figure(\ref{fig:subfigureExample505}) we have explicitly shown the behavior of the extra dimension
dependent first excited state for all possible signatures of Gauss-Bonnet coupling for $k_{+}$ and $k_{-}$ branch.
Additionally in figure(\ref{fig:subfigureExample515}) we have plotted the graphical behavior 
of the extra dimension dependent wave function for $k_{+}$ and $k_{-}$ branch corresponding the first excited 
state for two possible signatures of bulk cosmological constant $\Lambda_{(5)}$ and negative two -loop conformal coupling $A_{1}$
for three distinct values of Gauss-Bonnet coupling $\alpha_{(5)}$.
 
For massless $n=0$ mode the solution of the equation(\ref{dif1}) turns out to be
\be\begin{array}{lllll}\label{jklo}
   \displaystyle  \chi^{(0)}_{\pm;\cal A}=-\frac{C_{1}}{2k_{\pm}r_{c}}e^{-2A_{\pm}(y)}+C_{2}.
   \end{array}\ee
Here $C_{1}$ and $C_{2}$ are arbitrary integration constants. Now applying the boundary condition through the continuity
of the wave function we get $C_{1}=0$. As a result the zero mode solution turns out to be $ \chi^{(0)}_{\pm;\cal A}=C_{2}$.
Now applying the normalization condition the ground state massless zero mode wave function turns out to be
\be\begin{array}{llll}\label{ml}
   \displaystyle  \chi^{(0)}_{\pm;\cal A}=C_{2}=\frac{1}{\sqrt{2\pi}}.
   \end{array}\ee

\subsection{\bf ${\cal SU}({\cal N})$ Non-Abelian Gauge Field}
\label{f2}

The five dimensional action for the pure ${\cal SU}({\cal N})$ non-abelian gauge theory can be written as
\be\begin{array}{llll}\label{absnon}
   \displaystyle S_{{\cal A}}=-\frac{1}{4}\int d^{5}x\sqrt{-g_{(5)}}~{\cal F}^{a}_{AB}(x,y){\cal F}^{AB}_{a}(x,y)
   \end{array}\ee
where the five dimensional rank-2 antisymmetric ${\cal SU}({\cal N})$ non-abelian gauge field strength tensor is given by
\be\begin{array}{llll}\label{abse}
   \displaystyle {\cal F}^{a}_{MN}:=\overrightarrow{\partial}_{[M}{\cal A}^{a}_{N]} (x,y)+{ g}_{{\cal A}}{ f}^{abc}{\cal A}_{Mb}(x,y){\cal A}_{Nc}(x,y)
   \end{array}\ee
with the matrix valued five dimensional non-abelian gauge field is defined as ${\cal A}^{a}_{M}:=\left({\cal A}^{a}_{\alpha},{\cal A}^{a}_{4}\right)$. 
Here the superscript $a$ is used for ${\cal SU}({\cal N})$ non-abelian gauge index runs from $a=1,2,.....,{\cal N}^{2}-1$. Next applying the gauge constraint
the action reduces to the following form
\be\begin{array}{llll}\label{nonab}
   \displaystyle S_{{\cal A}}=-\frac{1}{4}\int d^{5}x\left[\eta^{\mu\kappa}\eta^{\nu\lambda}
{\cal F}^{a}_{\kappa\lambda}(x,y){\cal F}_{\mu\nu;a}(x,y)-2\eta^{\nu\lambda}{\cal A}_{\lambda a}(x,y)
\overrightarrow{{\cal D}_{y}}\left(e^{-2A_{\pm}(y)}\overrightarrow{{\cal D}_{y}}{\cal A}^{a}_{\nu}(x,y)\right)\right]. 
   \end{array}\ee
In this context the Kaluza-Klien decomposition of the ${\cal SU}({\cal N})$ non-abelian gauge field ${\cal A}^{a}_{\mu}(x,y)$ can be written as
\be\begin{array}{lllll}\label{KK2}
   \displaystyle {\cal A}^{a}_{\mu}(x,y)=\sum^{\infty}_{n=0}{\cal A}^{a;(n)}_{\mu}(x)~\frac{\chi^{(n)}_{\pm;\cal A}(y)}{\sqrt{r_{c}}}. 
   \end{array}\ee
Substituting equation(\ref{KK2}) in equation(\ref{nonab}) we get 
\be\begin{array}{llll}\label{nonab}
   \displaystyle S_{{\cal A}}=\int d^{4}x\sum^{\infty}_{n=0}\left[-\frac{1}{4}\left\{\left(\overrightarrow{\partial}_{[\mu}{\cal A}^{a;(n)}_{\nu]}(x)\right)^{2}
+2\frac{{ g}_{{\cal A}}}{r^{\frac{3}{2}}_{c}}f^{abc}\sum^{\infty}_{p=0}\sum^{\infty}_{q=0}{\cal C}^{(npq)}_{1}\left(\overrightarrow{\partial}_{[\mu}{\cal A}^{a;(n)}_{\nu]}(x)\right)
{\cal A}_{b}^{\mu;(p)}(x){\cal A}_{c}^{\mu;(q)}(x)\right.\right.\\ \left.\left
.~~~~~~~~~~~~~~~~\displaystyle 
+\frac{{ g}^{2}_{{\cal A}}}{r^{2}_{c}}f^{abc}f_{ade}\sum^{\infty}_{m=0}\sum^{\infty}_{p=0}\sum^{\infty}_{q=0}{\cal C}^{(nmpq)}_{2}
{\cal A}_{\mu b}^{(n)}(x){\cal A}_{\mu c}^{(m)}(x){\cal A}^{d\mu;(p)}(x){\cal A}^{e\nu;(q)}(x)\right\}-\frac{1}{2}\left(m^{{\cal A}}_{n}\right)^{2}_{\pm}
\eta^{\nu\lambda}{\cal A}^{(n)}_{\nu}(x){\cal A}^{a;(n)}_{\lambda a}(x)
\right] 
\end{array}\ee
where
\be\begin{array}{llll}\label{cup1}
   \displaystyle {\cal C}^{(npq)}_{1}:=\int^{+\pi}_{-\pi} dy~\chi^{(n)}_{\pm;\cal A}(y)~\chi^{(p)}_{\pm;\cal A}(y)~\chi^{(q)}_{\pm;\cal A}(y)\\
  \displaystyle {\cal C}^{(nmpq)}_{2}:=\int^{+\pi}_{-\pi} dy~\chi^{(n)}_{\pm;\cal A}(y)~\chi^{(m)}_{\pm;\cal A}(y)~\chi^{(p)}_{\pm;\cal A}(y)~\chi^{(q)}_{\pm;\cal A}(y)  
\end{array}\ee
characterizes trilinear and quartic self interaction of ${\cal SU}({\cal N})$ non-abelian gauge field. Most importantly the terms appearing in equation(\ref{cup1}) terms breaks the 
${\cal SU}({\cal N})$ non-abelian gauge invariance in the four dimensional effective field theory. But the amount of ${\cal SU}({\cal N})$ gauge breaking 
can be considerably small if the non-abelian ${\cal SU}({\cal N})$ gauge coupling ${ g}_{{\cal A}}$ is very small. In table(\ref{tab1})
- table(\ref{tab2a}) we have tabulated the numerical values of 
the trilinear and quartic interaction for zero and lowest lying modes for $k_{-}$ and $k_{+}$ branch. During the evaluation of these interaction terms we use the previous results for ${\cal U}(1)$ abelian gauge field
theory since for both of the cases the results are exactly same.

\begin{table}[h]
\begin{tabular}{|c|c|c|c|c|c|c|c|c|c|c|c|c|}
\hline ${\cal C}^{(000)}_{1}$ & ${\cal C}^{(001)}_{1}$ & ${\cal C}^{(010)}_{1}$ &
${\cal C}^{(011)}_{1}$&${\cal C}^{(100)}_{1}$&${\cal C}^{(101)}_{1}$&${\cal C}^{(110)}_{1}$&${\cal C}^{(111)}_{1}$
 \\
 \hline
0.398&0.214 &0.214 &0.113 &0.214 &0.113 &0.113 &0.068\\
\hline
\end{tabular}
\caption{Numerical values of ${\cal C}^{(npq)}_{1}$ for lowest lying modes of the trilinear ${\cal SU}({\cal N})$ non-abelian gauge interaction for $k_{-}$ branch.}\label{tab1}
\end{table}

\begin{table}[h]
\begin{tabular}{|c|c|c|c|c|c|c|c|c|c|c|c|c|}
\hline ${\cal C}^{(000)}_{1}$ & ${\cal C}^{(001)}_{1}$ & ${\cal C}^{(010)}_{1}$ &
${\cal C}^{(011)}_{1}$&${\cal C}^{(100)}_{1}$&${\cal C}^{(101)}_{1}$&${\cal C}^{(110)}_{1}$&${\cal C}^{(111)}_{1}$
 \\
 \hline
0.398&0.378 &0.378 &0.234 &0.378 &0.234 &0.234 &0.123\\
\hline
\end{tabular}
\caption{Numerical values of ${\cal C}^{(npq)}_{1}$ for lowest lying modes of the trilinear ${\cal SU}({\cal N})$ non-abelian gauge interaction for $k_{+}$ branch.}\label{tab1a}
\end{table}

\begin{table}[h]
\begin{tabular}{|c|c|c|c|c|c|c|c|c|c|c|c|c|c|c|c|c|c|c|c|c|}
\hline ${\cal C}^{(0000)}_{2}$ & ${\cal C}^{(0001)}_{2}$ & ${\cal C}^{(0010)}_{2}$ &
${\cal C}^{(0011)}_{2}$&${\cal C}^{(0100)}_{2}$&${\cal C}^{(0101)}_{2}$&${\cal C}^{(0110)}_{2}$&${\cal C}^{(0111)}_{2}$&${\cal C}^{(1000)}_{2}$&${\cal C}^{(1001)}_{2}$
&${\cal C}^{(1010)}_{2}$&${\cal C}^{(1011)}_{2}$&${\cal C}^{(1100)}_{2}$&${\cal C}^{(1101)}_{2}$&${\cal C}^{(1110)}_{2}$&${\cal C}^{(1111)}_{2}$
 \\
 \hline
0.159&0.115 &0.115 &0.094 &0.115 &0.094 &0.094 &0.045 &0.115 &0.094 &0.094 &0.045 & 0.094&0.045 &0.045 &0.007 \\
\hline
\end{tabular}
\caption{Numerical values of ${\cal C}^{(nmpq)}_{2}$ for lowest lying modes of the quartic ${\cal SU}({\cal N})$ non-abelian gauge interaction for $k_{-}$ branch.}\label{tab2}
\end{table}
\begin{table}[h]
\begin{tabular}{|c|c|c|c|c|c|c|c|c|c|c|c|c|c|c|c|c|c|c|c|c|}
\hline ${\cal C}^{(0000)}_{2}$ & ${\cal C}^{(0001)}_{2}$ & ${\cal C}^{(0010)}_{2}$ &
${\cal C}^{(0011)}_{2}$&${\cal C}^{(0100)}_{2}$&${\cal C}^{(0101)}_{2}$&${\cal C}^{(0110)}_{2}$&${\cal C}^{(0111)}_{2}$&${\cal C}^{(1000)}_{2}$&${\cal C}^{(1001)}_{2}$
&${\cal C}^{(1010)}_{2}$&${\cal C}^{(1011)}_{2}$&${\cal C}^{(1100)}_{2}$&${\cal C}^{(1101)}_{2}$&${\cal C}^{(1110)}_{2}$&${\cal C}^{(1111)}_{2}$
 \\
 \hline
0.159&0.142 &0.142 &0.108 &0.142 &0.108 &0.108 &0.087 &0.142 &0.108 &0.108 &0.087 & 0.108&0.087 &0.087 &0.056 \\
\hline
\end{tabular}
\caption{Numerical values of ${\cal C}^{(nmpq)}_{2}$ for lowest lying modes of the quartic ${\cal SU}({\cal N})$ non-abelian gauge interaction for $k_{+}$ branch.}\label{tab2a}
\end{table}

\subsection{\bf Massive Fermionic Field}
\label{f3}

In this subsection we explore the profile of bulk fermion wave function where we begin with an action of such fermions
coupled to a ${\cal U}(1)$ bulk gauge field as described in the previous subsection.
The five dimensional action for the massive fermionic field theory $\left(spin~\frac{1}{2}~ type\right)$ can be written as
\be\begin{array}{llll}\label{fer}
 \displaystyle S_{f}=\int d^{5}x\left[Det({\cal V})\right]~\left\{i\bar{\Psi}_{{\bf L},{\bf R}}(x,y)\gamma^{\alpha}
{\cal V}_{\alpha}^{M}\overleftrightarrow{{\large\bf D}_{\mu}}\Psi_{{\bf L},{\bf R}}(x,y)\delta^{\mu}_{M}
-sgn(y)m_{f}\bar{\Psi}_{{\bf L},{\bf R}}(x,y)\Psi_{{\bf R},{\bf L}}(x,y)+~h.c.\right\}
   \end{array}\ee
where $\overleftrightarrow{{\large\bf D}_{\mu}}:=\left(\overleftrightarrow{\partial_{\mu}}
+\Omega_{\mu}+ig_{f}{\cal A}_{\mu}\right)$ represents the covariant derivative in presence
${\cal U}(1)$ abelian gauge field and fermionic spin connection 
$\Omega_{\mu}=\frac{1}{8}\omega_{\mu}^{\hat{A}\hat{B}}\left[\Gamma_{\hat{A}},\Gamma_{\hat{B}}\right]$. 
Here $\omega_{\mu}^{\hat{A}\hat{B}}$ represents the gauge field respecting ${\cal SO}(3,1)$ transformation on the vierbein
coordinate. Most importantly the 5D Gamma matrix is given by $\Gamma^{\hat{A}}=\left(\gamma^{\mu},\gamma_{5}:=
\frac{i}{4!}{\bf \epsilon}_{\mu\nu\alpha\beta}\gamma^{\mu}\gamma^{\nu}\gamma^{\alpha}\gamma^{\beta}=i\gamma_{4}\right)$ satisfies
the Clifford algebra
anti-commutation relation $\{\Gamma^{\hat{A}},\Gamma^{\hat{B}}\}=2\eta^{\hat{A}\hat{B}}$ with $\eta^{\hat{A}\hat{B}}=diag\left(-1,+1,+1,+1,+1\right)$.
In this context 
\be\begin{array}{llll}\label{vbcx}
    \displaystyle g_{MN}:=\left({\cal V}_{M}^{\hat{A}}\otimes{\cal V}_{N}^{\hat{B}}\right)\eta_{\hat{A}\hat{B}}
   \end{array}\ee
where ${\cal V}_{M}^{\hat{A}}$ represents vierbein  (inverse of f$\ddot{u}$nfbein) characterized by the following conditions:
\be\begin{array}{llll}\label{vbcx}
    \displaystyle {\cal V}_{4}^{4}=1,~~~{\cal V}_{\mu}^{\hat{A}}=e^{A_{\pm}(y)}\delta_{\mu}^{\hat{A}},~~Det({\cal V})=e^{-4A_{\pm}(y)}
   \end{array}\ee
and $\hat{A},\hat{B}$ being tangent space indices. 
For our set up ${\cal SO}(3,1)$ gauge field can be written in terms of the christoffel connection and vierbein  degrees of freedom as
\be\begin{array}{llll}\label{scon}
\displaystyle    \omega_{\mu}^{\hat{A}\hat{B}}:=\frac{1}{2}g^{NP}\left({\cal V}_{N}^{[\hat{A}}\partial_{[\mu}{\cal V}_{P]}^{\hat{B}]}
+\frac{1}{2}g^{TS}{\cal V}^{[\hat{A}}_{N}{\cal V}^{\hat{B}]}_{T}\partial_{[S}{\cal V}_{P]}^{\hat{C}}{\cal V}^{\hat{D}}_{\mu}\eta_{\hat{C}\hat{D}}\right)
={\cal V}^{\hat{A}}_{N}\left({\cal V}^{\hat{B}P}\Gamma^{N}_{MP}+\partial_{M}{\cal V}^{\hat{B}N}\right)
   \end{array}\ee
In presence of Gauss-bonnet coupling $\Omega_{4}=0$ and $\Omega_{\mu}=-\frac{1}{2}e^{-A_{\pm}(y)}k_{\pm}r_{c}\gamma_{5}\gamma_{\mu}$.
It is important to mention here that the contribution to the action from the spin connection vanishes due to the presence of the hermitian 
conjugate counterpart is included. Here $\Psi_{{\bf L},{\bf R}}$ represents the left and right chiral fermionic field defined as
\be\begin{array}{lll}\label{proj}
  \displaystyle  \Psi_{{\bf L},{\bf R}}(x,y)\equiv{\cal P}_{{\bf L},{\bf R}}\Psi(x,y)
   \end{array}\ee
where the left/right chiral projection operator is defined as ${\cal P}_{{\bf L},{\bf R}}=\frac{1}{2}\left(1\mp\gamma_{5}\right)$ which satisfies
${\cal P}_{{\bf R}}+{\cal P}_{{\bf L}}=1$ and ${\cal P}_{{\bf R}}{\cal P}_{{\bf L}}={\cal P}_{{\bf L}}{\cal P}_{{\bf R}}=0$.
To find the effective four dimensional action the Kaluza-Klien decomposition of the massive left/right chiral fermionic spin $\frac{1}{2}$ type of field 
is given by 
\be\begin{array}{lllll}\label{opium}
    \displaystyle \Psi_{{\bf L},{\bf R}}\left(x,y\right)=\sum^{\infty}_{n=0}\Psi^{(n)}_{{\bf L},{\bf R}}(x)\frac{e^{2A_{\pm}}(y)}{\sqrt{r_{c}}}\hat{f}^{(n)}_{{\bf L},{\bf R}}(y)
   \end{array}\ee
where ${{\bf L},{\bf R}}$ represent the chirality of the massive fermionic fields and $\hat{f}^{(n)}_{{\bf L},{\bf R}}(y)$ characterizes two distinct set of complete orthonormal
 function satisfies the following orthonormalization criteria:
\be\begin{array}{llll}\label{orth}
\displaystyle    \int^{+\pi}_{-\pi}dy~e^{A_{\pm}(y)}\hat{f}^{(m)\star}_{{\bf L}}(y)\hat{f}^{(n)}_{{\bf L}}(y)=\delta^{mn},\\
\displaystyle    \int^{+\pi}_{-\pi}dy~e^{A_{\pm}(y)}\hat{f}^{(m)\star}_{{\bf R}}(y)\hat{f}^{(n)}_{{\bf R}}(y)=\delta^{mn}.
   \end{array}\ee
Due to the requirement of ${\bf Z_{2}}$ symmetry of the action, $\hat{f}^{(n)}_{{\bf R}}(y)$ and $\hat{f}^{(n)}_{{\bf L}}(y)$ necessarily
have opposite ${\bf Z_{2}}$ parity. Without loosing any physical information we choose $\hat{f}^{(n)}_{{\bf L}}(y)$ to be ${\bf Z_{2}}$ even and 
$\hat{f}^{(n)}_{{\bf R}}(y)$ to be ${\bf Z_{2}}$ odd. Then the matter fields then refers to the zero mode fermionic function $\hat{f}^{(0)}_{{\bf L}}$. 
Consequently the action for the fermionic fields takes the following form
\be\begin{array}{lll}\label{redac}
 \displaystyle S_{f}=\int d^{4}x\int^{+\pi}_{-\pi}dy \left[e^{-3A_{\pm}(y)}\left(\bar{\Psi}_{{\bf L},{\bf R}}(x,y)
i\overleftrightarrow{{\partial}\slashed}{\Psi}_{{\bf L},{\bf R}}(x,y)\right)
-e^{-4A_{\pm}(y)}sgn(y)m_{f}\bar{\Psi}_{{\bf L},{\bf R}}(x,y)\Psi_{{\bf R},{\bf L}}(x,y)\right.\\ \left. \displaystyle~~~~~~~~~~~~
-\bar{\Psi}_{{\bf L},{\bf R}}(x,y)\left(e^{-4A_{\pm}(y)}
\overrightarrow{\partial}_{4}+\overrightarrow{\partial}_{4}e^{-4A_{\pm}(y)}\right)\Psi_{{\bf R},{\bf L}}(x,y)
-g_{f}e^{-3A_{\pm}(y)}\left(\bar{\Psi}_{{\bf L},{\bf R}}(x,y){\cal A}\slashed{\Psi}_{{\bf L},{\bf R}}(x,y)\right)+h.c.\right]\\
\displaystyle~~=\int d^{4}x\sum^{\infty}_{n=0}\left[\bar{\Psi}^{(n)}_{{\bf L},{\bf R}}(x)i\overleftrightarrow{{\partial}\slashed}{\Psi}^{(n)}_{{\bf L},{\bf R}}(x)
-m^{{\bf L},{\bf R}}_{n}\bar{\Psi}^{(n)}_{{\bf L},{\bf R}}(x)\Psi^{(n)}_{{\bf R},{\bf L}}(x)
+\frac{ig_{f}}{\sqrt{r_{c}}}\sum^{\infty}_{m=0}\sum^{\infty}_{p=0}{\cal I}^{(nmp)}_{{\bf L},{\bf R}}\bar{\Psi}^{(n)}_{{\bf L},{\bf R}}(x)
i{\cal A}\slashed^{(m)}(x)\Psi^{(p)}_{{\bf R},{\bf L}}(x)
\right]  
   \end{array}\ee
where the trilinear interaction term between massive fermeonic field and ${\cal U}(1)$ abelian gauge field is given by
\be\begin{array}{lllll}\label{ggcuyt}
    \displaystyle {\cal I}^{(nmp)}_{{\bf L},{\bf R}}:=\int^{+\pi}_{-\pi}dy e^{A_{\pm}(y)}\hat{f}^{(n)\star}_{{\bf L},
{\bf R}}(y)\chi^{(m)}_{\pm;\cal A}(y)\hat{f}^{(p)}_{{\bf L},{\bf R}}(y).
   \end{array}\ee
 In table(\ref{tab3}) and table(\ref{tab3a}) we have tabulated the numerical values of 
the trilinear interaction for zero and lowest lying modes.

\begin{table}[h]
\begin{tabular}{|c|c|c|c|c|c|c|c|c|c|c|c|c|}
\hline ${\cal I}^{(000)}_{{\bf L},{\bf R}}$ & ${\cal I}^{(001)}_{{\bf L},{\bf R}}$ & ${\cal I}^{(010)}_{{\bf L},{\bf R}}$ &
${\cal I}^{(011)}_{{\bf L},{\bf R}}$&${\cal I}^{(100)}_{{\bf L},{\bf R}}$&${\cal I}^{(101)}_{{\bf L},{\bf R}}$&${\cal I}^{(110)}_{{\bf L},{\bf R}}$&${\cal I}^{(111)}_{{\bf L},{\bf R}}$
 \\
 \hline
0.053/0.562&0.031/0.412 &0.028/0.246 &0.014/0.206 &0.031/0.412 &0.009/0.128 &0.014/0.206 &0.001/0.105\\
\hline
\end{tabular}
\caption{Numerical values of ${\cal I}^{(nmp)}_{{\bf L},{\bf R}}$ for lowest lying modes of the trilinear interaction 
between massive fermeonic field and ${\cal U}(1)$ abelian gauge fields for $k_{-}$ branch.}\label{tab3}
\end{table}

\begin{table}[h]
\begin{tabular}{|c|c|c|c|c|c|c|c|c|c|c|c|c|}
\hline ${\cal I}^{(000)}_{{\bf L},{\bf R}}$ & ${\cal I}^{(001)}_{{\bf L},{\bf R}}$ & ${\cal I}^{(010)}_{{\bf L},{\bf R}}$ &
${\cal I}^{(011)}_{{\bf L},{\bf R}}$&${\cal I}^{(100)}_{{\bf L},{\bf R}}$&${\cal I}^{(101)}_{{\bf L},{\bf R}}$&${\cal I}^{(110)}_{{\bf L},{\bf R}}$&${\cal I}^{(111)}_{{\bf L},{\bf R}}$
 \\
 \hline
0.053/0.562&0.043/0.502 &0.035/0.341 &0.027/0.271 &0.043/0.502 &0.026/0.197 &0.027/0.271 &0.012/0.165\\
\hline
\end{tabular}
\caption{Numerical values of ${\cal I}^{(nmp)}_{{\bf L},{\bf R}}$ for lowest lying modes of the trilinear interaction 
between massive fermeonic field and ${\cal U}(1)$ abelian gauge fields for $k_{+}$ branch.}\label{tab3a}
\end{table}

\begin{figure}[ht]
\centering
\subfigure[]{
    \includegraphics[width=8.5cm,height=7cm] {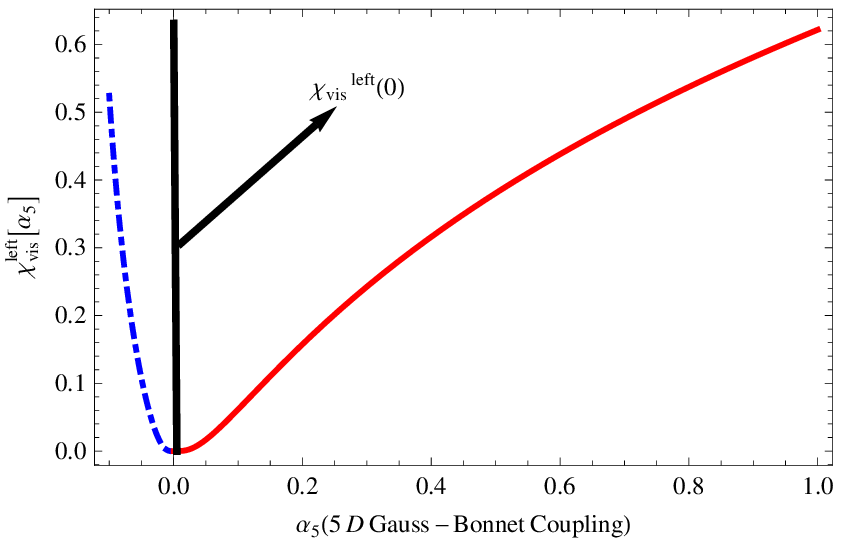}
    \label{fig:subfig32}
}
\subfigure[]{
    \includegraphics[width=8.5cm,height=7cm] {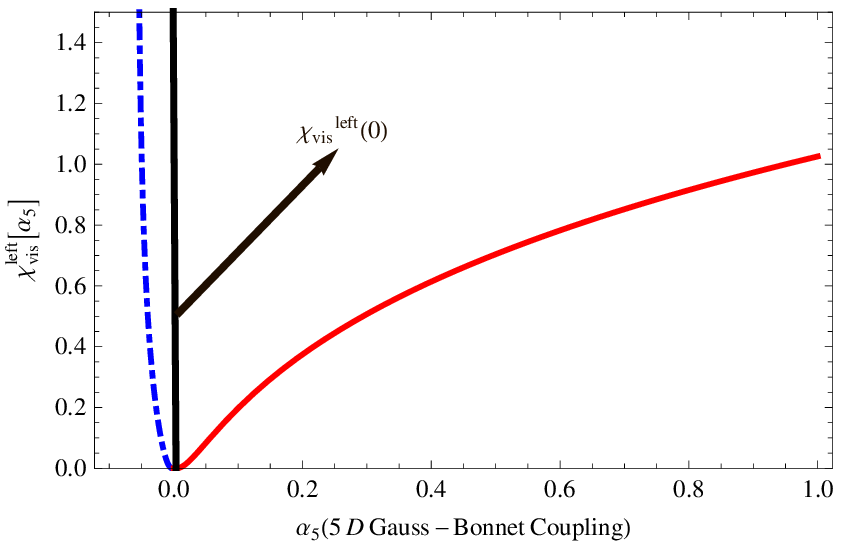}
    \label{fig:subfig33}
}
\subfigure[]{
    \includegraphics[width=8.5cm,height=7cm] {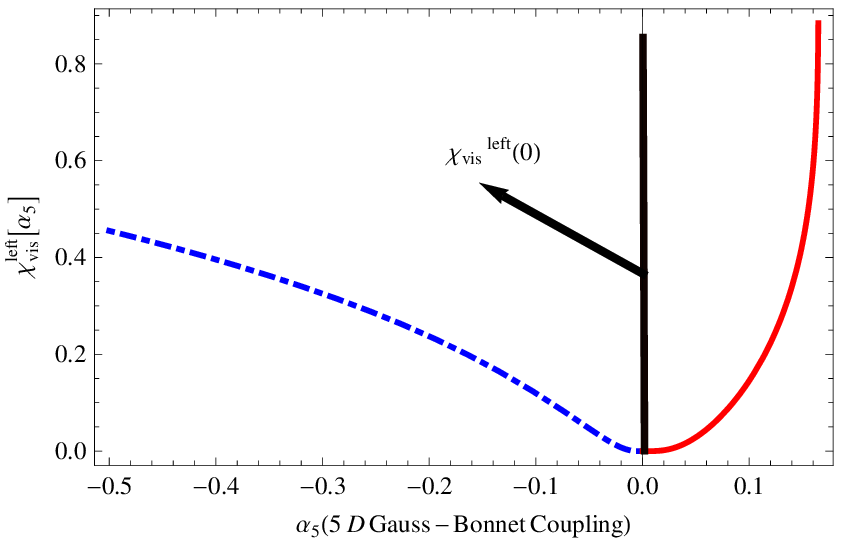}
    \label{fig:subfig34}
}
\subfigure[]{
    \includegraphics[width=8.5cm,height=7cm] {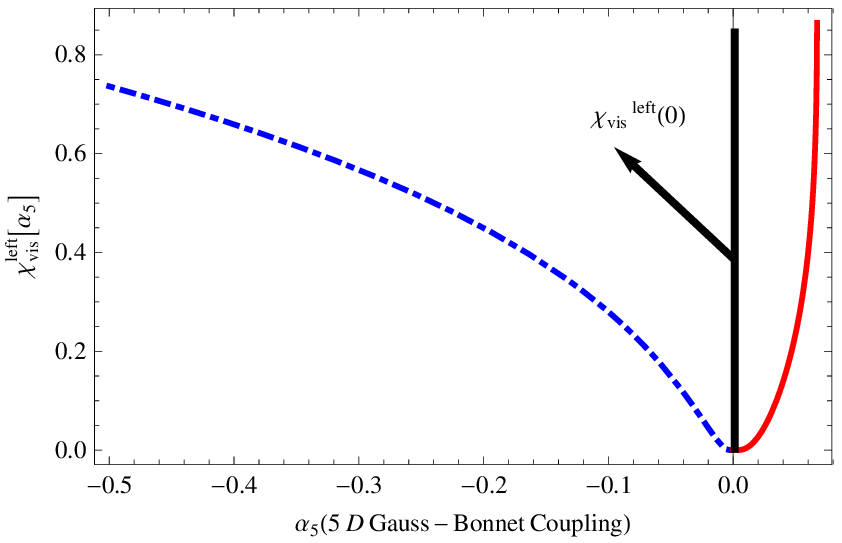}
    \label{fig:subfig35}
}
\caption[Optional caption for list of figures]{Variation
 of $\chi^{(1)}_{\pm;\bf L}(\pi)(=\chi^{(1)}_{vis})$ 
vs Gauss-Bonnet coupling $\alpha_{(5)}$ for \subref{fig:subfig32} $\Lambda_{(5)}>0$ and $A_{1}>0$, \subref{fig:subfig33} $\Lambda_{(5)}>0$ and $A_{1}<0$, 
\subref{fig:subfig34} $\Lambda_{(5)}<0$ and $A_{1}>0$ and
\subref{fig:subfig35} $\Lambda_{(5)}<0$ and $A_{1}<0$. In this context $B_{0}=0.002$, $r_{c}=0.996\sim 1$, $|A_{1}|=0.04$, $\theta_{1}=0.05$ and $\theta_{2}=0.04$.}
\label{fig:subfigureExample5156}
\end{figure}

\begin{figure}[ht]
\centering
\subfigure[]{
   \includegraphics[width=8.5cm,height=7cm] {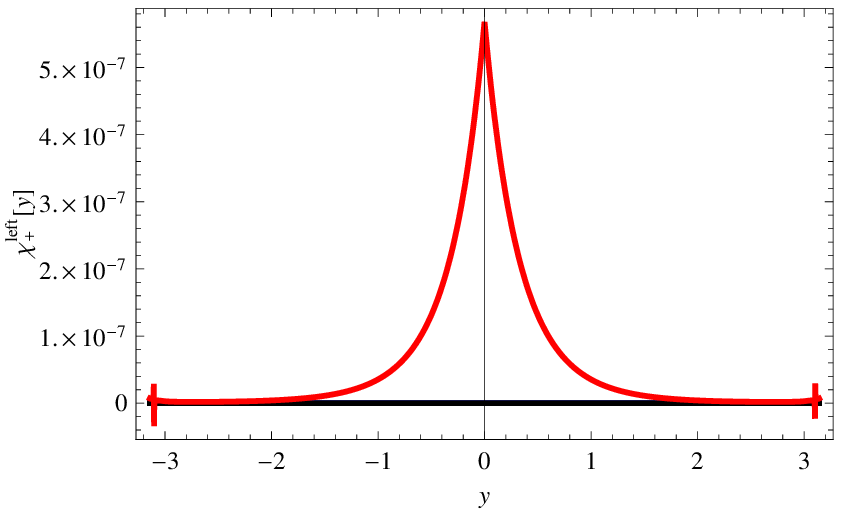}
    \label{fig:subfig38}
}
\subfigure[]{
    \includegraphics[width=8.5cm,height=7cm] {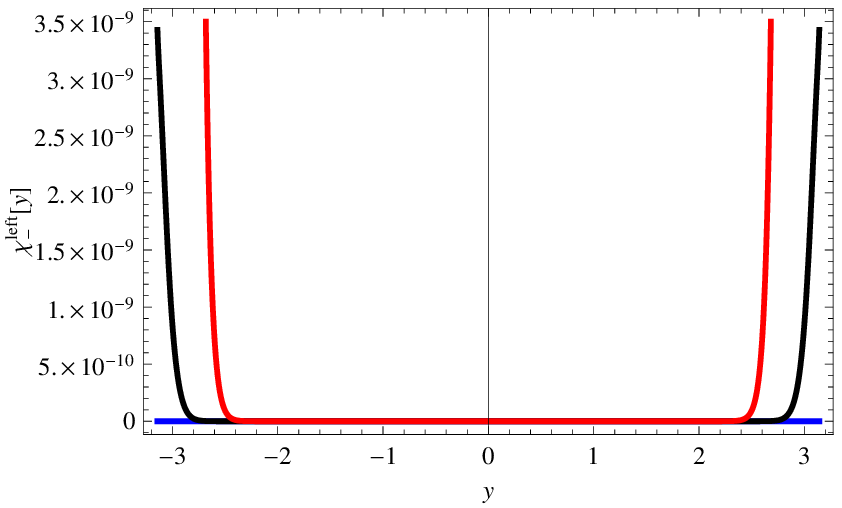}
    \label{fig:subfig39}
}
\caption[Optional caption for list of figures]{Variation
 of $\chi^{(1)}_{+;\bf L}(y)(= \hat{f}^{(1)}_{+,{\bf L}}(y))$ and  $\chi^{(1)}_{-;\bf L}(y)(= \hat{f}^{(1)}_{-,{\bf L}}(y))$
vs extra dimensional coordinate $y$ for \subref{fig:subfig38} $\Lambda_{(5)}<0$ and $A_{1}<0$  
and \subref{fig:subfig39} $\Lambda_{(5)}>0$ and $A_{1}<0$ respectively. In this context we use $B_{0}=0.002$, $r_{c}=0.996\sim 1$, $|A_{1}|=0.04$, $\theta_{1}=0.05$ and $\theta_{2}=0.04$
three distinct values of Gauss-Bonnet Coupling $\alpha_{(5)}$.}
\label{fig:subfigureExample5457}
\end{figure}


\begin{figure}[ht]
\centering
\subfigure[]{
    \includegraphics[width=8.5cm,height=7cm] {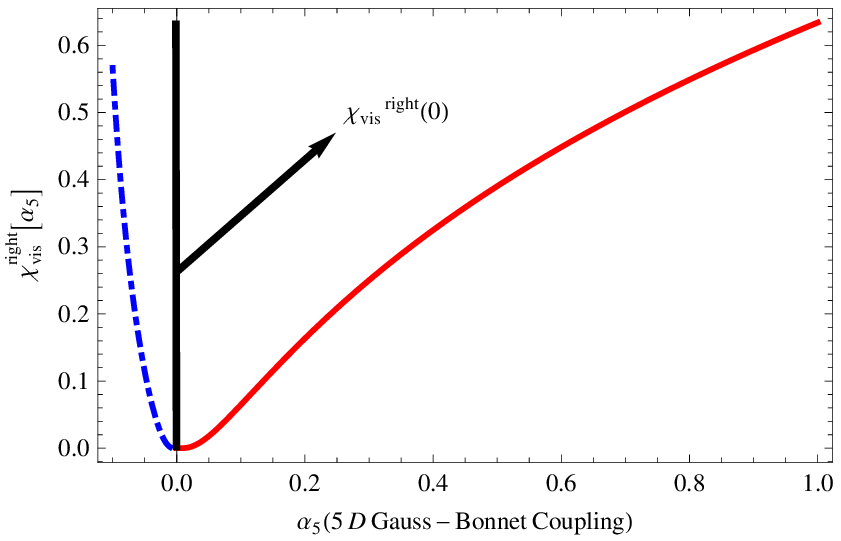}
    \label{fig:subfig321}
}
\subfigure[]{
    \includegraphics[width=8.5cm,height=7cm] {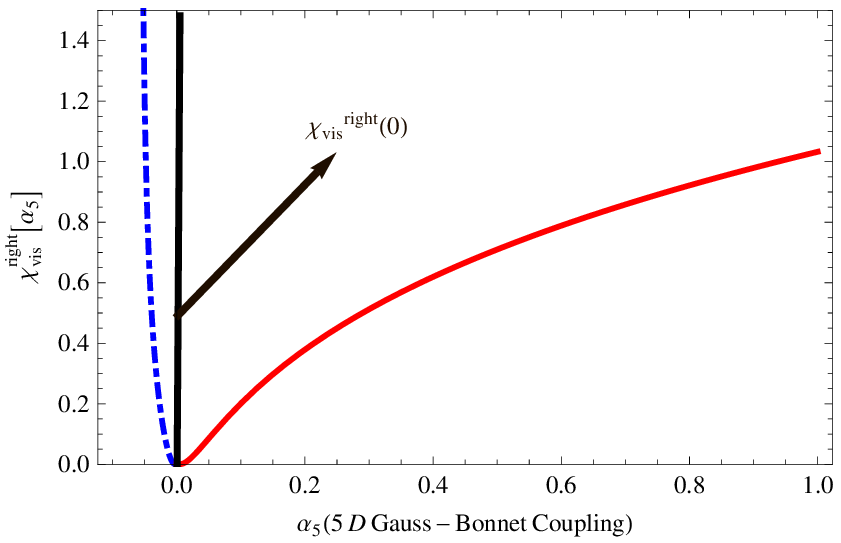}
    \label{fig:subfig331}
}
\subfigure[]{
    \includegraphics[width=8.5cm,height=7cm] {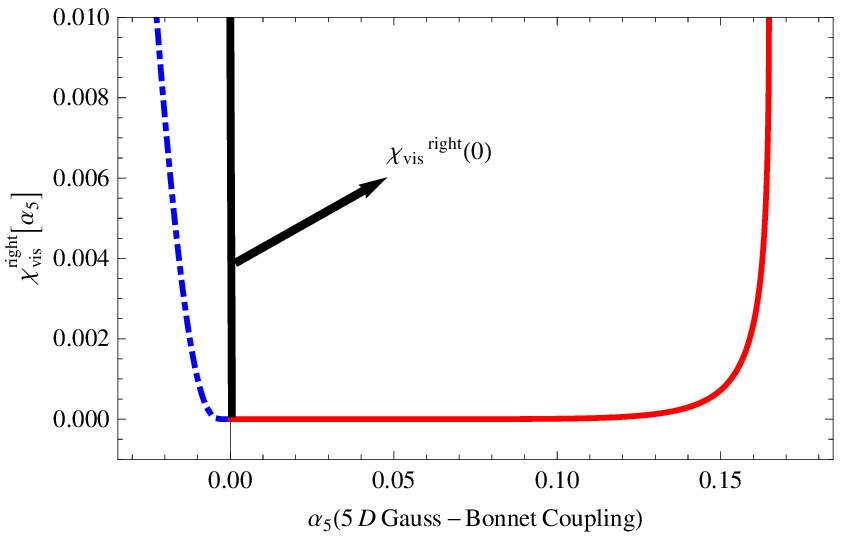}
    \label{fig:subfig341}
}
\subfigure[]{
    \includegraphics[width=8.5cm,height=7cm] {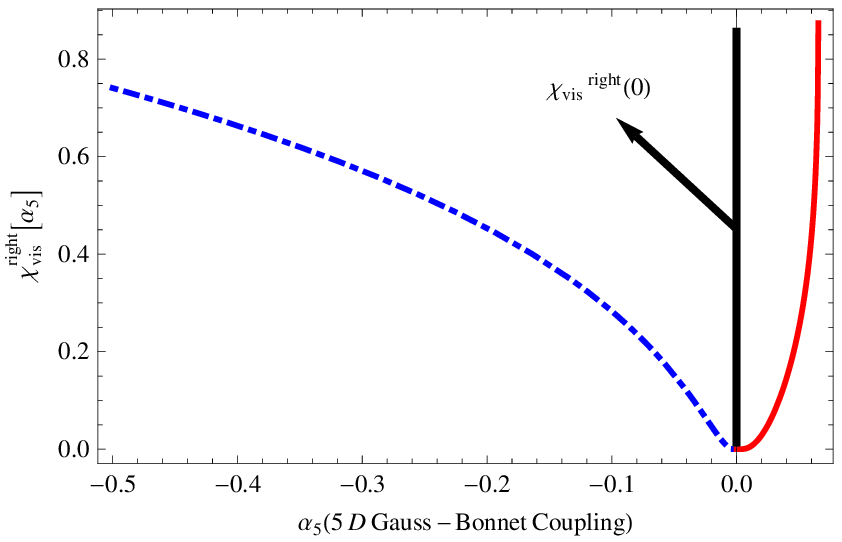}
    \label{fig:subfig351}
}
\caption[Optional caption for list of figures]{Variation
 of $\chi^{(1)}_{\bf R}(\pi)(=\chi^{(1)}_{\pm;vis})$ 
vs Gauss-Bonnet coupling $\alpha_{(5)}$ for \subref{fig:subfig321} $\Lambda_{(5)}>0$ and $A_{1}>0$, \subref{fig:subfig331} $\Lambda_{(5)}>0$ and $A_{1}<0$, 
\subref{fig:subfig341} $\Lambda_{(5)}<0$ and $A_{1}>0$ and
\subref{fig:subfig351} $\Lambda_{(5)}<0$ and $A_{1}<0$. In this context $B_{0}=0.002$, $r_{c}=0.996\sim 1$, $|A_{1}|=0.04$, $\theta_{1}=0.05$ and $\theta_{2}=0.04$.}
\label{fig:subfigureExample51519}
\end{figure}


\begin{figure}[h]
\centering
\subfigure[]{
    \includegraphics[width=8.5cm,height=7cm] {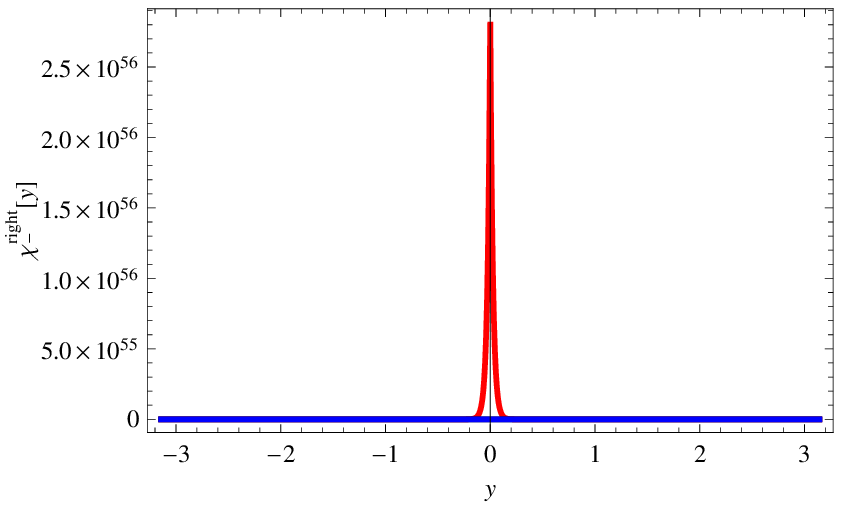}
    \label{fig:subfig401}
}
\subfigure[]{
    \includegraphics[width=8.5cm,height=7cm] {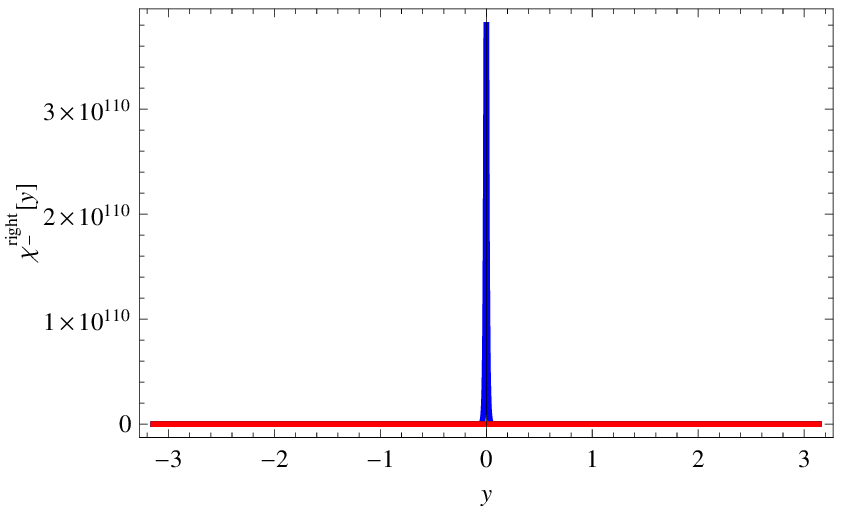}
    \label{fig:subfig411}
}
\subfigure[]{
    \includegraphics[width=8.5cm,height=7cm] {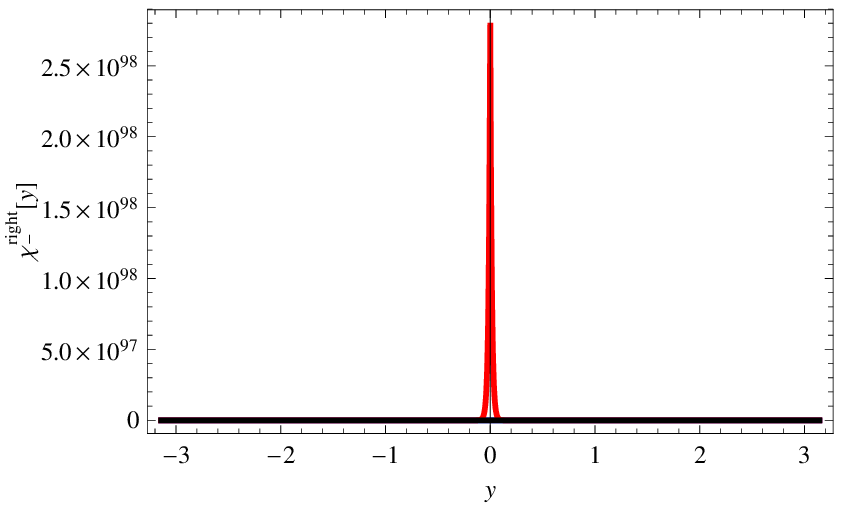}
    \label{fig:subfig421}
}
\subfigure[]{
    \includegraphics[width=8.5cm,height=7cm] {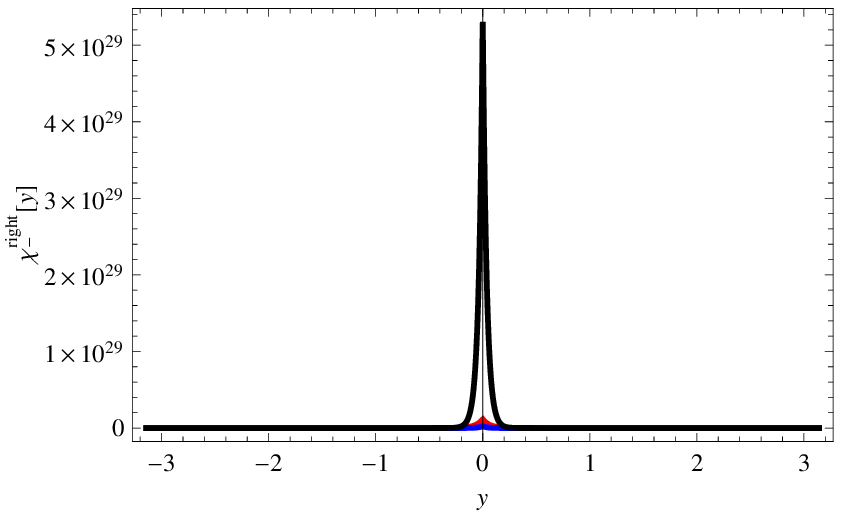}
    \label{fig:subfig431}
}
\caption[Optional caption for list of figures]{Variation
 of $\chi^{(1)}_{-;\bf R}(y)(= \hat{f}^{(1)}_{-,{\bf R}}(y))$ 
vs extra dimensional coordinate $y$ for \subref{fig:subfig401} $\Lambda_{(5)}>0$ and $A_{1}>0$, \subref{fig:subfig411} $\Lambda_{(5)}>0$ and $A_{1}<0$, 
\subref{fig:subfig421} $\Lambda_{(5)}<0$ and $A_{1}>0$ and
\subref{fig:subfig431} $\Lambda_{(5)}<0$ and $A_{1}<0$.
 In this context we use $B_{0}=0.002$, $r_{c}=0.996\sim 1$, $|A_{1}|=0.04$, $\theta_{1}=0.05$ and $\theta_{2}=0.04$
three distinct values of Gauss-Bonnet Coupling $\alpha_{(5)}$.}
\label{fig:subfigureExample55512}
\end{figure}

Finally clubbing the contributions from the first and last term of the action stated in the second line of the equation(\ref{redac}) the compact form of the effective four dimensional 
action can be recast as 
\be\begin{array}{lll}\label{finac}
 \displaystyle S_{f}=\int d^{4}x\sum^{\infty}_{n=0}\sum^{\infty}_{p=0}
\bar{\Psi}^{(n)}_{{\bf L},{\bf R}}(x)\overleftrightarrow{{\bf \Delta}}^{(np)}_{DIRAC} {\Psi}^{(p)}_{{\bf L},{\bf R}}(x) 
   \end{array}\ee

where the four dimensional covariant Dirac operator for the effective massive fermionic field theory interacting via electromagnetic (photon) ${\cal U}(1)$ abelian
gauge field is given by 
\be\begin{array}{llll}\label{bvxcz}
\displaystyle\overleftrightarrow{{\bf \Delta}}^{(np)}_{DIRAC}:=\left[i\overleftrightarrow{{\bf D}\slashed}^{(np)}_{COV}
-m^{{\bf L},{\bf R}}_{n}\delta^{np}
\right],\\
  \displaystyle  \overleftrightarrow{{\bf D}\slashed}^{(np)}_{COV}:=\left(\delta^{np}\overleftrightarrow{{\partial}\slashed}
+\frac{ig_{f}}{\sqrt{r_{c}}}\sum^{\infty}_{m=0}{\cal I}^{(nmp)}_{{\bf L},{\bf R}}
{\cal A}\slashed^{(m)}(x)\right).
   \end{array}\ee
Throughout the analysis we assume that the Majorana fermions do not contribute in the effective action.
The fermionic Kaluza-Klien mass spectrum is determined from the following two fold differential equation 
\be\begin{array}{lll}\label{fmass}
    \displaystyle \left(\pm\frac{1}{r_{c}}\overrightarrow{\cal D}_{y}-m_{f}\right)\hat{f}^{(n)}_{{\bf L},{\bf R}}(y)=-\left(m^{{\bf L},{\bf R}}_{n}\right)_{\pm}\hat{f}^{(n)}_{{\bf L},{\bf R}}(y)
   \end{array}\ee
subject to the following boundary conditions:
\be\begin{array}{lll}\label{bvchj}
    \displaystyle \hat{f}^{(n)}_{{\bf L}}(y_{i})\hat{f}^{(n)}_{{\bf R}}(y_{i})=0,\\
 \displaystyle\hat{f}^{(n)}_{{\bf L}}(y_{i})=0,~~~\hat{f}^{(n)}_{{\bf R}}(y_{i})=0
   \end{array}\ee
where at $y_{i}=0,\pi$ the ${\bf Z_{2}}$ orbifolding symmetry is imposed. This follows from the fact that 
left-handed or all right-handed fermionic wave functions are ${\bf Z_{2}}$ odd. In this context the fermionic differential operator 
$ \left(\pm\frac{1}{r_{c}}\overrightarrow{\cal D}_{y}-m_{f}\right)$ are hermitian and the mass eigen values are real. Consequently $\hat{f}^{(n)}_{{\bf L},{\bf R}}(y)$
is chosen to be real.

Now introducing a new variable $z^{\pm;{\bf L},{\bf R}}_{n}:=\frac{\left(m^{{\bf L},{\bf R}}_{n}\right)_{\pm}}{k_{\pm}}e^{A_{\pm}(y)}$ 
equation(\ref{fmass}) can be recast in terms of Bessel differential equation as
\be\begin{array}{llll}\label{dif4}
     \displaystyle \left(\pm z^{\pm;{\bf L},{\bf R}}_{n}\overrightarrow{{\cal D}}_{z^{\pm;{\bf L},{\bf R}}_{n}}-\nu^{\pm}
\right)\hat{f}^{(n)}_{{\bf L},{\bf R}}(z^{\pm;{\bf L},{\bf R}}_{n})=-z^{\pm;{\bf L},{\bf R}}_{n}\hat{f}^{(n)}_{{\bf L},{\bf R}}(z^{\pm;{\bf L},{\bf R}}_{n})\\
    \displaystyle \Rightarrow \left[\left(z^{\pm;{\bf L},{\bf R}}_{n}\right)^{2}\overrightarrow{{\cal D}^{2}}_{z^{\pm;{\bf L},{\bf R}}_{n}}
+\left(z^{\pm;{\bf L},{\bf R}}_{n}\right)^{2}-\nu^{\pm}\left(\nu^{\pm}\mp 1\right)\right]\hat{f}^{(n)}_{{\bf L},{\bf R}}(z^{\pm;{\bf L},{\bf R}}_{n})
=0
   \end{array}\ee
and the analytical solution for $n\neq 0$ turns out to be
\be\begin{array}{llll}\label{sol6}
\displaystyle    \hat{f}^{(n)}_{{\bf L},{\bf R}}(z^{\pm;{\bf L},{\bf R}}_{n})=\frac{z^{\pm;{\bf L},{\bf R}}_{n}}{{{\cal N}^{\pm;{\bf L},{\bf R}}}_{(n)}
\sqrt{\frac{\left(m^{{\bf L},{\bf R}}_{n}\right)_{\pm}}{k_{\pm}}}}\left[{\cal J}_{\mp\left(\frac{1}{2}+\nu^{\pm}\right)}
(z^{\pm;{\bf L},{\bf R}}_{n})+\beta^{\pm;{\bf L},{\bf R}}_{n}{\cal Y}_{\mp\left(\frac{1}{2}+\nu^{\pm}\right)}(z^{\pm;{\bf L},{\bf R}}_{n})\right].
   \end{array}\ee
Here ${\cal N}^{\pm;{\bf L},{\bf R}}_{(n)}$ be the normalization constant of the extra dimension dependent wave function and $\beta^{\pm;{\bf L},{\bf R}}_{n}$ is the
integration constant determined from the orthonormalization condition and the continuity conditions at the orbifold fixed point.
 In this context we use $\nu^{\pm}:=\frac{m_{f}}{k_{\pm}}$.

 Now applying the boundary condition on equation(\ref{sol6}) we get
\be\begin{array}{llll}\label{cond16}
 \displaystyle  \hat{f}^{(n)}_{{\bf L},{\bf R}}|_{y_{i}=0}=0~~\implies \beta^{\pm;{\bf L},{\bf R}}_{n}=
-\frac{{\cal J}_{\mp\left(\frac{1}{2}+\nu^{\pm}\right)}\left(\frac{\left(m^{{\bf L},{\bf R}}_{n}\right)_{\pm}}{k_{\pm}}\right)}
{{\cal Y}_{\mp\left(\frac{1}{2}+\nu^{\pm}\right)}\left(\frac{\left(m^{{\bf L},{\bf R}}_{n}\right)_{\pm}}{k_{\pm}}\right)}.
   \end{array}\ee

\be\begin{array}{llll}\label{cond21}
 \displaystyle  \hat{f}^{(n)}_{{\bf L},{\bf R}}|_{y_{i}=\pi}=0~~\implies \beta^{\pm;{\bf L},{\bf R}}_{n}=
-\frac{{\cal J}_{\mp\left(\frac{1}{2}+\nu^{\pm}\right)}\left(x^{\pm;{\bf L},{\bf R}}_{n}\right)
}{{\cal Y}_{\mp\left(\frac{1}{2}+\nu^{\pm}\right)}\left(x^{\pm;{\bf L},{\bf R}}_{n}\right)}
   \end{array}\ee
where $z^{\pm;{\bf L},{\bf R}}_{n}(\pi):=x^{\pm;{\bf L},{\bf R}}_{n}=\frac{\left(m^{{\bf L},{\bf R}}_{n}\right)_{\pm}}{k_{\pm}}e^{k_{\pm}r_{c}\pi}$.
Now using equation(\ref{cond16}) and equation(\ref{cond21}) we get
\be\begin{array}{llllll}\label{root}
 \displaystyle   \frac{{\cal J}_{\mp\left(\frac{1}{2}+\nu^{\pm}\right)}\left(x^{\pm;{\bf L},{\bf R}}_{n}e^{-k_{\pm}r_{c}\pi}\right)}
{{\cal Y}_{\mp\left(\frac{1}{2}+\nu^{\pm}\right)}\left(x^{\pm;{\bf L},{\bf R}}_{n}e^{-k_{\pm}r_{c}\pi}\right)}
= \frac{{\cal J}_{\mp\left(\frac{1}{2}+\nu^{\pm}\right)}\left(x^{\pm;{\bf L},{\bf R}}_{n}\right)
}{{\cal Y}_{\mp\left(\frac{1}{2}+\nu^{\pm}\right)}\left(x^{\pm;{\bf L},{\bf R}}_{n}\right)} 
   \end{array}\ee
which is an transcendental equation of $x^{\pm;{\bf L},{\bf R}}_{n}$ and the roots of this equation gives the left and right chiral 
fermionic field mass spectrum $\left(m^{{\bf L},{\bf R}}_{n}\right)_{\pm}$
 in presence of perturbative 
Gasuss-Bonnet coupling $\alpha_{(5)}$.
This leads to approximately
\be\begin{array}{llll}\label{massasdfer}
   \displaystyle  \left(m^{{\bf L},{\bf R}}_{n}\right)_{\pm}\approx \left(n+\frac{1}{2}\left[\nu^{\pm}\pm\frac{1}{2}\right]-\frac{1}{4}\right)\pi k_{\pm}e^{-k_{\pm}r_{c}\pi}.
   \end{array}\ee
This again shows a similar feature as of the graviton modes.

 Now using equation(\ref{orth}) the normalization constant for $n\neq 0$ mode reduces to the following expression
\be\begin{array}{llll}\label{vcvcasa}
    \displaystyle {\cal N}^{\pm;{\bf L},{\bf R}}_{(n)}=\frac{e^{k_{\pm}r_{c}\pi}}{
\sqrt{k_{\pm}r_{c}}}\left\{\left[{\cal J}_{\mp\left(\frac{1}{2}+\nu^{\pm}\right)}\left(x^{\pm;{\bf L},{\bf R}}_{n}\right)
+\beta^{\pm;\bf L,R}_{n}{\cal Y}_{\mp\left(\frac{1}{2}+\nu^{\pm}\right)}\left(x^{\pm;{\bf L},{\bf R}}_{n}\right)
\right]^{2}\right.\\ \left.~~~~~~~~~~~~~~~~~~~~~~~~~~~\displaystyle-e^{-2k_{\pm}r_{c}}\left[{\cal J}_{\mp\left(\frac{1}{2}+\nu^{\pm}\right)}\left(x^{\pm;{\bf L},{\bf R}}_{n}e^{-k_{\pm}r_{c}\pi}
\right)+\beta^{\pm;\bf L,R}_{n}{\cal Y}_{\mp\left(\frac{1}{2}+\nu^{\pm}\right)}\left(x^{\pm;{\bf L},{\bf R}}_{n}e^{-k_{\pm}r_{c}\pi}\right)
\right]^{2}\right\}^{\frac{1}{2}}.\end{array}\ee

 For $e^{k_{\pm}r_{c}\pi}\gg 1,~\frac{\left(m^{{\bf L},{\bf R}}_{n}\right)_{\pm}}{k_{\pm}}\ll 1$ the integration constant $\beta^{\pm;{\bf L},{\bf R}}_{n}\ll 1$.
Consequently ${\cal Y}_{\mp\left(\frac{1}{2}+\nu^{\pm}\right)}(z^{\pm;{\bf L},{\bf R}}_{n})$
 is neglected compared to ${\cal J}_{\mp\left(\frac{1}{2}+\nu^{\pm}\right)}(z^{\pm;{\bf L},{\bf R}}_{n})$ in equation(\ref{sol6}) 
and then the normalization constant for $n\neq 0$ mode
turns out to be

\be\begin{array}{llll}\label{vcvczx}
    \displaystyle {\cal N}^{\pm;{\bf L},{\bf R}}_{(n)}=\frac{e^{k_{\pm}r_{c}\pi}}{
\sqrt{k_{\pm}r_{c}}}{\cal J}_{\mp\left(\frac{1}{2}+\nu^{\pm}\right)}\left(x^{\pm;{\bf L},{\bf R}}_{n}\right). 
   \end{array}\ee
Consequently the extra dimension dependent wave function for $n\neq 0$ mode turns out to be 

\be\begin{array}{llll}\label{sol61}
\displaystyle    \hat{f}^{(n)}_{{\bf L},{\bf R}}(z^{\pm;{\bf L},{\bf R}}_{n})=\frac{z^{\pm;{\bf L},{\bf R}}_{n}}{e^{k_{\pm}r_{c}\pi}
\sqrt{\left(m^{{\bf L},{\bf R}}_{n}\right)_{\pm}r_{c}}}\frac{{\cal J}_{\mp\left(\frac{1}{2}+\nu^{\pm}\right)}
(z^{\pm;{\bf L},{\bf R}}_{n})}{{\cal J}_{\mp\left(\frac{1}{2}+\nu^{\pm}\right)}\left(x^{\pm;{\bf L},{\bf R}}_{n}\right)}.
   \end{array}\ee

In figure(\ref{fig:subfigureExample5156}) we have explicitly shown the behavior of the extra dimension
dependent left chiral fermionic first excited state for all possible signatures of Gauss-Bonnet coupling for $k_{+}$ and $k_{-}$ branch.
For $\alpha_{(5)}\rightarrow 0$ the left chiral wave function corresponding to the $k_{-}$ branch falls faster than compared to the $k_{+}$ branch
for bulk cosmological constant $\Lambda_{(5)}>0$ and all possible signatures of two -loop conformal coupling $A_{1}$. On the other hand just 
exactly opposite behavior 
is observed in the case of $\Lambda_{(5)}<0$ including the information from all possible signatures of $A_{1}$.
Additionally in figure(\ref{fig:subfigureExample5457}) we have plotted the graphical behavior 
of the extra dimension dependent wave function for $k_{+}$ and $k_{-}$ branch corresponding the first excited 
state for two signatures of $\Lambda_{(5)}$ and negative two-loop conformal coupling $A_{1}$
including three distinct values of Gauss-Bonnet coupling $\alpha_{(5)}$. It is clearly observed from
 the plot that the left chiral wave function for $k_{-}$ branch is more localized than the $k_{+}$ branch
at the boundary of the visible brane. So from the phenomenological point of view figure(\ref{fig:subfig38}) is not desirable.
Similarly in figure(\ref{fig:subfigureExample51519}) we have explicitly depicted the behavior of the extra dimension
dependent right chiral fermionic first excited state for all possible signatures of Gauss-Bonnet coupling for $k_{+}$ and $k_{-}$ branch which follows subsequently
different behavior from its left chiral counterpart.
Additionally in figure(\ref{fig:subfigureExample55512}) we have plotted the graphical behavior 
of the extra dimension dependent wave function for $k_{-}$ branch corresponding the first excited 
state for all possible signatures of bulk cosmological constant $\Lambda_{(5)}$ and two -loop conformal coupling $A_{1}$
for three distinct values of Gauss-Bonnet coupling $\alpha_{(5)}$. The figure(\ref{fig:subfig39}) and  figure(\ref{fig:subfigureExample55512})
depicts that to achieve the localization of the left handed chiral mode on the boundary of the visible brane the branch $k_{-}$ is a favored choice
as parameter. Additionally in $\alpha_{(5)}\rightarrow 0$ limit the left/right chiral solution for $k_{-}$ branch exactly reproduces the
 Randall-Sundrum behavior compared to the rest of the physical situations.

For massless $n=0$ mode the solution of the equation(\ref{fmass}) turns out to be
\be\begin{array}{lllll}\label{j1}
   \displaystyle  \hat{f}^{(0)}_{{\bf L},{\bf R}}(y)=\frac{e^{\pm\nu^{\pm}A_{\pm}(y)}}{ {\cal N}^{\pm;{\bf L},{\bf R}}_{(0)}}.
   \end{array}\ee
Here ${\cal N}^{\pm;{\bf L},{\bf R}}_{(0)}$ normalization constant for zero mode. Now applying the normalization condition 
we get ${\cal N}^{\pm;{\bf L},{\bf R}}_{(0)}=\sqrt{\frac{\left(1\pm 2\nu^{\pm}\right)k_{\pm}r_{c}}{2\left[e^{(1\pm 2\nu^{\pm})k_{\pm}r_{c}\pi}-1
\right]}}$ and the ground state massless zero mode wave function for fermionic species turns out to be
\be\begin{array}{llll}\label{mlkl}
   \displaystyle  \hat{f}^{(0)}_{{\bf L},{\bf R}}(y)=\sqrt{\frac{2\left[e^{(1\pm 2\nu^{\pm})k_{\pm}r_{c}\pi}-1
\right]}{\left(1\pm 2\nu^{\pm}\right)k_{\pm}r_{c}}}e^{\pm\nu^{\pm}A_{\pm}(y)}.
\end{array}\ee


\subsection{\bf Bulk Kalb-Rammond Antisymmetric Tensor Field}
\label{kr1}
In the context of string theory closed string modes include antisymmetric tensor fields of different rank.
The five dimensional action for rank-3 antisymmetric pure Kalb-Rammond tensor field can be written as \cite{ssg10}
\be\begin{array}{llll}\label{abshg}
   \displaystyle S_{{\cal H}}=\int d^{5}x\sqrt{-g_{(5)}}~{\cal H}_{MNL}(x,y){\cal H}^{MNL}(x,y)
   \end{array}\ee
where five dimensional action for rank-3 antisymmetric pure Kalb-Rammond field strength tensor is given by
\be\begin{array}{llll}\label{abse}
   \displaystyle {\cal H}_{MNL}:=\overrightarrow{\partial}_{[M}{\cal B}_{NL]} (x,y)
   \end{array}\ee
with antisymmetric tensor potential
${\cal B}_{NL}=-{\cal B}_{LN}$, usually called ``{\it Neveu-Schwarz–
Neveu-Schwarz}'' (NS-NS) two-form. For historical reasons the field B is also called ``torsion'' since, to lowest order, it
can be identified with the antisymmetric part
of the affine connection, in the
context of a non-Riemannian geometric structure. An alternative, often used, name is
``Kalb-Ramond axion'', in reference to the pseudo-scalar axionic field related to the Kalb-Rammond antisymmetric tensor field via space-time ``duality'' transformation is elaborately 
discussed in the next subsection. 

 Now applying the gauge fixing condition ${\cal B}_{4\mu}=0$ the action stated in equation(\ref{abshg}) takes the following form
 \be\begin{array}{llll}\label{abskb}
   \displaystyle S_{{\cal H}}=\int d^{5}x~ r_{c}~e^{2A_{\pm}(y)}\left[\eta^{\mu\alpha}\eta^{\nu\beta}\eta^{\lambda\gamma}
{\cal H}_{\mu\nu\lambda}(x,y){\cal H}_{\alpha\beta\gamma}(x,y)-\frac{3}{r^{2}_{c}}e^{-2A_{\pm}(y)}\eta^{\mu\alpha}\eta^{\nu\beta}{\cal B}_{\mu\nu}(x,y)
\overrightarrow{{\cal D}_{y}}^{2}{\cal B}_{\alpha\beta}(x,y)\right]
   \end{array}\ee
where we introduce a new symbol $\overrightarrow{{\cal D}_{y}}:=\frac{d}{dy}$. Let the Kaluza-Klien expansion
of the Kalb-Rammond antysmmetric NS-NS two form potential field is given by
\be\begin{array}{lllll}\label{KK4}
   \displaystyle {\cal B}_{\mu\nu}(x,y)=\sum^{\infty}_{n=0}{\cal B}^{(n)}_{\mu\nu}(x)~\frac{\chi^{(n)}_{\pm;\cal H}(y)}{\sqrt{r_{c}}}. 
   \end{array}\ee
 
Now plugging equation(\ref{KK4}) in equation(\ref{abskb}) the effective four dimensional action reduces to the following form:
\be\begin{array}{llll}\label{abs232}
   \displaystyle S_{{\cal H}}=\int d^{4}x\sum^{\infty}_{n=0}\left[\eta^{\mu\alpha}\eta^{\nu\beta}\eta^{\lambda\gamma}
{\cal H}^{(n)}_{\mu\nu\lambda}(x){\cal H}^{(n)}_{\alpha\beta\gamma}(x)+\left(M^{{\cal H}}_{n}\right)^{2}_{\pm}
\eta^{\mu\alpha}\eta^{\nu\beta}{\cal B}^{(n)}_{\mu\nu}(x){\cal B}^{(n)}_{\alpha\beta}(x)\right]\end{array}\ee

where the effective four dimensional Kalb-Rammond field strength is defined as ${\cal H}^{(n)}_{\mu\nu\lambda}(x):=\overrightarrow{\partial}_{[\mu}{\cal B}^{(n)}_{\nu\lambda]}(x)$.
In this context we impose the following orthonormalization condition of extra dimension dependent wave functions
\be\begin{array}{llll}\label{no12}
 \displaystyle   \int^{+\pi}_{-\pi}dy~e^{2A_{\pm}(y)}~\chi^{(m)}_{\pm;\cal H}(y)~\chi^{(n)}_{\pm;\cal H}(y)=\delta^{mn}
   \end{array}\ee
and the mass term of the gauge field is defined through the following differential equation as
\be\begin{array}{llll}\label{dif18}
  \displaystyle   -\frac{1}{r^{2}_{c}}\overrightarrow{{\cal D}^{2}_{y}}\chi^{(n)}_{\pm;\cal H}(y)=e^{2A_{\pm}(y)}\left(m^{{\cal H}}_{n}\right)^{2}_{\pm}\chi^{(n)}_{\pm;\cal H}(y).
   \end{array}\ee
Here the mass of the nth mode Kalb-Rammond antisymmetric field is given by
$\left(M^{{\cal H}}_{n}\right)_{\pm}=\sqrt{3}\left(m^{{\cal H}}_{n}\right)_{\pm}$.
Now introducing a new variable $z^{\pm;{\cal H}}_{n}:=\frac{\left(m^{{\cal H}}_{n}\right)_{\pm}}{k_{\pm}}e^{A_{\pm}(y)}$ 
equation(\ref{dif1}) can be recast in terms of Bessel differential equation of order zero as
\be\begin{array}{llll}\label{dif28}
    \displaystyle  \left[\left(z^{\pm;{\cal H}}_{n}\right)^{2}\overrightarrow{{\cal D}^{2}}_{z^{\pm;{\cal H}}_{n}}
+z^{\pm;{\cal H}}_{n}\overrightarrow{{\cal D}}_{z^{\pm;{\cal H}}_{n}}+\left(z^{\pm;{\cal H}}_{n}\right)^{2}\right]\chi^{(n)}_{\pm;\cal H}
=0
   \end{array}\ee
and the analytical solution turns out to be
\be\begin{array}{llll}\label{sol1xc}
\displaystyle    \chi^{(n)}_{\pm;\cal H}(y)=\frac{1}{{\cal N}^{\pm;\cal H}_{(n)}}\left[{\cal J}_{0}
(z^{\pm;{\cal H}}_{n})+\alpha^{\pm;\cal H}_{n}{\cal Y}_{0}(z^{\pm;{\cal H}}_{n})\right].
   \end{array}\ee
Here ${\cal N}^{\pm;\cal H}_{(n)}$ be the normalization constant of the extra dimension dependent wave function and $\alpha^{\pm;\cal H}_{n}$ is the
integration constant determined from the orthonormalization condition and the continuity conditions at the orbifold fixed point.
Self-adjointness and hermiticity of the differential operator appearing in equation(\ref{dif28})
demands that $\overrightarrow{{\cal D}_{y}}\chi^{(n)}_{\pm;\cal H}(y)$ is
 continious at the orbifold fixed points $y_{i}=0,\pi$. Consequently we have
\be\begin{array}{llll}\label{cond1zx}
 \displaystyle  \overrightarrow{{\cal D}_{y}}\chi^{(n)}_{\pm;\cal H}|_{y_{i}=0}=0~~\implies \alpha^{\pm;\cal H}_{n}=
-\frac{\left[{\cal J}_{0}\left(\frac{\left(m^{{\cal H}}_{n}\right)_{\pm}}{k_{\pm}}\right)
+\frac{\left(m^{{\cal H}}_{n}\right)_{\pm}}{k_{\pm}}{\cal J}^{'}_{0}\left(\frac{\left(m^{{\cal H}}_{n}\right)_{\pm}}{k_{\pm}}\right)\right]}{\left[{\cal Y}_{0}\left(\frac{\left(m^{{\cal H}}_{n}\right)_{\pm}}{k_{\pm}}\right)
+\frac{\left(m^{{\cal H}}_{n}\right)_{\pm}}{k_{\pm}}{\cal Y}^{'}_{0}\left(\frac{\left(m^{{\cal H}}_{n}\right)_{\pm}}{k_{\pm}}\right)\right]}.
   \end{array}\ee

\be\begin{array}{llll}\label{cond2zx}
 \displaystyle  \overrightarrow{{\cal D}_{y}}\chi^{(n)}_{\pm;{\cal H}}|_{y_{i}=\pi}=0~~\implies \alpha^{\pm;{\cal H}}_{n}=
-\frac{\left[{\cal J}_{0}\left(x^{\pm;{\cal H}}_{n}\right)
+x^{\pm;{\cal H}}_{n}{\cal J}^{'}_{0}\left(x^{\pm;{\cal H}}_{n}\right)\right]}{\left[{\cal Y}_{0}\left(x^{\pm;{\cal H}}_{n}\right)
+x^{\pm;{\cal H}}_{n}{\cal Y}^{'}_{0}\left(x^{\pm;{\cal H}}_{n}\right)\right]}
   \end{array}\ee
where $z^{\pm;H}_{n}(\pi):=x^{\pm;H}_{n}=\frac{\left(m^{{\cal H}}_{n}\right)_{\pm}}{k_{\pm}}e^{k_{\pm}r_{c}\pi}$.
For $e^{k_{\pm}r_{c}\pi}\gg 1,~\frac{\left(m^{{\cal H}}_{n}\right)_{\pm}}{k_{\pm}}\ll 1$ the mass spectrum for the Kalb-Rammond fields is expected to be of the order of TeV scale i.e.
\be\begin{array}{llll}\label{approxcv}
 \displaystyle    \alpha^{\pm;{\cal H}}_{n}\simeq x^{\pm;{\cal H}}_{n}e^{-2k_{\pm}r_{c}\pi}. 
   \end{array}\ee

Now using equation(\ref{approxcv}) and equation(\ref{cond1zx}) we get
\be\begin{array}{llllll}\label{rootzx}
 \displaystyle    x^{\pm;{\cal H}}_{n}e^{-2k_{\pm}r_{c}\pi}
=  \frac{\left[{\cal J}_{0}\left(x^{\pm;{\cal H}}_{n}\right)
-x^{\pm;{\cal H}}_{n}{\cal J}_{1}\left(x^{\pm;{\cal H}}_{n}\right)\right]}{\left[{\cal Y}_{0}\left(x^{\pm;{\cal H}}_{n}\right)
-x^{\pm;{\cal H}}_{n}{\cal Y}_{1}\left(x^{\pm;{\cal H}}_{n}\right)\right]} ~~~~~~~~~~~
\displaystyle \Rightarrow {\cal J}_{1}\left(x^{\pm;{\cal H}}_{n}\right)\simeq \frac{\pi}{2}x^{\pm;{\cal H}}_{n}e^{-2k_{\pm}r_{c}\pi}\approx 0
   \end{array}\ee
which is an transcendental equation of $x^{\pm;{\cal H}}_{n}$ and the roots of this equation gives the Kalb-Rammond field mass spectrum $\left(m^{{\cal H}}_{n}\right)_{\pm}$ in presence of perturbative 
Gasuss-Bonnet coupling $\alpha_{(5)}$. Now using equation(\ref{no12}) the normalization constant for $n\neq 0$ mode reduces to the following expression
\be\begin{array}{llll}\label{vcvc}
    \displaystyle {\cal N}^{\pm;{\cal H}}_{(n)}=\frac{e^{k_{\pm}r_{c}\pi}}{
\sqrt{k_{\pm}r_{c}}}\sqrt{\left\{\left[{\cal J}_{0}\left(x^{\pm;{\cal H}}_{n}\right)+\alpha^{\pm;{\cal H}}_{n}{\cal Y}_{0}\left(x^{\pm;{\cal H}}_{n}\right)
\right]^{2}-e^{-2k_{\pm}r_{c}}\left[{\cal J}_{0}\left(x^{\pm;{\cal H}}_{n}e^{-k_{\pm}r_{c}\pi}
\right)+\alpha^{\pm;{\cal H}}_{n}{\cal Y}_{0}\left(x^{\pm;{\cal H}}_{n}e^{-k_{\pm}r_{c}\pi}\right)
\right]^{2}\right\}}. 
   \end{array}\ee
For $e^{k_{\pm}r_{c}\pi}\gg 1,~\frac{\left(m^{{\cal H}}_{n}\right)_{\pm}}{k_{\pm}}\ll 1$ the integration constant $\alpha^{\pm;{\cal H}}_{n}\ll 1$.
Consequently ${\cal Y}_{0}(z^{\pm;{\cal H}}_{n})$ is neglected compared to ${\cal J}_{0}(z^{\pm;{\cal H}}_{n})$ in equation(\ref{sol1xc}) 
and then the normalization constant for $n\neq 0$ mode
turns out to be
\be\begin{array}{llll}\label{vcvc}
    \displaystyle {\cal N}^{\pm;{\cal H}}_{(n)}=\frac{e^{k_{\pm}r_{c}\pi}}{
\sqrt{k_{\pm}r_{c}}}{\cal J}_{0}\left(x^{\pm;{\cal H}}_{n}\right)\approx \frac{\pi}{2}\frac{x^{\pm;{\cal H}}_{n}}{\sqrt{k_{\pm}r_{c}}}e^{-k_{\pm}r_{c}\pi}. 
   \end{array}\ee
Consequently the extra dimensional dependent wave function for $n\neq 0$ turns out to be 
\be\begin{array}{llll}\label{wfvbion}
   \displaystyle    \chi^{(n)}_{\pm;\cal H}(y)=\frac{2\sqrt{k_{\pm}r_{c}}}{\pi x^{\pm;{\cal H}}_{n}}e^{k_{\pm}r_{c}\pi}{\cal J}_{0}
(z^{\pm;{\cal H}}_{n}). 
   \end{array}\ee

For massless $n=0$ mode the solution of the equation(\ref{dif18}) turns out to be
\be\begin{array}{lllll}\label{jkloss}
   \displaystyle  \chi^{(0)}_{\pm;{\cal H}}=C_{1}|y|+C_{2}.
   \end{array}\ee
Here $C_{1}$ and $C_{2}$ are arbitrary integration constants. Now applying the boundary condition through the continuity
of the wave function we get $C_{1}=0$. As a result the zero mode solution turns out to be $ \chi^{(0)}_{\pm;{\cal H}}=C_{2}$.
Now applying the normalization condition the ground state massless zero mode wave function turns out to be
\be\begin{array}{llll}\label{mlsx}
   \displaystyle  \chi^{(0)}_{\pm;{\cal H}}=C_{2}=\sqrt{\frac{k_{\pm}r_{c}}{e^{2k_{\pm}r_{c}\pi}-1}}\approx \sqrt{k_{\pm}r_{c}}e^{-k_{\pm}r_{c}\pi}.
   \end{array}\ee

This give zero mode is heavily suppressed in the visible brane, though the warping will be reduced if one choses large Gauss-Bonnet coupling $\alpha_{(5)}$.


\subsection{\bf Bulk Kalb-Rammond Antisymmetric Tensor Field with parity violating extension}
\label{kr2}

The five dimensional action for rank-3 antisymmetric pure Kalb-Rammond tensor field can be extended with a parity violating term as \cite{ssgn},
\be\begin{array}{llll}\label{abshgjbj}
   \displaystyle S_{{\cal H}}=\int d^{5}x\sqrt{-g_{(5)}}~\left[{\cal H}_{MNL}(x,y){\cal H}^{MNL}(x,y)+2{ \Theta}_{0}~{\epsilon}^{ABMNL}{\cal B}_{AB}(x,y){\cal H}_{MNL}(x,y)\right]
   \end{array}\ee
where five dimensional action for rank-3 antisymmetric pure Kalb-Rammond field strength tensor is given by
\be\begin{array}{llll}\label{absejbj}
   \displaystyle {\cal H}_{MNL}:=\overrightarrow{\partial}_{[M}{\cal B}_{NL]} (x,y)
   \end{array}\ee
with antisymmetric tensor potential
${\cal B}_{NL}=-{\cal B}_{LN}$. Here $\Theta_{0}$ represents the axion Kalb-Rammond interaction strength. 
The parity violating term is a topological term invariant under the Kalb-Rammond gauge transformation: $\delta{\cal B}_{MN}=\partial_{[M}\Lambda_{N]}$, 
where $\Lambda$ is the gauge parameter.
Now to get rid of massive vector modes on the brane we applying the gauge fixing condition ${\cal B}_{4\mu}=0$ and consequently 
the action stated in equation(\ref{abshgjbj}) takes the following form
 \be\begin{array}{llll}\label{abskbjbj}
   \displaystyle S_{{\cal H}}=\int d^{5}x~ r_{c}~e^{2A_{\pm}(y)}\left[\eta^{\mu\alpha}\eta^{\nu\beta}\eta^{\lambda\gamma}
{\cal H}_{\mu\nu\lambda}(x,y){\cal H}_{\alpha\beta\gamma}(x,y)-\frac{3}{r^{2}_{c}}e^{-2A_{\pm}(y)}\eta^{\mu\alpha}\eta^{\nu\beta}{\cal B}_{\mu\nu}(x,y)
\overrightarrow{{\cal D}_{y}}^{2}{\cal B}_{\alpha\beta}(x,y)\right.\\ \left.~~~~~~~~~~~~~~~~~~~~~~~~~~~~~~~~~~~~~~~~~~~~~~~~~~~~~~~~~
~~~~~~~~~~~~~~~~~~~~~~~~~\displaystyle +\frac{6\Theta_{0}}{r_{c}}e^{-2A_{\pm}(y)}{\cal E}^{4\mu\nu\alpha\beta}{\cal B}_{\alpha\beta}(x,y)
\overrightarrow{{\cal D}_{y}}{\cal B}_{\mu\nu}(x,y)\right]
   \end{array}\ee
where we introduce a new symbol $\overrightarrow{{\cal D}_{y}}:=\frac{d}{dy}$ and the five dimensional Levi-Civita tensor (${\epsilon}^{ABMNL} $) is defined 
in terms of five dimensional Levi-Civita tensor density (${\cal E}^{ABMNL}$) as
\be\begin{array}{lllll}\label{levcvi}
    \displaystyle {\epsilon}^{ABMNL}:=\frac{{\cal E}^{ABMNL}}{\sqrt{-g_{(5)}}}.
   \end{array}\ee
Let the Kaluza-Klien expansion
of the Kalb-Rammond antisymmetric NS-NS two form potential field is given by
\be\begin{array}{lllll}\label{KK7}
   \displaystyle {\cal B}_{\mu\nu}(x,y)=\sum^{\infty}_{n=0}{\cal B}^{(n)}_{\mu\nu}(x)~\frac{\chi^{(n)}_{\pm;\cal H}(y)}{\sqrt{r_{c}}}. 
   \end{array}\ee
 
Now plugging equation(\ref{KK7}) in equation(\ref{abskbjbj}) the effective four dimensional action reduces to the following form:
\be\begin{array}{llll}\label{abskkht}
   \displaystyle S_{{\cal H}}=\int d^{4}x\sum^{\infty}_{n=0}\left[\eta^{\mu\alpha}\eta^{\nu\beta}\eta^{\lambda\gamma}
{\cal H}^{(n)}_{\mu\nu\lambda}{\cal H}^{(n)}_{\alpha\beta\gamma}+
3\eta^{\mu\alpha}\eta^{\nu\beta}{\cal B}^{(n)}_{\mu\nu}(x){\cal B}^{(n)}_{\alpha\beta}(x)\left(\frac{e^{-2A_{\pm}(y)}}{r^{2}_{c}\chi^{(n)}_{\pm;\cal H}}
\overrightarrow{{\cal D}_{y}}^{2}\chi^{(n)}_{\pm;\cal H}\right)\right.\\ \left.~~~~~~~~~~~~~~~~~~~~~~~~~~~~~~~~~~~~~~~~~~~~~~~~~~~~~~~~~~~\displaystyle 
+6\epsilon^{\mu\nu\alpha\beta}{\cal B}^{(n)}_{\mu\nu}(x){\cal B}^{(n)}_{\alpha\beta}(x)
\left(\frac{e^{-2A_{\pm}(y)}\Theta_{0}}{r^{2}_{c}\chi^{(n)}_{\pm;\cal H}}
\overrightarrow{{\cal D}_{y}}\chi^{(n)}_{\pm;\cal H}\right)\right].\end{array}\ee

In this context we impose the following orthonormalization condition of extra dimension dependent wave functions
\be\begin{array}{llll}\label{no12cv}
 \displaystyle   \int^{+\pi}_{-\pi}dy~e^{2A_{\pm}(y)}~\chi^{(m)}_{\pm;\cal H}(y)~\chi^{(n)}_{\pm;\cal H}(y)=\delta^{mn}
   \end{array}\ee
It is interesting to mention here that the four dimensional
effective action contains, apart from the kinetic term 
and the mass term (${\cal B}^{(n)}_{\mu\nu}{\cal B}^{(n)~\mu\nu}$) for the Kalb-Rammond field, an 
additional term of the form ${\cal B}^{(n)}_{\mu\nu}\tilde{{\cal B}}^{(n)~\mu\nu}$ where 
$\tilde{{\cal B}}^{(n)~\mu\nu}~=\epsilon^{\mu\nu\alpha\beta}{\cal B}^{(n)}_{\alpha\beta}$ 
is the {\it dual of Kalb-Rammond ~field}. On solving the equation of motion from this effective four dimensional 
action stated in equation(\ref{abskkht}), it is quite straightforward to find the solution for the 
Kalb-Rammond field. It is pointed out in cite{} that the only non-trivial solution corresponds to 
self-dual or anti-dual Kalb-Rammond fields i.e, 
${\cal B}^{(n)}_{\mu\nu}=\tilde{{\cal B}}_{(n)}^{\mu\nu}$ or ${\cal B}^{(n)}_{\mu\nu}=
-\tilde{{\cal B}}_{(n)}^{\mu\nu}$. Such self-dual or anti self-dual conditions
imply the reduction in the degrees of freedom of the Kalb-Rammond field  has a five dimensional topological quantum 
field theoretic origin (TQFT). Then the effective four dimensional action reduces to the following expression:

\be\begin{array}{llll}\label{abskkkk}
   \displaystyle S_{{\cal H}}=\int d^{4}x\sum^{\infty}_{n=0}\left[\eta^{\mu\alpha}\eta^{\nu\beta}\eta^{\lambda\gamma}
{\cal H}^{(n)}_{\mu\nu\lambda}(x){\cal H}^{(n)}_{\alpha\beta\gamma}(x)+\left(M^{{\cal H}}_{n}\right)^{2}_{\pm;{\bf SD/AD}}
\eta^{\mu\alpha}\eta^{\nu\beta}{\cal B}^{(n)}_{\mu\nu}(x){\cal B}^{(n)}_{\alpha\beta}(x)\right]\end{array}\ee
  
where the effective four dimensional Kalb-Rammond field strength is defined as ${\cal H}^{(n)}_{\mu\nu\lambda}(x):=\overrightarrow{\partial}_{[\mu}{\cal B}^{(n)}_{\nu\lambda]}(x)$.
Most importantly the mass term of the Kalb-Rammond field is defined through the following two fold differential equations as
\be\begin{array}{llll}\label{dif178}
  \displaystyle  {\bf Self-Dual ~KR}:~~~~ 
-\frac{1}{r^{2}_{c}}\overrightarrow{{\cal D}^{2}_{y}}\chi^{(n)}_{\pm;\cal H}(y)
+\frac{2\Theta_{0}}{r_{c}}\overrightarrow{{\cal D}_{y}}\chi^{(n)}_{\pm;\cal H}(y)=e^{2A_{\pm}(y)}\left(m^{{\cal H}}_{n}\right)^{2}_{\pm;{\bf SD}}\chi^{(n)}_{\pm;\cal H}(y)\\
\displaystyle  {\bf Anti-Dual~ KR}: ~~~~-\frac{1}{r^{2}_{c}}\overrightarrow{{\cal D}^{2}_{y}}\chi^{(n)}_{\pm;\cal H}(y)
-\frac{2\Theta_{0}}{r_{c}}\overrightarrow{{\cal D}_{y}}\chi^{(n)}_{\pm;\cal H}(y)=e^{2A_{\pm}(y)}\left(m^{{\cal H}}_{n}\right)^{2}_{\pm;{\bf AD}}\chi^{(n)}_{\pm;\cal H}(y).
   \end{array}\ee
It may be observed that now apart from the two possible branches $k_{+}$ and $k_{-}$, the axion term $\Theta_{0}$ has 
resulted into the decomposition of Kalb-Rammond field into self-dual and anti-self dual parts with different equation of motion.
Here the mass of the nth mode self-dual and anti-dual Kalb-Rammond antisymmetric field is given by
$\left(M^{{\cal H}}_{n}\right)_{\pm;{\bf SD/AD}}=\sqrt{3}\left(m^{{\cal H}}_{n}\right)_{\pm;{\bf SD/AD}}$.
Now introducing a new variable $z^{\pm;{\cal H};{\bf SD/AD}}_{n}:=\frac{\left(m^{{\cal H}}_{n}\right)_{\pm;{\bf SD/AD}}}{k_{\pm}}e^{A_{\pm}(y)}$ 
equation(\ref{dif178}) can be recast in terms of Bessel differential equation of order $\nu:=\frac{2\Theta_{0}}{k_{\pm}}$ as
\be\begin{array}{llll}\label{dif28}
    \displaystyle  {\bf Self-Dual ~KR}:~~~~  \left[\left(z^{\pm;{\cal H};{\bf SD}}_{n}\right)^{2}\overrightarrow{{\cal D}^{2}}_{z^{\pm;{\cal H};{\bf SD}}_{n}}
+\left(1-\nu\right)z^{\pm;{\cal H};{\bf SD}}_{n}\overrightarrow{{\cal D}}_{z^{\pm;{\cal H};{\bf SD}}_{n}}+\left(z^{\pm;{\cal H};{\bf SD}}_{n}\right)^{2}\right]\chi^{(n)}_{\pm;\cal H;{\bf SD}}
=0\\
\displaystyle  {\bf Anti-Dual ~KR}:~~~~ \left[\left(z^{\pm;{\cal H};{\bf AD}}_{n}\right)^{2}\overrightarrow{{\cal D}^{2}}_{z^{\pm;{\cal H};{\bf AD}}_{n}}
+\left(1+\nu\right)z^{\pm;{\cal H};{\bf AD}}_{n}\overrightarrow{{\cal D}}_{z^{\pm;{\cal H};{\bf AD}}_{n}}+\left(z^{\pm;{\cal H};{\bf AD}}_{n}\right)^{2}\right]\chi^{(n)}_{\pm;\cal H;{\bf AD}}
=0
   \end{array}\ee
and the analytical solution turns out to be
\be\begin{array}{llll}\label{sol1xcbn}
\displaystyle {\bf Self-Dual ~KR}:~~~~    \chi^{(n)}_{\pm;\cal H;{\bf SD}}(y)=\frac{\left(z^{\pm;{\cal H};{\bf SD}}_{n}\right)^{\nu}}{{\cal N}^{\pm;\cal H; {\bf SD}}_{(n)}}\left[{\cal J}_{\nu}
(z^{\pm;{\cal H};{\bf SD}}_{n})+\left(\alpha^{\pm;\cal H}_{n}\right)_{{\bf SD}}{\cal Y}_{\nu}(z^{\pm;{\cal H};{\bf SD}}_{n})\right]\\
\displaystyle  {\bf Anti-Dual ~KR}:~~~~     \chi^{(n)}_{\pm;\cal H;{\bf AD}}(y)=\frac{\left(z^{\pm;{\cal H};{\bf AD}}_{n}\right)^{-\nu}}{{\cal N}^{\pm;\cal H; {\bf AD}}_{(n)}}\left[{\cal J}_{\nu}
(z^{\pm;{\cal H};{\bf AD}}_{n})+\left(\alpha^{\pm;\cal H}_{n}\right)_{{\bf AD}}{\cal Y}_{\nu}(z^{\pm;{\cal H};{\bf AD}}_{n})\right].
   \end{array}\ee
Here ${\cal N}^{\pm;\cal H;{\bf SD/AD}}_{(n)}$ be the normalization constant of the extra dimension dependent wave function and $\left(\alpha^{\pm;\cal H}_{n}\right)_{{\bf SD/AD}}$ is the
integration constant determined from the orthonormalization condition and the continuity conditions at the orbifold fixed point.
Self-adjointness and hermiticity of the differential operator appearing in equation(\ref{dif28})
demands that $\overrightarrow{{\cal D}_{y}}\chi^{(n)}_{\pm;\cal H;{\bf SD/AD}}(y)$ is
 continuous at the orbifold fixed points $y_{i}=0,\pi$. Consequently we have  
\be\begin{array}{llll}\label{cond1zxsd}
\displaystyle {\bf Self-Dual ~KR}:~~~~  \\
 \displaystyle ~~~~~~~~~~~~~~~~~~~~~~~~~~~~~ \overrightarrow{{\cal D}_{y}}\chi^{(n)}_{\pm;\cal H;{\bf SD}}|_{y_{i}=0}=0~~\implies \left(\alpha^{\pm;\cal H}_{n}\right)_{\bf SD}=
-\frac{\left[\nu{\cal J}_{\nu}\left(\frac{\left(m^{{\cal H}}_{n}\right)_{\pm;\bf SD}}{k_{\pm}}\right)
+{\cal J}^{'}_{\nu}\left(\frac{\left(m^{{\cal H}}_{n}\right)_{\pm;\bf SD}}{k_{\pm}}\right)\right]}{\left[\nu{\cal Y}_{\nu}\left(\frac{\left(m^{{\cal H}}_{n;\bf SD}\right)_{\pm;\bf SD}}{k_{\pm}}\right)
+{\cal Y}^{'}_{\nu}\left(\frac{\left(m^{{\cal H}}_{n}\right)_{\pm;\bf SD}}{k_{\pm}}\right)\right]}.
   \end{array}\ee

\be\begin{array}{llll}\label{cond2zxsd}
 \displaystyle~~~~~~~~~~~~~~~~~~~~~~~~~~~~~ \overrightarrow{{\cal D}_{y}}\chi^{(n)}_{\pm;{\cal H};{\bf SD}}|_{y_{i}=\pi}=0~~\implies \left(\alpha^{\pm;{\cal H}}_{n}\right)_{\bf SD}=
-\frac{\left[\nu{\cal J}_{\nu}\left(x^{\pm;{\cal H};{\bf SD}}_{n}\right)
+{\cal J}^{'}_{\nu}\left(x^{\pm;{\cal H};{\bf SD}}_{n}\right)\right]}{\left[\nu{\cal Y}_{\nu}\left(x^{\pm;{\cal H};{\bf SD}}_{n}\right)
+{\cal Y}^{'}_{\nu}\left(x^{\pm;{\cal H};{\bf SD}}_{n}\right)\right]}
   \end{array}\ee

\be\begin{array}{llll}\label{cond1zxsd}
\displaystyle   {\bf Anti-Dual ~KR}:~~~~  \\
 \displaystyle ~~~~~~~~~~~~~~~~~~~~~~~~~~~~~  \overrightarrow{{\cal D}_{y}}\chi^{(n)}_{\pm;\cal H;{\bf AD}}|_{y_{i}=0}=0~~\implies \left(\alpha^{\pm;\cal H}_{n}\right)_{\bf AD}=
-\frac{\left[-\nu{\cal J}_{\nu}\left(\frac{\left(m^{{\cal H}}_{n}\right)_{\pm;\bf AD}}{k_{\pm}}\right)
+{\cal J}^{'}_{\nu}\left(\frac{\left(m^{{\cal H}}_{n}\right)_{\pm;\bf AD}}{k_{\pm}}\right)\right]}{\left[-\nu{\cal Y}_{\nu}\left(\frac{\left(m^{{\cal H}}_{n}\right)_{\pm;\bf AD}}{k_{\pm}}\right)
+{\cal Y}^{'}_{\nu}\left(\frac{\left(m^{{\cal H}}_{n}\right)_{\pm;\bf AD}}{k_{\pm}}\right)\right]}.
   \end{array}\ee

\be\begin{array}{llll}\label{cond2zxsd}
 \displaystyle ~~~~~~~~~~~~~~~~~~~~~~~~~~~~~  \overrightarrow{{\cal D}_{y}}\chi^{(n)}_{\pm;{\cal H};{\bf AD}}|_{y_{i}=\pi}=0~~\implies \left(\alpha^{\pm;{\cal H}}_{n}\right)_{\bf AD}=
-\frac{\left[-\nu{\cal J}_{\nu}\left(x^{\pm;{\cal H};{\bf AD}}_{n}\right)
+{\cal J}^{'}_{\nu}\left(x^{\pm;{\cal H};{\bf AD}}_{n}\right)\right]}{\left[-\nu{\cal Y}_{\nu}\left(x^{\pm;{\cal H};{\bf AD}}_{n}\right)
+{\cal Y}^{'}_{\nu}\left(x^{\pm;{\cal H};{\bf AD}}_{n}\right)\right]}
   \end{array}\ee

where $z^{\pm;{\cal H};{\bf SD/AD}}_{n}(\pi):=x^{\pm;{\cal H};{\bf SD/AD}}_{n}=\frac{\left(m^{{\cal H}}_{n}\right)_{\pm;{\bf SD/AD}}}{k_{\pm}}e^{k_{\pm}r_{c}\pi}$.
For $e^{k_{\pm}r_{c}\pi}\gg 1,~\frac{\left(m^{{\cal H}}_{n}\right)_{\pm;{\bf SD/AD}}}{k_{\pm}}\ll 1$ the mass spectrum for the Kalb-Rammond fields is expected to be of the order of TeV scale i.e.
\be\begin{array}{llll}\label{approxcv}
 \displaystyle  {\bf Self-Dual ~KR}:~~~~   \left(\alpha^{\pm;{\cal H}}_{n}\right)_{\bf SD}\simeq \left[\frac{1}{\sqrt{\nu-2}\left(\nu-1\right)!}
\left(\frac{x^{\pm;{\cal H};{\bf SD}}_{n}}{2}e^{-2k_{\pm}r_{c}\pi}\right)^{\nu-1}\right]^{2}\\
\displaystyle  {\bf Anti-Dual ~KR}:~~~~  \left(\alpha^{\pm;{\cal H}}_{n}\right)_{\bf AD}\simeq\left[\frac{1}{\nu!}
\left(\frac{x^{\pm;{\cal H};{\bf AD}}_{n}}{2}e^{-2k_{\pm}r_{c}\pi}\right)^{\nu +1}\right]^{2}. 
   \end{array}\ee

Now using equation(\ref{approxcv}) and equation(\ref{cond1zx}) we get
\be\begin{array}{llllll}\label{rootzxmn}
\displaystyle  {\bf Self-Dual ~KR}:~~~~ \\
 \displaystyle    \left[\frac{1}{\sqrt{\nu-2}\left(\nu-1\right)!}
\left(\frac{x^{\pm;{\cal H};{\bf SD}}_{n}}{2}e^{-2k_{\pm}r_{c}\pi}\right)^{\nu-1}\right]^{2}
= -\frac{\left[\nu{\cal J}_{\nu}\left(x^{\pm;{\cal H};{\bf SD}}_{n}e^{-k_{\pm}r_{c}\pi}\right)
+{\cal J}^{'}_{\nu}\left(x^{\pm;{\cal H};{\bf SD}}_{n}e^{-k_{\pm}r_{c}\pi}\right)\right]}{\left[\nu{\cal Y}_{\nu}\left(x^{\pm;{\cal H};{\bf SD}}_{n}e^{-k_{\pm}r_{c}\pi}\right)
+{\cal Y}^{'}_{\nu}\left(x^{\pm;{\cal H};{\bf SD}}_{n}e^{-k_{\pm}r_{c}\pi}\right)\right]}\\
\displaystyle \Rightarrow {\cal J}_{\nu-1}\left(x^{\pm;{\cal H};{\bf SD}}_{n}\right)\approx 0
   \end{array}\ee

\be\begin{array}{llllll}\label{rootzxmn1}
\displaystyle  {\bf Anti-Dual ~KR}:~~~~~~~~~~~~~~~~~~~~~~ \\
 \displaystyle  ~~~~~~~~~~~~~~  \left[\frac{1}{\nu!}
\left(\frac{x^{\pm;{\cal H};{\bf AD}}_{n}}{2}e^{-2k_{\pm}r_{c}\pi}\right)^{\nu +1}\right]^{2}
= -\frac{\left[-\nu{\cal J}_{\nu}\left(x^{\pm;{\cal H};{\bf AD}}_{n}e^{-k_{\pm}r_{c}\pi}\right)
+{\cal J}^{'}_{\nu}\left(x^{\pm;{\cal H};{\bf AD}}_{n}e^{-k_{\pm}r_{c}\pi}\right)\right]}{\left[-\nu{\cal Y}_{\nu}\left(x^{\pm;{\cal H};{\bf AD}}_{n}e^{-k_{\pm}r_{c}\pi}\right)
+{\cal Y}^{'}_{\nu}\left(x^{\pm;{\cal H};{\bf AD}}_{n}e^{-k_{\pm}r_{c}\pi}\right)\right]}\\
\displaystyle \Rightarrow {\cal J}_{\nu +1}\left(x^{\pm;{\cal H};{\bf AD}}_{n}\right)\approx 0
   \end{array}\ee

which are transcendental equations of $x^{\pm;{\cal H};{\bf SD/AD}}_{n}$ and the roots of these equations give the Kalb-Rammond field mass spectrum
 $\left(m^{{\cal H}}_{n}\right)_{\pm;{\bf SD/AD}}$ in presence of perturbative 
Gauss-Bonnet coupling $\alpha_{(5)}$. Now using equation(\ref{no12cv}) the normalization constant for $n\neq 0$ mode reduces to the following expression
\be\begin{array}{llll}\label{vcvc1}
\displaystyle  {\bf Self-Dual ~KR}:\\
    \displaystyle {\cal N}^{\pm;{\cal H};{\bf SD}}_{(n)}=\sqrt{\left\{\int^{+\pi}_{-\pi}dy~e^{2A_{\pm}(y)}\left(z^{\pm;{\cal H};{\bf SD}}_{n}\right)^{2\nu}
\left[{\cal J}_{\nu}\left(z^{\pm;{\cal H};{\bf SD}}_{n}\right)+\left(\alpha^{\pm;{\cal H}}_{n}\right)_{\bf SD}{\cal Y}_{\nu}\left(z^{\pm;{\cal H};{\bf SD}}_{n}\right)
\right]^{2}\right\}} 
   \end{array}\ee
\be\begin{array}{llll}\label{vcvc2}
\displaystyle  {\bf Anti-Dual ~KR}:\\
    \displaystyle {\cal N}^{\pm;{\cal H};{\bf AD}}_{(n)}=\sqrt{\left\{\int^{+\pi}_{-\pi}dy~e^{2A_{\pm}(y)}\left(z^{\pm;{\cal H};{\bf AD}}_{n}\right)^{-2\nu}
\left[{\cal J}_{\nu}\left(z^{\pm;{\cal H};{\bf AD}}_{n}\right)+\left(\alpha^{\pm;{\cal H}}_{n}\right)_{\bf AD}{\cal Y}_{\nu}\left(z^{\pm;{\cal H};{\bf AD}}_{n}\right)
\right]^{2}\right\}}. 
   \end{array}\ee
For $e^{k_{\pm}r_{c}\pi}\gg 1,~\frac{\left(m^{{\cal H}}_{n}\right)_{\pm;{\bf SD/AD}}}{k_{\pm}}\ll 1$ the integration constant $\left(\alpha^{\pm;{\cal H}}_{n}\right)_{\bf SD/AD}\ll 1$.
Consequently ${\cal Y}_{\nu}(z^{\pm;{\cal H};{\bf SD/AD}}_{n})$ is neglected compared to ${\cal J}_{\nu}(z^{\pm;{\cal H};{\bf SD/AD}}_{n})$ in equation(\ref{sol1xcbn}) 
and then the normalization constant for $n\neq 0$ mode
turns out to be
\be\begin{array}{llll}\label{vcvc3}
\displaystyle  {\bf Self-Dual ~KR}:\\
    \displaystyle ~~~~~~~~~~~~~~~~~~~~~~~~~~{\cal N}^{\pm;{\cal H};{\bf SD}}_{(n)}=\sqrt{\left\{\int^{+\pi}_{-\pi}dy~e^{2A_{\pm}(y)}\left(z^{\pm;{\cal H};{\bf SD}}_{n}\right)^{2\nu}
\left[{\cal J}_{\nu}\left(z^{\pm;{\cal H};{\bf SD}}_{n}\right)
\right]^{2}\right\}} 
   \end{array}\ee
\be\begin{array}{llll}\label{vcvc4}
\displaystyle  {\bf Anti-Dual ~KR}:\\
    \displaystyle ~~~~~~~~~~~~~~~~~~~~~~~~~~{\cal N}^{\pm;{\cal H};{\bf AD}}_{(n)}=\sqrt{\left\{\int^{+\pi}_{-\pi}dy~e^{2A_{\pm}(y)}\left(z^{\pm;{\cal H};{\bf AD}}_{n}\right)^{-2\nu}
\left[{\cal J}_{\nu}\left(z^{\pm;{\cal H};{\bf AD}}_{n}\right)
\right]^{2}\right\}}. 
   \end{array}\ee
Consequently the extra dimensional dependent wave function for $n\neq 0$ turns out to be 
\be\begin{array}{llll}\label{sol1xcbn1}
\displaystyle {\bf Self-Dual ~KR}:~~~~    \chi^{(n)}_{\pm;\cal H;{\bf SD}}(y)=\frac{\left(z^{\pm;{\cal H};{\bf SD}}_{n}\right)^{\nu}}
{\sqrt{\left\{\int^{+\pi}_{-\pi}dy~e^{2A_{\pm}(y)}\left(z^{\pm;{\cal H};{\bf SD}}_{n}\right)^{2\nu}
\left[{\cal J}_{\nu}\left(z^{\pm;{\cal H};{\bf SD}}_{n}\right)
\right]^{2}\right\}}}{\cal J}_{\nu}
(z^{\pm;{\cal H};{\bf SD}}_{n})\\
\displaystyle  {\bf Anti-Dual ~KR}:~~~~     \chi^{(n)}_{\pm;\cal H;{\bf AD}}(y)=\frac{\left(z^{\pm;{\cal H};{\bf AD}}_{n}\right)^{-\nu}}
{\sqrt{\left\{\int^{+\pi}_{-\pi}dy~e^{2A_{\pm}(y)}\left(z^{\pm;{\cal H};{\bf AD}}_{n}\right)^{-2\nu}
\left[{\cal J}_{\nu}\left(z^{\pm;{\cal H};{\bf AD}}_{n}\right)
\right]^{2}\right\}}}{\cal J}_{\nu}
(z^{\pm;{\cal H};{\bf AD}}_{n}).\end{array}\ee

For massless $n=0$ mode the solution of the equation(\ref{dif178}) turns out to be
\be\begin{array}{llll}\label{sol1xcbn1}
\displaystyle {\bf Self-Dual ~KR}:~~~~    \chi^{(0)}_{\pm;\cal H;{\bf SD}}=\frac{C_{2}}{2\Theta_{0}r_{c}}e^{2\Theta_{0}r_{c}|y|}+C_{1}\\
\displaystyle  {\bf Anti-Dual ~KR}:~~~~     \chi^{(0)}_{\pm;\cal H;{\bf AD}}=-\frac{C_{2}}{2\Theta_{0}r_{c}}e^{-2\Theta_{0}r_{c}|y|}+C_{1}.\end{array}\ee
Here $C_{1}$ and $C_{2}$ are arbitrary integration constants. Now applying the boundary condition through the continuity
of the wave function we get $C_{2}=0$. As a result the zero mode solution turns out to be $ \chi^{(0)}_{\pm;{\cal H};{\bf SD}}= \chi^{(0)}_{\pm;{\cal H};{\bf AD}}=C_{1}$.
Now applying the normalization condition the ground state massless zero mode wave function turns out to be
\be\begin{array}{llll}\label{mlsx}
   \displaystyle   \chi^{(0)}_{\pm;{\cal H};{\bf SD}}= \chi^{(0)}_{\pm;{\cal H};{\bf AD}}
=C_{1}=\sqrt{\frac{k_{\pm}r_{c}}{e^{2k_{\pm}r_{c}\pi}-1}}\approx \sqrt{k_{\pm}r_{c}}e^{-k_{\pm}r_{c}\pi}.
   \end{array}\ee
This is again heavily suppressed on the visible brane.

\subsection{\bf Bulk Rank-4 Antisymmetric Tensor Field}
\label{kr3}

In five dimension we can have at most rank-3 antisymmetric tensor field with rank-4 antisymmetric tensor field strength
whose five dimensional action can be written as \cite{ssg12}
\be\begin{array}{llll}\label{abshg1c}
   \displaystyle S_{{\cal Z}}=\int d^{5}x\sqrt{-g_{(5)}}~{\cal Z}_{MNAB}(x,y){\cal Z}^{MNAB}(x,y)
   \end{array}\ee
where five dimensional action for rank-4 antisymmetric field strength tensor is given by
\be\begin{array}{llll}\label{abse1c}
   \displaystyle {\cal Z}_{MNAB}:=\overrightarrow{\partial}_{[M}{\cal X}_{NAB]} (x,y)
   \end{array}\ee
with antisymmetric rank-3 tensor potential
${\cal X}_{NAB}$, under the exchange of any two indices . It is usually called ``{\it Rammond-
Rammond}'' (R-R) differential three-form generated from the Rammond-Rammond sector of the closed string excitation. 
 Now applying the gauge fixing condition ${\cal X}_{\mu\nu4}=0$ the action stated in equation(\ref{abshg1c}) takes the following form
 \be\begin{array}{llll}\label{abskb1c}
   \displaystyle S_{{\cal Z}}=\int d^{5}x~\left[ e^{4A_{\pm}(y)}\eta^{\mu\lambda}\eta^{\nu\rho}\eta^{\alpha\gamma}\eta^{\beta\delta}
{\cal Z}_{\mu\nu\alpha\beta}(x,y){\cal Z}_{\lambda\rho\gamma\delta}(x,y)+\frac{4}{r_{c}}e^{2A_{\pm}(y)}\eta^{\mu\lambda}\eta^{\nu\rho}\eta^{\alpha\gamma}
\left(\overrightarrow{{\cal D}_{y}}{\cal X}_{\mu\nu\alpha}(x,y)\right)
\left(\overrightarrow{{\cal D}_{y}}{\cal X}_{\lambda\rho\gamma}(x,y)\right)\right]
   \end{array}\ee
where we introduce a new symbol $\overrightarrow{{\cal D}_{y}}:=\frac{d}{dy}$. Let the Kaluza-Klien expansion
of the rank-4 antisymmetric R-R three form potential field is given by
\be\begin{array}{lllll}\label{KK41c}
   \displaystyle {\cal X}_{\mu\nu\alpha}(x,y)=\sum^{\infty}_{n=0}{\cal X}^{(n)}_{\mu\nu\alpha}(x)~\frac{\chi^{(n)}_{\pm;\cal Z}(y)}{\sqrt{r_{c}}}. 
   \end{array}\ee
 
Now plugging equation(\ref{KK41c}) in equation(\ref{abskb1c}) the effective four dimensional action reduces to the following form:
\be\begin{array}{llll}\label{abs2321}
   \displaystyle S_{{\cal Z}}=\int d^{4}x\sum^{\infty}_{n=0}\left[\eta^{\mu\lambda}\eta^{\nu\rho}\eta^{\alpha\gamma}\eta^{\beta\delta}
{\cal Z}^{(n)}_{\mu\nu\alpha\beta}(x){\cal Z}^{(n)}_{\lambda\rho\gamma\delta}(x)+\left(M^{{\cal Z}}_{n}\right)^{2}_{\pm}
\eta^{\mu\lambda}\eta^{\nu\rho}\eta^{\alpha\gamma}{\cal X}^{(n)}_{\mu\nu}(x){\cal X}^{(n)}_{\alpha\beta}(x)\right]\end{array}\ee

where the effective four dimensional Rammond-Rammond field strength is defined as ${\cal Z}^{(n)}_{\mu\nu\alpha\beta}(x):=\overrightarrow{\partial}_{[\mu}{\cal X}^{(n)}_{\nu\alpha\beta]}(x)$.
In this context we impose the following orthonormalization condition of extra dimension dependent wave functions
\be\begin{array}{llll}\label{no121}
 \displaystyle   \int^{+\pi}_{-\pi}dy~e^{4A_{\pm}(y)}~\chi^{(m)}_{\pm;\cal Z}(y)~\chi^{(n)}_{\pm;\cal Z}(y)=\delta^{mn}
   \end{array}\ee
and the mass term of the gauge field is defined through the following differential equation as
\be\begin{array}{llll}\label{dif181c}
  \displaystyle   -\frac{1}{r^{2}_{c}}\overrightarrow{{\cal D}_{y}}\left(e^{2A_{\pm}(y)}\overrightarrow{{\cal D}_{y}}\chi^{(n)}_{\pm;\cal Z}(y)\right)=
e^{4A_{\pm}(y)}\left(m^{{\cal Z}}_{n}\right)^{2}_{\pm}\chi^{(n)}_{\pm;\cal Z}(y).
   \end{array}\ee
Here the mass of the nth mode Rammond-Rammond antisymmetric field is given by
$\left(M^{{\cal Z}}_{n}\right)_{\pm}=2\left(m^{{\cal Z}}_{n}\right)_{\pm}$.
Now introducing a new variable $z^{\pm;{\cal Z}}_{n}:=\frac{\left(m^{{\cal Z}}_{n}\right)_{\pm}}{k_{\pm}}e^{A_{\pm}(y)}$ 
equation(\ref{dif181c}) can be recast in terms of Bessel differential equation of order one as
\be\begin{array}{llll}\label{dif281cc}
    \displaystyle  \left[\left(z^{\pm;{\cal Z}}_{n}\right)^{2}\overrightarrow{{\cal D}^{2}}_{z^{\pm;{\cal Z}}_{n}}
+z^{\pm;{\cal Z}}_{n}\overrightarrow{{\cal D}}_{z^{\pm;{\cal Z}}_{n}}+\left\{\left(z^{\pm;{\cal Z}}_{n}\right)^{2}-1\right\}\right]\chi^{(n)}_{\pm;\cal Z}
=0
   \end{array}\ee
and the analytical solution turns out to be
\be\begin{array}{llll}\label{sol1xc11}
\displaystyle    \chi^{(n)}_{\pm;\cal Z}(y)=\frac{e^{-A_{\pm}(y)}}{{\cal N}^{\pm;\cal Z}_{(n)}}\left[{\cal J}_{1}
(z^{\pm;{\cal Z}}_{n})+\alpha^{\pm;\cal Z}_{n}{\cal Y}_{1}(z^{\pm;{\cal Z}}_{n})\right].
   \end{array}\ee
Here ${\cal N}^{\pm;\cal Z}_{(n)}$ be the normalization constant of the extra dimension dependent wave function and $\alpha^{\pm;\cal Z}_{n}$ is the
integration constant determined from the orthonormalization condition and the continuity conditions at the orbifold fixed point.
Self-adjointness and hermiticity of the differential operator appearing in equation(\ref{dif281cc})
demands that $\overrightarrow{{\cal D}_{y}}\chi^{(n)}_{\pm;\cal Z}(y)$ is
 continuous at the orbifold fixed points $y_{i}=0,\pi$. Consequently we have
\be\begin{array}{llll}\label{cond1zx1c}
 \displaystyle  \overrightarrow{{\cal D}_{y}}\chi^{(n)}_{\pm;\cal Z}|_{y_{i}=0}=0~~\implies \alpha^{\pm;\cal Z}_{n}=
\frac{\left[\frac{\left(m^{{\cal Z}}_{n}\right)_{\pm}}{k_{\pm}}{\cal J}^{'}_{1}\left(\frac{\left(m^{{\cal Z}}_{n}\right)_{\pm}}{k_{\pm}}\right)
-{\cal J}_{1}\left(\frac{\left(m^{{\cal Z}}_{n}\right)_{\pm}}{k_{\pm}}\right)\right]}{\left[{\cal Y}_{1}\left(\frac{\left(m^{{\cal Z}}_{n}\right)_{\pm}}{k_{\pm}}\right)
-\frac{\left(m^{{\cal Z}}_{n}\right)_{\pm}}{k_{\pm}}{\cal Y}^{'}_{1}\left(\frac{\left(m^{{\cal Z}}_{n}\right)_{\pm}}{k_{\pm}}\right)\right]}.
   \end{array}\ee

\be\begin{array}{llll}\label{cond2zx1c}
 \displaystyle  \overrightarrow{{\cal D}_{y}}\chi^{(n)}_{\pm;{\cal Z}}|_{y_{i}=\pi}=0~~\implies \alpha^{\pm;{\cal Z}}_{n}=
\frac{\left[{\cal J}_{1}\left(x^{\pm;{\cal Z}}_{n}\right)
-x^{\pm;{\cal Z}}_{n}{\cal J}^{'}_{1}\left(x^{\pm;{\cal Z}}_{n}\right)\right]}{\left[x^{\pm;{\cal Z}}_{n}{\cal Y}^{'}_{1}\left(x^{\pm;{\cal Z}}_{n}\right)
-{\cal Y}_{1}\left(x^{\pm;{\cal Z}}_{n}\right)\right]}
   \end{array}\ee
where $z^{\pm;{\cal Z}}_{n}(\pi):=x^{\pm;{\cal Z}}_{n}=\frac{\left(m^{{\cal Z}}_{n}\right)_{\pm}}{k_{\pm}}e^{k_{\pm}r_{c}\pi}$.
For $e^{k_{\pm}r_{c}\pi}\gg 1,~\frac{\left(m^{{\cal Z}}_{n}\right)_{\pm}}{k_{\pm}}\ll 1$ the mass spectrum for the Rammond-Rammond fields is expected to be of the order of TeV scale i.e.
\be\begin{array}{llll}\label{approxcv1c}
 \displaystyle    \alpha^{\pm;{\cal Z}}_{n}\simeq\frac{\pi}{32}\left(x^{\pm;{\cal Z}}_{n}\right)^{4}e^{-4k_{\pm}r_{c}\pi}. 
   \end{array}\ee

Now using equation(\ref{approxcv1c}) and equation(\ref{cond1zx1c}) we get
\be\begin{array}{llllll}\label{rootzx}
 \displaystyle    \frac{\pi}{32}\left(x^{\pm;{\cal Z}}_{n}\right)^{4}e^{-4k_{\pm}r_{c}\pi}
= \frac{\left[{\cal J}_{1}\left(x^{\pm;{\cal Z}}_{n}\right)
-x^{\pm;{\cal Z}}_{n}{\cal J}^{'}_{1}\left(x^{\pm;{\cal Z}}_{n}\right)\right]}{\left[x^{\pm;{\cal Z}}_{n}{\cal Y}^{'}_{1}\left(x^{\pm;{\cal Z}}_{n}\right)
-{\cal Y}_{1}\left(x^{\pm;{\cal Z}}_{n}\right)\right]} ~~~~~~~~~~~
\displaystyle \Rightarrow {\cal J}_{2}\left(x^{\pm;{\cal Z}}_{n}\right)\simeq \frac{\pi}{32}\left(x^{\pm;{\cal H}}_{n}\right)^{4}e^{-4k_{\pm}r_{c}\pi}
{\cal Y}^{'}_{1}\left(x^{\pm;{\cal Z}}_{n}\right)\approx 0
   \end{array}\ee
which is an transcendental equation of $x^{\pm;{\cal Z}}_{n}$ and the roots of this equation gives the gauge field mass spectrum 
$\left(m^{{\cal Z}}_{n}\right)_{\pm}$ in presence of perturbative 
Gauss-Bonnet coupling $\alpha_{(5)}$. Now using equation(\ref{no121}) the normalization constant for $n\neq 0$ mode reduces to the following expression
\be\begin{array}{llll}\label{vcvc1}
    \displaystyle {\cal N}^{\pm;{\cal Z}}_{(n)}=\frac{e^{k_{\pm}r_{c}\pi}}{
\sqrt{k_{\pm}r_{c}}}\sqrt{\left\{\left[{\cal J}_{1}\left(x^{\pm;{\cal Z}}_{n}\right)+\alpha^{\pm;{\cal Z}}_{n}{\cal Y}_{1}\left(x^{\pm;{\cal Z}}_{n}\right)
\right]^{2}-e^{-2k_{\pm}r_{c}}\left[{\cal J}_{1}\left(x^{\pm;{\cal Z}}_{n}e^{-k_{\pm}r_{c}\pi}
\right)+\alpha^{\pm;{\cal Z}}_{n}{\cal Y}_{1}\left(x^{\pm;{\cal Z}}_{n}e^{-k_{\pm}r_{c}\pi}\right)
\right]^{2}\right\}}. 
   \end{array}\ee
For $e^{k_{\pm}r_{c}\pi}\gg 1,~\frac{\left(m^{{\cal Z}}_{n}\right)_{\pm}}{k_{\pm}}\ll 1$ the integration constant $\alpha^{\pm;{\cal Z}}_{n}\ll 1$.
Consequently ${\cal Y}_{1}(z^{\pm;{\cal Z}}_{n})$ is neglected compared to ${\cal J}_{1}(z^{\pm;{\cal Z}}_{n})$ in equation(\ref{sol1xc11}) 
and then the normalization constant for $n\neq 0$ mode
turns out to be
\be\begin{array}{llll}\label{vcvc12}
    \displaystyle {\cal N}^{\pm;{\cal Z}}_{(n)}=\frac{e^{k_{\pm}r_{c}\pi}}{
\sqrt{k_{\pm}r_{c}}}{\cal J}_{1}\left(x^{\pm;{\cal Z}}_{n}\right). 
   \end{array}\ee
Consequently the extra dimensional dependent wave function for $n\neq 0$ turns out to be 
\be\begin{array}{llll}\label{sol1xc1}
\displaystyle    \chi^{(n)}_{\pm;\cal Z}(y)=\frac{\sqrt{k_{\pm}r_{c}}~e^{-A_{\pm}(y)}}{e^{k_{\pm}r_{c}\pi}}\frac{{\cal J}_{1}
(z^{\pm;{\cal Z}}_{n})}{{\cal J}_{1}
(x^{\pm;{\cal Z}}_{n})}.
   \end{array}\ee

For massless $n=0$ mode the solution of the equation(\ref{dif181c}) turns out to be
\be\begin{array}{lllll}\label{jkloss1}
   \displaystyle  \chi^{(0)}_{\pm;{\cal Z}}=-\frac{C_{1}}{2k_{\pm}r_{c}}e^{-2A_{\pm}(y)}+C_{2}.
   \end{array}\ee
Here $C_{1}$ and $C_{2}$ are arbitrary integration constants. Now applying the boundary condition through the continuity
of the wave function we get $C_{1}=0$. As a result the zero mode solution turns out to be $ \chi^{(0)}_{\pm;{\cal Z}}=C_{2}$.
Now applying the normalization condition the ground state massless zero mode wave function turns out to be
\be\begin{array}{llll}\label{mlsx}
   \displaystyle  \chi^{(0)}_{\pm;{\cal Z}}=C_{2}=\sqrt{\frac{2k_{\pm}r_{c}}{e^{4k_{\pm}r_{c}\pi}-1}}\approx \sqrt{2k_{\pm}r_{c}}e^{-2k_{\pm}r_{c}\pi}.
   \end{array}\ee

This give zero mode is heavily suppressed in the visible brane, though the warping will be reduced if one choses large Gauss-Bonnet coupling $\alpha_{(5)}$.
Moreover the zeroth mode is function of extra dimensional coordinate $y$ appearing through the dilatonic contribution. It is interesting to note that
the suppression of the zero mode on the visible brane increases with the rank of the field. This explains the reason of invisibility of these fields in our universe.

\section{\bf Bulk-brane Interaction in presence of Gauss-bonnet Coupling}
In this section we elaborately discuss about the possible interaction picture between brane - bulk fields
in the context of ${\bf dS_{5}/AdS_{5}\otimes S^{5}}$ warped phenomenology and there consequences in presence of the five dimensional bulk Gauss-Bonnet coupling. 

\subsection{\bf Fermion interaction}


\subsubsection{\bf Brane Standard Model fields with bulk Gravitons}

The five dimensional action describing the interaction between bulk graviton and visible Standard Model
fields dominated by fermionic contribution on the brane is given by
\be\begin{array}{llll}\label{delq1}
 \displaystyle {\cal S}_{\bf SM-\bf G}=-\frac{{\cal K}_{(5)}}{2}\int d^{5}x\sqrt{-g_{(5)}}{\bf T}^{\alpha\beta}_{\bf SM}(x){\bf h}_{\alpha\beta}(x,y)\delta(y-\pi)   
   \end{array}\ee
where ${\bf T}^{\alpha\beta}_{\bf SM}(x)$ represents the energy momentum or stress energy tensor containing all informations of Standard Model matter fields 
on the visible brane. In this context ${\cal K}_{(5)}$ is the coupling strength describing the tensor fluctuation in the context of graviton phenomenology.
After substituting the Kaluza-Klien expansion for graviton degrees of freedom and rescaling the fields appropriately, the effective four dimensional action turns out to be
\be\begin{array}{llll}\label{delq2}
 \displaystyle {\cal S}_{\bf SM-\bf G}=-\frac{{\cal K}_{(5)}}{2}\int d^{4}x~r_{c}~e^{-4A_{\pm}(y)}
{\bf T}^{\alpha\beta}_{\bf SM}(x)\sum^{\infty}_{n=0}{\bf h}^{(n)}_{\alpha\beta}(x)\frac{\chi^{(n)}_{\pm;\bf G}(y)}{\sqrt{r_{c}}}\delta(y-\pi)\\
\displaystyle ~~~~~~~~~~=-\frac{\sqrt{r_{c}}{\cal K}_{(5)}}{2}\int d^{4}x~e^{-4A_{\pm}(\pi)}
{\bf T}^{\alpha\beta}_{\bf SM}(x)\sum^{\infty}_{n=0}{\bf h}^{(n)}_{\alpha\beta}(x)\chi^{(n)}_{\pm;\bf G}(\pi)\\
\displaystyle ~~~~~~~~~~=-\frac{\sqrt{k_{\pm}}r_{c}{\cal K}_{(5)}}{2}\int d^{4}x~
{\bf T}^{\alpha\beta}_{\bf SM}(x)\left[{\bf h}^{(0)}_{\alpha\beta}(x)
+ e^{k_{\pm}r_{c}\pi}\sum^{\infty}_{n=1}{\bf h}^{(n)}_{\alpha\beta}(x)\right].   
   \end{array}\ee
It is evident from equation(\ref{delq2}) that while the zero mode couples to the brane fields with usual gravitational coupling $\sim 1/M_{PL}$ which we have taken as unity,
the coupling of the KK modes are $\sim  e^{k_{\pm}r_{c}\pi} / M_{PL} \sim ~ TeV^{-1}$ which is much larger than the coupling of massless graviton.
Though such feature is also observed for the graviton KK modes in the usual RS model, here due to GB coupling $\alpha_{(5)}$, the $k_{\pm}$ will change. It may be seen from 
the figures that the values of  $k_{\pm}$ decrease with $\alpha_{(5)}$ and hence the graviton KK mode couplings decrease due to GB interaction leading to the decrease in 
their detection signature in collider experiments unless one modifies the value of $r_c$ to resolve the gauge hierarchy problem. 
Moreover equation(\ref{massasdgrav}) and figure(\ref{fig:subfigureExample51a}) indicate that the decrease in $k_{\pm}$ lead to increase in the masses for
the graviton KK modes. Thus the absence of any signature of graviton KK modes, as reported by ATLAS data in dilepton decay processes, may be
the result of GB  coupling rather than any negative result for the warped geometry models. However in an alternative scenario if one modifies the value
of $r_c$ to obtain the desired Planck to TeV scale warping, then the KK mode graviton couplings with brane fields do not change from the RS values, but the
graviton KK mode masses decrease from their counter part in RS model. In that case the non-vanishing GB coupling make the detectability of the signature 
of KK mode graviton through dilepton decay process more pronounced. Absence of any such signature, as reported by ATLAS collaborations, put question on
the validity of GB extension in RS like warped geometry model.  


\subsubsection{\bf Brane fermions with bulk Kalb-Rammond field}
The interaction between bulk Kalb-Rammond field with the fermions localized at visible brane is described by the following
action:
\be\begin{array}{lllll}\label{int1}
    \displaystyle {\cal S}_{\bf \bar{\Psi}\Psi{\cal H}}=-ig\int d^{5}x ~Det({\cal V})~\bar{\Psi}_{\bf L,R}(x)\gamma^{\alpha}
{\cal V}_{\alpha}^{M}\sigma^{NL}{\cal H}_{\mu\nu\lambda}(x,y){\bf \Psi}_{\bf L,R}(x)\delta_{M}^{\mu}\delta_{N}^{\nu}\delta_{L}^{\lambda}\delta(y-\pi)
   \end{array}\ee
where $\sigma^{NL}:=\frac{i}{4}\left[\Gamma^{N},\Gamma^{L}\right]$. Substituting Kaluza-Klein expansion of the 
bulk Kalb-Rammond field in equation(\ref{int1}) we get
\be\begin{array}{lllll}\label{int2}
  \displaystyle  {\cal S}_{\bf \bar{\Psi}\Psi{\cal H}}=-\frac{ig}{\sqrt{r_{c}}}\int d^{4}x \sum^{\infty}_{n=0}e^{-\frac{3}{2}A_{\pm}(\pi)} 
\bar{\Psi}_{\bf L,R}(x)\gamma^{\mu}\sigma^{\nu\lambda}{\cal H}^{(n)}_{\mu\nu\lambda}(x)\chi^{(n)}_{\pm;{\cal H}}(\pi){\bf \Psi}_{\bf L,R}(x)e^{-\frac{3}{2}A_{\pm}(\pi)} 
   \end{array}\ee
Now rescaling the fermionic fields via $\bar{\Psi}_{\bf L,R}(x)\rightarrow e^{-\frac{3}{2}A_{\pm}(\pi)}\bar{\Psi}_{\bf L,R}(x)$ equation(\ref{int2})
takes the following form
\be\begin{array}{lllll}\label{int3}
\displaystyle \underline{\bf Pure~KR}:-\\
    \displaystyle {\cal S}^{Pure}_{\bf \bar{\Psi}\Psi{\cal H}}=-i\int d^{4}x\bar{\Psi}_{\bf L,R}(x)\gamma^{\mu}\sigma^{\nu\lambda}
\left[\frac{1}{\frac{M_{PL}}{g}e^{k_{\pm}r_{c}\pi}}{\cal H}^{(0)}_{\mu\nu\lambda}(x)+
\frac{2}{\pi}\frac{1}{\frac{\Lambda_{\pi}}{g}}\sum^{\infty}_{n=1}{\cal H}^{(n)}_{\mu\nu\lambda}(x)
\frac{{\cal J}_{0}(x^{\pm;\cal H}_{n})}{x^{\pm;\cal H}_{n}}\right]{\bf \Psi}_{\bf L,R}(x)
   \end{array}\ee

\be\begin{array}{lllll}\label{int31}
\displaystyle \underline{\bf Topologically ~~extended~~KR}:-\\
\displaystyle \underline{\bf (a)Self~Dual~KR}:-\\
    \displaystyle {\cal S}^{\bf SD}_{\bf \bar{\Psi}\Psi{\cal H}}=-i\int d^{4}x\bar{\Psi}_{\bf L,R}(x)\gamma^{\mu}\sigma^{\nu\lambda}
\left[\frac{1}{\frac{M_{PL}}{g}e^{k_{\pm}r_{c}\pi}}{\cal H}^{(0)}_{\mu\nu\lambda}(x)\right.\\ \left. \displaystyle ~~~~~~~~~~~~~~~~~~~~~~~~~~~~~~~~~~~~~~ +
\frac{1}{\sqrt{k_{\pm}r_{c}}M_{PL}}\sum^{\infty}_{n=1}
\frac{{\cal H}^{(n)}_{\mu\nu\lambda}(x){\cal J}_{\nu}
(x^{\pm;{\cal H};{\bf SD}}_{n})\left(x^{\pm;{\cal H};{\bf SD}}_{n}\right)^{\nu}}
{\sqrt{\left\{\int^{+\pi}_{-\pi}dy~e^{2A_{\pm}(y)}\left(z^{\pm;{\cal H};{\bf SD}}_{n}\right)^{2\nu}
\left[{\cal J}_{\nu}\left(z^{\pm;{\cal H};{\bf SD}}_{n}\right)
\right]^{2}\right\}}}\right]{\bf \Psi}_{\bf L,R}(x),\end{array}\ee
\be\begin{array}{lllll}\label{int31a}
\displaystyle \underline{\bf (b)Anti~Dual~KR}:-\\
     \displaystyle {\cal S}^{\bf AD}_{\bf \bar{\Psi}\Psi{\cal H}}=-i\int d^{4}x\bar{\Psi}_{\bf L,R}(x)\gamma^{\mu}\sigma^{\nu\lambda}
\left[\frac{1}{\frac{M_{PL}}{g}e^{k_{\pm}r_{c}\pi}}{\cal H}^{(0)}_{\mu\nu\lambda}(x)\right.\\ \left. \displaystyle ~~~~~~~~~~~~~~~~~~~~~~~~~~~~~~~~~~~~~~ +
\frac{1}{\sqrt{k_{\pm}r_{c}}M_{PL}}\sum^{\infty}_{n=1}
\frac{{\cal H}^{(n)}_{\mu\nu\lambda}(x){\cal J}_{\nu}
(x^{\pm;{\cal H};{\bf AD}}_{n})\left(x^{\pm;{\cal H};{\bf AD}}_{n}\right)^{-\nu}}
{\sqrt{\left\{\int^{+\pi}_{-\pi}dy~e^{2A_{\pm}(y)}\left(z^{\pm;{\cal H};{\bf AD}}_{n}\right)^{-2\nu}
\left[{\cal J}_{\nu}\left(z^{\pm;{\cal H};{\bf AD}}_{n}\right)
\right]^{2}\right\}}}\right]{\bf \Psi}_{\bf L,R}(x)
   \end{array}\ee

where $M_{PL}$ is defined earlier and $\Lambda_{\pi}:=M_{PL}e^{-k_{\pm}r_{c}\pi}$. It is evident from equation(\ref{int3})
that when pure Kalb-Rammond field is interacting with the fermions localized at the visible brane then
the zero mode is exponentially suppressed and the excited Kaluza-Klien modes of Kalb-Rammond field are stronger as the number of mode $n$ increases.
The remarkable point to note here is that the massive mode coupling to fermion, as
given by equations(\ref{int31}), are drastically reduced compared to the corresponding
case without the presence of the axionic contribution. It appears as if the large coefficient $\Theta_{0}$ in the
additional five dimensional topological term characterized by the axionic extra part in the action causes the Kalb-Rammond modes to decouple from
all visible physics on the brane, although a tower within the kinematic reach of accelerator
experiments is still around.


\subsubsection{\bf Brane fermions with bulk rank-4 antisymmetric tensor field}
The interaction between pure bulk rank-4 antisymmetric tensor field with the fermions localized at visible brane is described by the following
action:
\be\begin{array}{lllll}\label{int12}
    \displaystyle {\cal S}_{\bf \bar{\Psi}\Psi{\cal Z}}=-ig_{z}\int d^{5}x ~Det({\cal V})~\bar{\Psi}_{\bf L,R}(x)\gamma^{\alpha}
{\cal V}_{\alpha}^{M}\sigma^{NL}\Gamma^{S}{\cal Z}_{\mu\nu\rho\beta}(x,y){\bf \Psi}_{\bf L,R}(x)\delta_{M}^{\mu}\delta_{N}^{\nu}\delta_{L}^{\rho}\delta_{S}^{\beta}
\delta(y-\pi),\\
~~~~~~~~\displaystyle =-ig_{z}\int d^{4}x \int^{+\pi}_{-\pi}dy~\bar{\Psi}_{\bf L,R}(x)e^{-\frac{3}{2}A_{\pm}(y)}\gamma^{\alpha}
{\cal V}_{\alpha}^{\mu}\sigma^{\nu\lambda}\gamma^{\beta}{\cal Z}_{\mu\nu\lambda\beta}(x,y){\bf \Psi}_{\bf L,R}(x)e^{-\frac{3}{2}A_{\pm}(y)}
\delta(y-\pi)
   \end{array}\ee
Now rescaling the fermionic fields via $\bar{\Psi}_{\bf L,R}(x)\rightarrow e^{-\frac{3}{2}A_{\pm}(\pi)}\bar{\Psi}_{\bf L,R}(x)$ and
substituting the Kaluza-Klien expansion of the rank-4 antisymmetric tensor field in equation(\ref{int12}) we get
\be\begin{array}{lllll}\label{int33}
    \displaystyle {\cal S}_{\bf \bar{\Psi}\Psi{\cal Z}}=-i\int d^{4}x\bar{\Psi}_{\bf L,R}(x)\gamma^{\mu}\sigma^{\nu\lambda}\gamma^{\beta}
\left[\frac{1}{\frac{M_{PL}}{\sqrt{2}g_{z}}e^{2k_{\pm}r_{c}\pi}}{\cal Z}^{(0)}_{\mu\nu\lambda\beta}(x)+
\frac{1}{\frac{\Lambda_{\pi}}{g_{z}}e^{3k_{\pm}r_{c}\pi}}\sum^{\infty}_{n=1}{\cal Z}^{(n)}_{\mu\nu\lambda\beta}(x)
\right]{\bf \Psi}_{\bf L,R}(x).
   \end{array}\ee
This explicitly shows that both the zero mode and the excited mode of the Kaluza-Klien expansion of the bulk rank-4 antisymmetric tensor field 
are suppressed at the visible brane. But the amount of such suppression is larger for massive excited modes.

\begin{table}[hb]
\centering
\subtable[]{
\centering
\begin{tabular}{|c|c|c|}
\hline ${\bf H}^{(000)}_{9}$ & ${\bf H}^{(000)}_{10}$ & ${\bf H}^{(0000)}_{11}$ 
 \\
 \hline
0.452/4.348&0.219/3.456 &0.145/2.116\\
\hline
\end{tabular}
\label{tab1dilx1a}
}
$~~~~~~~~~~~~~~~~~~$\subtable[]{
\centering
\begin{tabular}{|c|c|c|}
\hline ${\bf H}^{(000)}_{9}$ & ${\bf H}^{(000)}_{10}$ & ${\bf H}^{(0000)}_{11}$ 
 \\
 \hline
0.489/4.248&0.244/3.566 &0.179/2.180\\
\hline
\end{tabular}
\label{tab1dilx1b}
}
\caption{Numerical values of different heterotypic couplings for lowest lying modes
 of the triliear fermionic interaction with dilatonic fields for \subref{tab1dilx1a} $k_{-}$ branch and \subref{tab1dilx1b} $k_{+}$ branch. }
\label{tab1dilx1}
\end{table}

\subsubsection{\bf Bulk fermions with bulk dilatons}

The five dimensional action describing the interaction between
 the massive fermionic field $\left(spin~\frac{1}{2}~ type\right)$ and dilaton field can be written as
\be\begin{array}{llll}\label{ferfhenit}
 \displaystyle S_{f-\phi}=\int d^{5}x\left[Det({\cal V})\right]~e^{\theta_{7}\phi(y)}\left\{i\bar{\Psi}_{{\bf L},{\bf R}}(x,y)\gamma^{\alpha}
{\cal V}_{\alpha}^{M}\overleftrightarrow{{\large\bf D}_{\mu}}\Psi_{{\bf L},{\bf R}}(x,y)\delta^{\mu}_{M}
-sgn(y)m_{f}\bar{\Psi}_{{\bf L},{\bf R}}(x,y)\Psi_{{\bf R},{\bf L}}(x,y)+~h.c.\right\}
   \end{array}\ee
where $\overleftrightarrow{{\large\bf D}_{\mu}}:=\left(\overleftrightarrow{\partial_{\mu}}
+\Omega_{\mu}+ig_{f}{\cal A}_{\mu}\right)$ represents the covariant derivative in presence
${\cal U}(1)$ abelian gauge field and fermionic spin connection 
$\Omega_{\mu}=\frac{1}{8}\omega_{\mu}^{\hat{A}\hat{B}}\left[\Gamma_{\hat{A}},\Gamma_{\hat{B}}\right]$.
Substituting  the Kaluza-Klien expansion for fermion and extra dimension dependent 
dilaton field (similar as the bulk scalar field)in the action stated in equation(\ref{ferfhenit}) 
we get 
\be\begin{array}{llll}\label{dftt1}
\displaystyle  S_{f-\phi}=\int d^{4}x \sum^{\infty}_{n=0}\sum^{\infty}_{m=0}
\left\{\left(1+\theta_{7}\sqrt{k_{\pm}}\right)\delta^{mn}+\frac{\theta_{7}}{\sqrt{r_{c}}}
\sum^{\infty}_{r=0}{\bf H}^{(mnr)}_{9}\right\}\left[\bar{\Psi}^{(n)}_{{\bf L},{\bf R}}(x)i\overleftrightarrow{{\partial}\slashed}{\Psi}^{(n)}_{{\bf L},{\bf R}}(x)
-m^{{\bf L},{\bf R}}_{n}\bar{\Psi}^{(n)}_{{\bf L},{\bf R}}(x)\Psi^{(n)}_{{\bf R},{\bf L}}(x)\right]\\
\displaystyle
~~~~~~~~~~~~+\frac{ig_{f}}{\sqrt{r_{c}}}\sum^{\infty}_{n=0}\sum^{\infty}_{m=0}\sum^{\infty}_{p=0}
\left\{\left(1+\theta_{7}\sqrt{k_{\pm}}\right){\bf H}^{(mnp)}_{10}+\frac{\theta_{7}}{\sqrt{r_{c}}}
\sum^{\infty}_{s=0}{\bf H}^{(mnps)}_{11}\right\}\bar{\Psi}^{(m)}_{{\bf L},{\bf R}}(x)
i{\cal A}\slashed^{(n)}(x)\Psi^{(p)}_{{\bf R},{\bf L}}(x) 
   \end{array}\ee
where the trilinear and quartic interaction between dilatonic field and fermionic fields are 
characterized by
\be\begin{array}{llll}\label{u1}
   \displaystyle  {\bf H}^{(mnr)}_{9}:=\int^{+\pi}_{-\pi}dy~e^{A_{\pm}(y)}\hat{f}^{(m)\star}_{{\bf L},{\bf R}}(z^{\pm;{\bf L},{\bf R}}_{m})
\hat{f}^{(n)}_{{\bf L},{\bf R}}(z^{\pm;{\bf L},{\bf R}}_{n})\chi^{(r)}_{\pm;\cal \phi}(z^{\pm;{\cal \phi}}_{r}),\\
\displaystyle  {\bf H}^{(mnp)}_{10}:=\int^{+\pi}_{-\pi}dy~e^{A_{\pm}(y)}\hat{f}^{(m)\star}_{{\bf L},{\bf R}}(z^{\pm;{\bf L},{\bf R}}_{m})
\hat{f}^{(n)}_{{\bf L},{\bf R}}(z^{\pm;{\bf L},{\bf R}}_{n})\chi^{(p)}_{\pm;\cal A}(z^{\pm;{\cal A}}_{p}),\\
\displaystyle  {\bf H}^{(mnps)}_{11}:=\int^{+\pi}_{-\pi}dy~e^{A_{\pm}(y)}\hat{f}^{(m)\star}_{{\bf L},{\bf R}}(z^{\pm;{\bf L},{\bf R}}_{m})
\hat{f}^{(p)}_{{\bf L},{\bf R}}(z^{\pm;{\bf L},{\bf R}}_{n})\chi^{(p)}_{\pm;\cal A}(z^{\pm;{\cal A}}_{p})\chi^{(s)}_{\pm;\cal \phi}(z^{\pm;{\cal \phi}}_{s}).
   \end{array}\ee
In table(\ref{tab1dilx1a}) and table(\ref{tab1dilx1b}) we have tabulated the numerical values of 
the trilinear and quartic interaction for zeroth mode.


\subsection{\bf Self interaction of  bulk scalar field }

In five dimension the m-th order self interaction for bulk scalar field (other than dilaton)
is described by 
\be\begin{array}{llll}\label{eq1}
\displaystyle {\cal S}_{\bf \Phi\Phi}=\frac{\lambda_{(5)}}{M^{3m-5}_{(5)}}\int d^{5}x\sqrt{-g_{(5)}}\left({\bf \Phi}(x,y)\right)^{2m}.    
   \end{array}\ee
 Substituting the Kaluza-Klien expansion for bulk scalar field the effective four dimensional contribution to the 
self interaction turns out to be

\be\begin{array}{llll}\label{eq2}
\displaystyle {\cal S}_{\bf \Phi\Phi}=\frac{\lambda_{(5)}}{M^{3m-5}_{(5)}r^{m}_{c}}\sum^{\infty}_{r=0}
\int d^{4}x~\left({\bf \Phi}^{(r)}(x)\right)^{2m}
\int^{+\pi}_{-\pi}dy ~r_{c}~e^{-4A_{\pm}(y)}\left(\chi^{(r)}_{\pm;\bf \Phi}(y)\right)^{2m}\\
\displaystyle~~~~~~=\sum^{\infty}_{r=0}
\int d^{4}x~\left({\bf \Phi}^{(r)}(x)\right)^{2m}\lambda^{\bf \Phi;(r)}_{(4)}    
   \end{array}\ee
where the effective four dimensional mth order self interaction coupling strength can be expressed in terms of its 
five dimensional counterpart as
\be\begin{array}{lllll}\label{fgd1}
\displaystyle  \lambda^{\bf \Phi;(r)}_{(4)}= \frac{\lambda_{(5)}}{M^{3m-5}_{(5)}r^{m}_{c}}\int^{+\pi}_{-\pi}dy ~r_{c}
~e^{-4A_{\pm}(y)}\left(\chi^{(r)}_{\pm;\bf \Phi}(y)\right)^{2m}\\
\displaystyle~~~~~=2\lambda_{(5)}\left(\frac{k_{\pm}}{M_{(5)}}\right)^{m-1}\left(M_{(5)}e^{-k_{\pm}r_{c}\pi}\right)^{4-2m}
 \int^{1}_{0}d{\bf \Pi}~{\bf\Pi}^{4m-5}\left[\frac{{\cal J}_{\nu^{\bf \Phi}_{\pm}}\left(x^{\pm;{\bf\Phi}}_{r}{\bf \Pi}\right)}
{{\cal J}_{\nu^{\bf \Phi}_{\pm}}\left(x^{\pm;{\bf\Phi}}_{r}\right)\sqrt{1+\frac{4-\left(\nu^{\bf \Phi}_{\pm}\right)^{2}}
{\left(x^{\pm;{\bf\Phi}}_{r}\right)^{2}}}}\right]^{2m}.  
   \end{array}\ee
It is important to mention here that the renormalizable scalar field theory only exist in the visible brane iff $m=2$. 
There may be other situation appears where the self interaction of the five dimensional bulk scalar 
field is characterized by the derivative m-th order self interaction. 
In the bulk the five dimensional action describing the effect of derivative self-interaction is characterized by 
\be\begin{array}{llll}\label{eq11}
\displaystyle {\cal S}_{\bf \Phi\Phi}=\frac{\lambda_{(5)}}{M^{5m-5}_{(5)}}\int d^{5}x\sqrt{-g_{(5)}}\left(g^{AB}\overrightarrow{\partial}_{A}{\bf \Phi}(x,y)
\overrightarrow{\partial}_{B}{\bf \Phi}(x,y)\right)^{m}.    
   \end{array}\ee
 Substituting the Kaluza-Klien expansion for bulk scalar field the effective four dimensional contribution to the 
derivative self interaction turns out to be
\be\begin{array}{llll}\label{eq2121}
\displaystyle {\cal S}_{\bf \Phi\Phi}=\frac{\lambda_{(5)}}{M^{5m-5}_{(5)}r^{m}_{c}}\sum^{\infty}_{r=0}\sum^{\infty}_{s=0}
\int d^{4}x~\left({\bf \Phi}^{(r)}(x)\right)^{m}\left({\bf \Phi}^{(s)}(x)\right)^{m}
\int^{+\pi}_{-\pi}dy ~r_{c}~e^{-4A_{\pm}(y)}\left(\overrightarrow{D}_{y}\chi^{(r)}_{\pm;\bf \Phi}(y)\right)^{m}\left(\overrightarrow{D}_{y}\chi^{(s)}_{\pm;\bf \Phi}(y)\right)^{m}\\
\displaystyle~~~~~~=\sum^{\infty}_{r=0}\sum^{\infty}_{s=0}
\int d^{4}x~\left({\bf \Phi}^{(r)}(x)\right)^{m}\left({\bf \Phi}^{(s)}(x)\right)^{m}\lambda^{\bf \Phi;(rs)}_{(4)}   
   \end{array}\ee
where the effective four dimensional mth order derivative self interaction coupling strength can be expressed in terms of its 
five dimensional counterpart as
\be\begin{array}{lllll}\label{fgd1zx11}
\displaystyle  \lambda^{\bf \Phi;(rs)}_{(4)}= \frac{\lambda_{(5)}}{M^{5m-5}_{(5)}r^{m}_{c}}\int^{+\pi}_{-\pi}dy ~r_{c}
~e^{-4A_{\pm}(y)}\left(\overrightarrow{D}_{y}\chi^{(r)}_{\pm;\bf \Phi}(y)\right)^{m}\left(\overrightarrow{D}_{y}\chi^{(s)}_{\pm;\bf \Phi}(y)\right)^{m}\\
\displaystyle~~~~~~~~=2\lambda_{(5)}\left(\frac{k_{\pm}}{M_{(5)}}\right)^{3m-1}\left(M_{(5)}e^{-k_{\pm}r_{c}\pi}\right)^{4-2m}
 \int^{1}_{0}d{\bf \Pi}~{\bf\Pi}^{2m-5}\left[\overrightarrow{D}_{\bf\Pi}\left(\frac{{\cal J}_{\nu^{\bf \Phi}_{\pm}}\left(x^{\pm;{\bf\Phi}}_{r}{\bf \Pi}\right)}
{{\cal J}_{\nu^{\bf \Phi}_{\pm}}\left(x^{\pm;{\bf\Phi}}_{r}\right)\sqrt{1+\frac{4-\left(\nu^{\bf \Phi}_{\pm}\right)^{2}}
{\left(x^{\pm;{\bf\Phi}}_{r}\right)^{2}}}}\right)\right]^{m}\\ \displaystyle~~~~~~~~~~~~~~~~~~~~~~~~~~~~~~~~~~~~~~~~~~~~~~~~~~~~~~~~~~~~~~~~~~~~~~~~~~~~~~~~~~~~~~~~~
\times\left[\overrightarrow{D}_{\bf\Pi}\left(\frac{{\cal J}_{\nu^{\bf \Phi}_{\pm}}\left(x^{\pm;{\bf\Phi}}_{s}{\bf \Pi}\right)}
{{\cal J}_{\nu^{\bf \Phi}_{\pm}}\left(x^{\pm;{\bf\Phi}}_{s}\right)\sqrt{1+\frac{4-\left(\nu^{\bf \Phi}_{\pm}\right)^{2}}
{\left(x^{\pm;{\bf\Phi}}_{s}\right)^{2}}}}\right)\right]^{m}.  
   \end{array}\ee


\subsection{\bf Bulk Gravidilatonic interaction}
The five dimensional action describing the interaction between two spin-2 graviton and the dilatonic 
field via Gauss-Bonnet perturbative coupling in the bulk is given by the following gravidilaton contribution
\be\begin{array}{lllll}\label{hj67}
   \displaystyle {\cal S}_{{\cal \phi}~h}=\alpha_{(5)}\int d^{5}x\sqrt{-g_{(5)}}e^{\theta_{8}\phi(y)}
{\bf h}_{\alpha\beta}(x,y){\bf h}^{\alpha\beta}(x,y).  
   \end{array}\ee
Throughout this analysis we assume that the graviton field non-interacting with other field contents in the bulk. 
Only self-interaction and gravidilatonic interaction are allowed in the bulk.
Now substituting the Kaluza-Klien expansion for graviton and dilaton equation(\ref{hj67}) reduces to the following form
 \be\begin{array}{llll}\label{hj78}
  \displaystyle  {\cal S}_{{\cal \phi}~h}=\alpha_{(5)}\int d^{4}x ~\eta^{\alpha\mu}\eta^{\beta\nu}
\sum^{\infty}_{p=0}\sum^{\infty}_{q=0}{\bf h}^{(p)}_{\alpha\beta}(x){\bf h}^{(q)}_{\mu\nu}(x)\left[\left(1+\theta_{8}\sqrt{k_{\pm}}\right)
{\bf X}^{(pq)}_{1}+\frac{\theta_{8}}{\sqrt{r_{c}}}\sum^{\infty}_{r=1}{\bf X}^{(pqr)}_{2}\right]
    \end{array}\ee
where the gravidilatonic interactions are characterized by the following integrals:
\be\begin{array}{llll}\label{hhhhxtttt}
 \displaystyle  {\bf X}^{(pq)}_{1}:={\bf G}^{pq}_{1}, ~~~~{\bf X}^{(pqr)}_{2}:=\int^{+\pi}_{-\pi}dy~e^{-2A_{\pm}(y)}\chi^{(p)}_{\pm;\bf G}(z^{\pm;{\bf G}}_{p}) 
 \chi^{(q)}_{\pm;\bf G}(z^{\pm;{\bf G}}_{q})\chi^{(r)}_{\pm;\cal \phi}(z^{\pm;{\cal \phi}}_{r}).
   \end{array}\ee
The numerical values of such contributions are estimated in Table(\ref{tab2121a}) and Table(\ref{tab2121b}).


\begin{table}[h!]
\centering
\subtable[]{
\centering
\begin{tabular}{|c|c|c|c|c|c|c|c|c|}
\hline ${\bf X}^{(00)}_{1}$ & ${\bf X}^{(01)}_{1}$ & ${\bf X}^{(10)}_{1}$ &
${\bf X}^{(11)}_{1}$
 \\
 \hline
0.500&0.367 &0.212 &0.189 \\
\hline
\end{tabular}
\label{tab2121a}
}
$~~~~~~~~~~~~~~~~~~$\subtable[]{
\centering
\begin{tabular}{|c|c|c|c|c|c|c|c|c|}
\hline ${\bf X}^{(00)}_{1}$ & ${\bf X}^{(01)}_{1}$ & ${\bf X}^{(10)}_{1}$ &
${\bf X}^{(11)}_{1}$
 \\
 \hline
0.411&0.256 &0.187 &0.009 \\
\hline
\end{tabular}
\label{tab2121b}
}
\caption{Numerical values of ${\bf X}^{(pq)}_{1}$ 
for lower lying modes of the nontrivial bilinear heterotypic gravidilatonic interaction for \subref{tab2121a} $k_{-}$ branch and \subref{tab2121b} $k_{+}$ branch. }
\label{tab2121}
\end{table}

\section{Summary and outlook}
In this article we have made a comprehensive study of string inspired warped geometry and it's phenomenological implications.
Our model is a perturbation of the RS model by Gauss-Bonnet coupling in five dimension which also includes the effect of string two loop 
correction in the gravity sector coming from the interaction with dilatonic degrees of freedom via the CFT disk amplitudes in the 
bulk ${\bf dS_{5}/AdS_{5}}$ geometry. Our study centered around three distinct aspects :-\\
\begin{itemize}
\item Determining the modified warp factor, the brane tensions and addressing the gauge hierarchy issue.\\
\item Study of different bulk fields and the profile of the wave functions to examine their overlap on the visible brane as well as various
KK mode masses for these bulk fields.\\
\item Examining the interaction with the brane fields to evaluate their possible signatures.\\ 
\end{itemize} 
We also compare our results with that obtained through the usual RS analysis.
Our results can be summarized as follows :\\
\begin{itemize}
\item For small GB coupling the warp factor turns out to be exponential with two different branches for the bulk parameters $k$ which we denote as $k_{\pm}$.
Moreover unlike the RS scenario, in our case $k_{\pm}$ depend on the bulk coordinate as well as the GB parameter $\alpha_{(5)}$. In addition a warped
solution can be obtained for both anti de-Sitter and de-Sitter bulk.\\
\item The gauge hierarchy problem can be resolved by appropriate choices of the parameters  $k_{\pm}$, $r_c$ and $\alpha_{(5)}$. The dependence of  $k_{\pm}$ on
$\alpha_{(5)}$ has been determined which indicates that the increase in the GB coupling decreases the value of $k_{\pm}$ leading to lesser warping between the
two branes unless one takes a larger value for the modulus $r_c$ to compensate the fall in the  value of  $k_{\pm}$. 
Also equation(\ref{planckmass}) implies that increase GB coupling causes increase in the effective 4-dimensional Planck scale from the pure RS scenario. 
The brane tensions also  increases with increase in $\alpha_{(5)}$ and finally reaches a saturation.\\

\item We have determined a stringent constraint on the GB coupling so that the required Planck to TeV
scale hierarchy can be achieved through the modified warp factor.  Most significantly for both the warping solutions the recently observed 
Higgs like boson at 125 GeV can be explained through our model for very small values of the Gauss-Bonnet coupling. 

\item We have evaluated and analyzed 
the zero mode and the KK mode excitations of bulk graviton along with the five dimensional ${\cal N}=1$ supergravity extension with bulk gravitino
from the bulk wave function. The characteristic features of graviton mass spectrum as well as the bulk wave functions 
are different for the two warping solutions for all possible signatures of Gauss-Bonnet coupling $\alpha_{(5)}$ and the string loop correction in two loop level.
In the limit $\alpha_{(5)}\rightarrow 0$ the negative warping branch produces the Randall-Sundrum features. The behavior of the 
mass spectrum for gravitino is almost similar to that of the graviton degrees of freedom.

\item We then extend our idea with bulk scalar as well as bulk gauge field by addressing both
abelian and non-abelian cases including dilaton coupling. It is a important finding of our model that while the zeroth mode bulk wave function for bulk scalar field and 
${\cal U}(1)$ abelian gauge field are exactly same as Randall-Sundrum model, the higher excited states are significantly different.
 Furthermore we have numerically estimated the values of the trilinear and quartic self interaction strength up to first excited state
in presence of Gauss-Bonnet coupling and string loop correction. 

\item Next we have studied the detailed features 
of the KK-spectrum of various higher rank antisymmetric tensor fields which are also possible candidates for bulk fields including the possible
dilaton-axion couplings which has a topological field theoretic origin. The bulk wave function for all such antisymmetric tensor fields follows
distinct features in presence of the two warping solutions. 

\item Following the similar prescription we
have analyzed the behavior of bulk fermions where the profile of both left
and right chiral modes are determined in presence of the GB extended gravity model in presence of dilaton and two loop conformal coupling. 
In this context we have estimated the trilinear interaction strength between the left/right chiral fermions and the ${\cal U}(1)$ abelian gauge fields.
Phenomenologically such values are very interesting and gives new informations in the context of TeV scale physics in presence of Gauss-Bonnet coupling and string loop correction. 
The behavior of left/right chiral fermions are significantly different for different warping solutions and different signatures of the couplings.
It is one of the important outcome of our model that the right and left chiral fermions are localized on the bulk and visible brane
respectively for the warping solution $A_{-}(y)$ . This establishes that among the two solutions, the $k_{-}$ branch is phenomenologically preferred over
the $k_{+}$ branch.
 
\item We have explicitly shown the detailed characteristic features of various
interactions among these bulk field contents by determining the numerical values of the coupling parameters. 
Such estimations are very very useful to understand the underlying physics of the
phenomenological model of a Einstein-Gauss-Bonnet warp geometry in presence of string loop corrections and dilaton couplings.

\item The profiles of different bulk fields apart from graviton are determined along with their KK mode masses. Since  $k_{\pm}$ decreases with
the GB coupling therefore the warping decreases and the KK mode masses of various bulk fields increase unless one introduces a little hierarchy
by taking a larger $r_c$ to resolve the gauge hierarchy problem.    
\item This brings out two possible scenario : 1) Due to fall in the value of  $k_{\pm}$, the warping decreases so that the requirement of Planck to TeV scale
warping to resolve the gauge hierarchy problem can not be met. However the couplings of the graviton KK modes with brane fields decreases and the masses of the
graviton KK mode increases which may lead to their escape from the present collider search, 2) If we change the value of $r_c$ to maintain the required hierarchy
then the coupling does not change from RS value but the KK mode masses decreases. The absence of any signature of graviton KK modes through their decay into dileptons
in ATLAS search at LHC therefore would signal the invalidity of the presence of GB couplings as a correction to RS warped geometry models.

\end{itemize}
   
Some interesting open issues in this context of the present study can be  to study the cosmological consequences of KK spectrum and detailed features of AdS/CFT correspondence 
for the GB coupled warped geometry  model.
The other possibility is to study the detailed bouncing cosmological features as well as its imprints on the Cosmic Microwave Background
via cosmological perturbation using the supergravity extension of our model. A detailed report
on this issue will be brought forth in future.


\section*{Acknowledgments}

SC thanks Council of Scientific and
Industrial Research, India for financial support through Senior
Research Fellowship (Grant No. 09/093(0132)/2010). SC also thanks 
Supratik Pal for his support and useful suggestions.





\begin{references}


\bibitem{lisa1} L. Randall and R. Sundrum, Phys. Rev. Lett. 83 (1999) 3370.

\bibitem{lisa2} L. Randall and R. Sundrum, Phys. Rev. Lett. 83 (1999) 4690.

\bibitem{lisa3} Y. Cui, L. Randall and B. Shuve, JHEP 1108 (2011) 073.

\bibitem{lisa4} C. Cheung, A. L. Fitzpatrick and L. Randall, JHEP 0801 (2008) 069.

\bibitem{lisa5} G. Perez and L. Randall, JHEP 01 (2009) 077.

\bibitem{lisa6} A. L. Fitzpatrick, G. Perez and L. Randall, Phys. Rev. Lett. 100 (2008) 171604.

\bibitem{lisa7} B. Lillie, L. Randall and Lian-Tao Wang, JHEP 0709 (2007) 074.

\bibitem{lisa8} A. L. Fitzpatrick, J. Kaplan, L. Randall and Lian-Tao Wang, JHEP 0709 (2007) 013.

\bibitem{lisa9} L. Randall, Y. Shadmi and N. Weiner,  JHEP 0301 (2003) 055.

\bibitem{lisa10} L. Randall and M. D. Schwartz, Phys. Rev. Lett. 88 (2002) 081801.

\bibitem{lisa11} L. Randall and M. D. Schwartz,  JHEP 0111 (2001) 003.

\bibitem{rizo1} T. G. Rizzo, Phys. Rev. D61 (2000) 055005.

\bibitem{rizo2} H. Davoudiasl, J. L. Hewett and T. G. Rizzo, Phys. Rev. Lett. 84 (2000) 2080.

\bibitem{rizo3} H. Davoudiasl, J. L. Hewett and T. G. Rizzo, Phys. Lett. B 473 (2000) 43.

\bibitem{rizo4} T. G. Rizzo, arXiv: hep-ph/0011139.

\bibitem{rizo5} H. Davoudiasl, J. L. Hewett and T. G. Rizzo, Phys. Rev. D 68 (2003) 045002.

\bibitem{rizo6} H. Davoudiasl, J. L. Hewett and T. G. Rizzo, JHEP 0304 (2003) 001.

\bibitem{rizo7} T. G. Rizzo, JHEP 0206 (2002) 056.

\bibitem{rizo8} J. L. Hewett, F. J. Petriello and T. G. Rizzo, JHEP 0209 (2002) 030.

\bibitem{rizo9} J. L. Hewett and T. G. Rizzo, JHEP 0308 (2003) 028.

\bibitem{rizo10} H. Davoudiasl, J. L. Hewett, T.G. Rizzo, JHEP 0308 (2003) 034.

\bibitem{rizo11} H. Davoudiasl, J. L. Hewett, B. Lillie and T. G. Rizzo, Phys. Rev. D 70 (2004) 015006.

\bibitem{rizo12} H. Davoudiasl, J. L. Hewett, B. Lillie and T. G. Rizzo,  JHEP 0405 (2004) 015.

\bibitem{rizo13} J. L. Hewett, B. Lillie and T. G. Rizzo, JHEP 0410 (2004) 014.

\bibitem{rizo14} T. G. Rizzo, JHEP 0501 (2005) 028.

\bibitem{rizo15} H. Davoudiasl, B. Lillie and T. G. Rizzo, JHEP 0608 (2006) 042.

\bibitem{rizo16} H. Davoudiasl, B. Lillie and T. G. Rizzo, arXiv: hep-ph/0509160.

\bibitem{rizo17} T. G. Rizzo, arXiv: hep-ph/0510420.

\bibitem{rizo18} T. G. Rizzo, Phys. Lett. B 647 (2007) 43.

\bibitem{rizo19} H. Davoudiasl, T. G. Rizzo and A. Soni,  Phys. Rev. D 77 (2008) 036001.

\bibitem{rizo20} H. Davoudiasl and T. G. Rizzo, JHEP 0811 (2008) 013.

\bibitem{rizo21} X. Calmet, P. de Aquino and T. G. Rizzo, Phys. Lett. B 682 (2010) 446.

\bibitem{rizo22} T. G. Rizzo, JHEP 1011 (2010) 156.

\bibitem{dal1} H. Davoudiasl, S. Gopalakrishna, E. Ponton and J. Santiago, New J. Phys. 12 (2010) 075011.

\bibitem{dal2} H. Davoudiasl, G. Perez and A. Soni, Phys. Lett. B 665 (2008) 67.

\bibitem{dal3} K. Agashe, H. Davoudiasl, S. Gopalakrishna, T. Han, Gui-Yu Huang, G. Perez, Zong-Guo Si and A. Soni,  Phys. Rev. D 76 (2007) 115015.

\bibitem{dal4} K. Agashe, H. Davoudiasl, G. Perez and A. Soni, Phys. Rev. D 76 (2007) 036006.

\bibitem{dal5} D. J. H. Chung, H. Davoudiasl and L. Everett, Phys. Rev. D 64 (2001) 065002.

\bibitem{sayan4} S. Choudhury and S. Pal, Nucl. Phys. B 857 (2012) 85.

\bibitem{gold1} W. D. Goldberger and M. B. Wise, Phys. Rev. D 60 (1999) 107505.

\bibitem{gold2} W. D. Goldberger and M. B. Wise, Phys. Rev. Lett. 83 (1999) 4922.

\bibitem{gold3} W. D. Goldberger and M. B. Wise, Phys. Lett. B 475 (2000) 275.

\bibitem{har} S. J. Huber and Q. Shafi, Phys. Lett. B 583 (2004) 293.

\bibitem{ssg4} R. S. Hundi, S. Roy and S. SenGupta, Phys. Rev. D 86 (2012) 036014.

\bibitem{ssg5} P. Dey, B. Mukhopadhyaya and S. SenGupta, Phys. Rev. D 81 (2010) 036011.

\bibitem{ssg6} S. SenGupta,  AIP Conf. Proc. 939 (2007) 236.

\bibitem{ssg7} S. Das, A. Dey and S. SenGupta, Europhys. Lett. 83 (2008) 51002.

\bibitem{ssg8} A. Dey, D. Maity and S. SenGupta, Phys. Rev. D 75 (2007) 107901.

\bibitem{ssg9} D. Maity, S. SenGupta and S. Sur, Phys. Lett. B 643 (2006) 348.

\bibitem{ssga} S. SenGupta,  arXiv: 0812.1092.

\bibitem{war} C. Csaki, Y. Grossman, P. Tanedo and Y. Tsai, Phys. Rev. D 83 (2011) 073002.

\bibitem{saki} C. Csaki, M. Graesser, L. Randall and J. Terning,  Phys. Rev. D 62 (2000) 045015.

\bibitem{kar} A. Karch and L. Randall, JHEP 0105 (2001) 008.

\bibitem{sayan1} S. Choudhury and S. Pal,  Phys. Rev. D 85 (2012) 043529.

\bibitem{sayan2} S. Choudhury and S. Pal, arXiv: 1209.5883. 

\bibitem{sayan3} S. Choudhury and S. Pal, arXiv: 1208.4433. 

\bibitem{safi1} S. J. Huber and Q. Shafi, Phys. Rev. D 68 (2003) 023503.

\bibitem{safi2} Q. Shafi, A. Sil and Siew-Phang Ng, Phys.Lett. B 620 (2005) 105.

\bibitem{safi3} B. Kyae and Q. Shafi, JHEP 0311 (2003) 036.

\bibitem{roy} R. Maartens and K. Koyama, Living Rev. Relativity 13, (2010), 5.

\bibitem{gary} G. N. Felder,  Int. J. Mod. Phys. A 17 (2002) 4297.

\bibitem{ssg1} P. Dey, B. Mukhopadhyaya and S. SenGupta, Phys. Rev. D 80 (2009) 055029.

\bibitem{ssg2} S. Das, D. Maity and S. SenGupta, JHEP 0805 (2008) 042.

\bibitem{ssg3} S. Lahiri and S. SenGupta, arXiv: 1204.0886.

\bibitem{fed1} G. N. Felder, A. Frolov and L. Kofman, Class. Quantum Grav. 19 (2002) 2983.

\bibitem{saba} C. Csaki, J. Erlich and C. Grojean, Nucl. Phys. B 604 (2001) 312.

\bibitem{gio} J. M. Cline, J. Descheneau, M. Giovannini and Jeremie Vinet, JHEP 06 (2003) 048.

\bibitem{hana} V. Barger, T. Han, T. Li, J. D. Lykkena and D. Marfatia, Phys. Lett. B 488 (2000) 97.

\bibitem{lhc} Large Hadron Collider collaboration, {\em http://lhc.web.cern.ch/lhc/}.

\bibitem{ssg10} B. Mukhopadhyaya, S. Sen and S. SenGupta, Phys. Rev. Lett. 89 (2002) 121101.

\bibitem{ssg11} B. Mukhopadhyaya, S. Sen and S. SenGupta, Phys. Rev. D 65 (2002) 124021.

\bibitem{ssg12} B. Mukhopadhyaya, S. Sen and S. SenGupta, Phys. Rev. D 76 (2007) 121501.

\bibitem{ssg13}  B. Mukhopadhyaya, S. Sen and S. SenGupta, Phys. Rev. D 79 (2009) 124029.

\bibitem{ssg14} R. Koley, J. Mitra and S. SenGupta, Europhys. Lett. 85 (2009) 41001.

\bibitem{ssg15} J. Mitra and S. SenGupta, Phys. Lett. B 683 (2010) 42.

\bibitem{ssg16} A. Das and S. SenGupta, Phys. Lett. B 698 (2011) 311.

\bibitem{ssg17} A. Das and S. SenGupta, arXiv: 1204.1154.

\bibitem{ssg18} R. S. Hundi and S. SenGupta,  arXiv: 1111.1106.

\bibitem{ssg19} N. Banerjee, S. Lahiri and S. SenGupta, arXiv: 1106.1735.

\bibitem{ssg20} B. Mukhopadhyaya, S. Sen and S. SenGupta, arXiv: 1106.1027.

\bibitem{ssg21} A. Das, R. S. Hundi and S. SenGupta, Phys. Rev. D 83 (2011) 116003.

\bibitem{ssg22} R. Koley, J. Mitra and S. SenGupta, Europhys. Lett. 91 (2010) 31001.

\bibitem{ssg23} R. Koley, J. Mitra and S. SenGupta, Phys. Rev. D 78 (2008) 045005.

\bibitem{ssg24} D. Choudhury and S. SenGupta, Phys. Rev. D 76 (2007) 064030.

\bibitem{klev} D. Baumann, A. Dymarsky, I. R. Klebanov, J. Maldacena, L. McAllister and A. Murugan, JHEP 0611 (2006) 031.

\bibitem{tony} T. Gherghetta, arXiv: 1008.2570.

\bibitem{rick} R. Rattazzi and A. Zaffaroni, JHEP 04 (2001) 021. 

\bibitem{sund} M. A. Luty and R. Sundrum, Phys. Rev. D 65 (2002) 066004.

\bibitem{anco}  L. Anchordoqui,, J. Edelstein, C. Nuñez, S. P. Bergliaffa, M. Schvellinger, M. Trobo5 and F. Zyserman, Phys. Rev. D 64 (2001) 084027. 

\bibitem{aga} K. Agashe and A. Delgado, Phys. Rev. D 67 (2003) 046003.  

\bibitem{hakg} S. W. Hawking, T. Hertog and H. S. Reall, Phys. Rev. D 62 (2000) 043501. 

\bibitem{brag} H. B. Filho, N. R. F. Braga and C. N. Ferreira,  Phys. Rev. D 73 (2006) 106006. 

\bibitem{kaya} N. S. Deger and A. Kaya, JHEP 05 (2001) 030.

\bibitem{lotu} M. A. Luty and T. Okui, JHEP 09 (2006) 070.

\bibitem{giov} M. Giovannini,  Phys. Rev. D 63 (2001) 064011.

\bibitem{andr} A. Karch and L. Randall,  Phys. Rev. Lett. 87 (2001) 061601.

\bibitem{pen} I. P. Neupane, Class. Quantum Grav. 28 (2011) 125015.

\bibitem{mav} N. E. Mavromatos and J. Rizos, Int. J. Mod. Phys. A 18 (2003) 57. 

\bibitem{dot} G. Dotti and R. J. Gleiser, Phys. Lett. B 627 (2005) 174.

\bibitem{sayan6}  S. Choudhury and S. Pal, arXiv: 1210.4478.

\bibitem{gari} Jaume Garriga and Takahiro Tanaka, Phys. Rev. Lett. 84 (2000) 2778.

\bibitem{gasp} M. Gasperini, Elements of String Cosmology, Cambridge University Press, New York, 2007.

\bibitem{aseem} A. Paranjape, S. Sarkar and T. Padmanabhan, Phys. Rev. D 74 (2006) 104015.

\bibitem{turner} S. M. Carroll1, A. De Felice, V. Duvvuri, D. A. Easson, M. Trodden and M. S. Turner, Phys. Rev. D 71 (2005) 063513.

\bibitem{cogo} G. Cognola1, E. Elizalde, S. Nojiri, S. D Odintsov and Sergio Zerbini, JCAP 02 (2005) 010.

\bibitem{fara} T. P. Sotiriou and V. Faraoni, Rev. Mod. Phys. 82 (2010) 451. 

\bibitem{boul} D. G. Boulware and S. Deser, Phys. Rev. Lett. 55 (1985) 2656.  

\bibitem{lid} James E. Lidsey and N. J. Nunes, Phys. Rev. D 67 (2003) 103510.

\bibitem{ish} I. P Neupane1 and B. M. N. Carter,  JCAP 06 (2006) 004.

\bibitem{sami1} J. F. Dufaux, J. E. Lidsey, R. Maartens and M. Sami, Phys. Rev. D 70 (2004) 083525.

\bibitem{sami2} S. Tsujikawa, M. Sami and R. Maartens, Phys. Rev. D 70 (2004) 063525.

\bibitem{whel}  J. T. Wheeler, Nucl. Phys. B 268 (1986) 737.

\bibitem{yaji}  T. Torii, H. Yajima and Kei-ichi Maeda, Phys. Rev. D 55 (1997) 739. 

\bibitem{ben}  B. M Leith and I. P. Neupane,  JCAP 05 (2007) 019.

\bibitem{kei} Kei-ichi Maeda and T. Torii, Phys. Rev. D 69 (2004) 024002.

\bibitem{tret} S. Nojiri, S. D. Odintsov and P. V. Tretyakov, Phys. Lett. B 651 (2007) 224.

\bibitem{kanti} P. Kanti, J. Rizos and K. Tamvakis, Phys. Rev. D 59 (1999) 083512. 

\bibitem{glei} G. Dotti and R. J. Gleiser, Class. Quantum Grav. 22 (2005) L1.

\bibitem{dota} G. Dotti and R. J. Gleiser, Phys. Rev. D 72 (2005) 044018.

\bibitem{peri}  M. Gasperini,  Phys. Rev. D 56 (1997) 4815.

\bibitem{zho} Y. Zhong, Yu-Xiao Liu and K. Yang, Phys. Lett. B 699 (2011) 398.

\bibitem{rami} R. G Daghigh, G. Kunstatter and J. Ziprick, Class. Quantum Grav. 24 (2007) 1981.

\bibitem{wmap9} WMAP collaboration, D. N. Spergel et al.,  Astrophys. J.  Suppl. {\bf 170}, 377 (2007);
for uptodate results on WMAP, see {\em
http://lambda.gsfc.nasa.gov/product/map/dr5}.

\bibitem{planck} Planck collaboration, {\em http://www.rssd.esa.int/index.php?project=Planck} ,
some early results are also available, see, for example, P. A. R.
Ade et.al., arXiv:1101.2022.

\bibitem{kony} K. Konya, Phys. Rev. D 75 (2007) 104003.

\bibitem{zkon} R. A. Konoplya and A. Zhidenko, Phys. Rev. D 82 (2010) 084003.

\bibitem{davbou} D. G. Boulware and S. Deser, Phys. Lett. B 175 (1986) 409.

\bibitem{pari} M. Parikh, Phys. Rev. D 84 (2011) 044048.

\bibitem{kasi} T. Kobayashi, General Relativity and Gravitation 37 (2005) 1869.

\bibitem{alex} A. Barrau, J. Grain and S. Alexeyeva, Phys. Lett B 584 (2004) 114.

\bibitem{marteen} R. Maartens, Prog. Theor. Phys. Suppl. 148 (2003) 213.

\bibitem{magmic} M. Maggiore, Nucl. Phys. B 525 (1998) 413.

\bibitem{cve} M. Cvetič, , S. Nojiri, , S. D. Odintsov, Nucl. Phys. B 628 (2002) 295.

\bibitem{roma} R. Konoplya, Phys. Rev. D 71 (2005) 024038.

\bibitem{marb}  M. Beroiz, G. Dotti and R. J. Gleiser, Phys. Rev. D 76 (2007) 024012. 

\bibitem{past} D. Anninos and G. Pastras, JHEP 07 (2009) 030.

\bibitem{rong} R. G. Cai, Phys. Lett. B 582 (2004) 237.

\bibitem{koko} R. A. Konoplya and A. Zhidenko, Phys. Rev. D 77 (2008) 104004. 

\bibitem{chan} H. C. Kim and R. G. Cai, Phys. Rev. D 77 (2008) 024045.

\bibitem{wil} D. L. Wiltshire, Phys. Rev. D 38 (1988) 2445.

\bibitem{sen1} A. Sen, JHEP 03 (2006) 008.

\bibitem{qi} R. G. Cai and Q. Guo, Phys. Rev. D 69 (2004) 104025. 

\bibitem{genr}R. G. Cai, Phys. Rev. D 65 (2002) 084014.

\bibitem{sen2} B. Sahoo and A. Sen, JHEP 01 (2007) 010.

\bibitem{alisa} M. Alishahiha and H. Ebrahim, JHEP 11 (2006) 017.

\bibitem{sen3} A. Sen, JHEP 07 (2005) 073.

\bibitem{james1} J. E. Lidsey, S. Nojiri and S. D. Odintsov, JHEP 06 (2002) 026.

\bibitem{fin} F. Larsena and F. Wilczek, Nucl. Phys. B 458 (1996) 249.

\bibitem{ohta} Kei-ichi Maeda, N. Ohta and Y. Sasagawa, Phys. Rev. D 80 (2009) 104032. 

\bibitem{tims} T. Clunan, S. F. Ross and D. J. Smith, Class. Quantum Grav. 21 (2004) 3447.

\bibitem{nijori} G. Cognola, E. Elizalde, S. Nojiri, S. D. Odintsov and S. Zerbini, Phys. Rev. D 73 (2006) 084007.

\bibitem{nijori1} S. Nojiri and S. D. Odintsov, Phys. Lett. B 631 (2005) 1.

\bibitem{luca} L. Amendola, C. Charmousis and S. C. Davis, JCAP 12 (2006) 020.

\bibitem{hide} H. Maeda and M. Nozawa, Phys. Rev. D 77 (2008) 064031. 

\bibitem{motu} T. Koivisto and D. F. Mota, Phys. Rev. D 75 (2007) 023518.

\bibitem{pres} M. Cvitan, S. Pallua and P. Prester, Phys. Lett. B 555 (2003) 248. 

\bibitem{pal} M. Cvitan, S. Pallua and P. Prester, Phys. Lett. B 546 (2002) 119.

\bibitem{binpan} Q. Pan and B. Wang, Phys. Lett. B 693 (2010) 159.

\bibitem{defeli} G. Calcagni, B. de Carlos and A. De Felice, Nucl. Phys. B 752 (2006) 404.

\bibitem{jili} Q. Pan, J. Jing and B. Wang, JHEP 11 (2011) 088.

\bibitem{jeans} P. Binétruy, C. Charmousis, Stephen Davis and J. F. Dufaux, Phys. Lett. B 544 (2002) 183.

\bibitem{rich} R. Easther and Kei-ichi Maeda, Phys. Rev. D 54 (1996) 7252.

\bibitem{nick} N. E. Mavromatos and J. Rizos, Phys. Rev. D 62 (2000) 124004.  

\bibitem{gava} I. Antoniadis, E. Gava and K. S. Narain, Phys. Lett. B 283 (1992) 209.

\bibitem{nem} R. Altendorfer, J. Bagger and D. Nemeschansky, Phys. Rev. D 63 (2001) 125025.

\bibitem{tony3} T. Gherghetta and A. Pomarol, Nucl. Phys. B 602 (2001) 3. 

\bibitem{poko} A. Falkowski, Zygmunt Lalak and Stefan Pokorski, Phys. Lett. B 491 (2000) 172.

\bibitem{ssgn} B. Mukhopadhyaya, S. Sen, S. Sen, S. SenGupta, Phys.Rev. D70 (2004) 066009.

\end{references}
\end{document}